\documentclass[longauth]{aa}
\usepackage{amsmath}	% Advanced maths commands
 \usepackage{amssymb}	% Extra maths symbols
\usepackage{natbib}
\usepackage{graphicx}

\newcommand{\HII}{\textrm{H~{\textsc{ii}}}}
\newcommand{\jwst}{\textit{JWST}}

\newcommand{\mum}{$\mu$m\xspace}

\usepackage{color}

\definecolor{cbpurple}{rgb}{0.47, 0.37, 0.94}

\definecolor{violet}{rgb}{1.0, 0., 0.80}

\begin{document}

\title{PDRs4All VI: Probing the Photochemical Evolution of PAHs in the Orion Bar Using Machine Learning Techniques}
\titlerunning{PDRs4All VI: PAH evolution in the Orion Bar}

   \author{Sofia Pasquini \inst{1} \and 
          Els Peeters \inst{1, 2, 3} \and
          Bethany Schefter \inst{1,2} \and
          Baria Khan \inst{1} \and
          Ameek Sidhu \inst{1,2} \and
          Ryan Chown \inst{1,2}\and
          Jan Cami \inst{1,2,3} \and 
          Alexander Tielens \inst{4,5} \and 
          Felipe Alarc\'on \inst{6} \and
          Am\'elie Canin \inst{7} \and
          Ilane Schroetter \inst{7} \and
          Boris Trahin \inst{8} \and
          Dries Van De Putte \inst{9} \and
          Christiaan Boersma \inst{10} \and
          Emmanuel Dartois \inst{11} \and
          Takashi Onaka \inst{12} \and
          Alessandra Candian \inst{13} \and 
          Patrick Hartigan \inst{14} \and
          Thomas S.-Y. Lai \inst{15} \and
          Ga\"el Rouill\'e \inst{16} \and
          Dinalva A. Sales \inst{17} \and
          Yong Zhang \inst{18} \and
          {E}milie Habart\inst{8} \and
          Olivier Bern\'{e} \inst{7} 
 }         
        
\institute{Department of Physics \& Astronomy, The University of Western Ontario, London ON N6A 3K7, Canada; \email{epeeters@uwo.ca}  \and 
Institute for Earth and Space Exploration, The University of Western Ontario, London ON N6A 3K7, Canada   \and 
Carl Sagan Center, SETI Institute, 339 Bernardo Avenue, Suite 200, Mountain View, CA 94043, USA \and
Leiden Observatory, Leiden University, P.O. Box 9513, 2300 RA Leiden, The Netherlands         \and
Astronomy Department, University of Maryland, College Park, MD 20742, USA        \and
Department of Astronomy, University of Michigan, 1085 South University Avenue, Ann Arbor, MI 48109, USA         \and
Institut de Recherche en Astrophysique et Plan\\'etologie, Universit\'e Toulouse III - Paul Sabatier, CNRS, CNES, 9 Av. du colonel Roche, 31028 Toulouse Cedex 04, France         \and 
Institut d'Astrophysique Spatiale, Universit\'e Paris-Saclay, CNRS,  B$\hat{a}$timent 121, 91405 Orsay Cedex, France        \and 
Space Telescope Science Institute, 3700 San Martin Drive, Baltimore, MD 21218, USA        \and
NASA Ames Research Center, MS 245-6, Moffett Field, CA 94035-1000, USA         \and
Institut des Sciences Mol\'eculaires d'Orsay, Universit\'e Paris-Saclay, CNRS,  B$\hat{a}$timent 520, 91405 Orsay Cedex, France         \and
Department of Astronomy, Graduate School of Science, The University of Tokyo, 7-3-1 Bunkyo-ku, Tokyo 113-0033, Japan         \and
Anton Pannekoek Institute for Astronomy, University of Amsterdam, The Netherlands \and
Department of Physics and Astronomy, Rice University, Houston TX, 77005-1892, USA          \and
IPAC, California Institute of Technology, Pasadena, CA, USA          \and
Laboratory Astrophysics Group of the Max Planck Institute for Astronomy at the Friedrich Schiller University Jena, Institute of Solid State Physics, Helmholtzweg 3, 07743 Jena, Germany         \and
Instituto de Matem\'atica, Estat\'istica e F\'isica, Universidade Federal do Rio Grande, 96201-900, Rio Grande, RS, Brazil         \and
School of Physics and Astronomy, Sun Yat-sen University, 2 Da Xue Road, Tangjia, Zhuhai 519000,  Guangdong Province, China      }

   \date{October 2023}

% \abstract{}{}{}{}{} 
% 5 {} token are mandatory
 
  \abstract
  %  context heading (optional)
   {Extraordinary observations of the Orion Bar by \jwst\ have shown, for the first time, the incredible richness of PAH emission bands and their variation on very small scales. These variations are the result of photochemical evolution of the PAH carrier.  
   }
  %{} leave it empty if necessary  
   % aims heading (mandatory)
    {We aim to probe the photochemical evolution of PAHs across the key zones of the ideal photodissociation region (PDR) that is the Orion Bar using unsupervised machine learning.}
   % methods heading (mandatory)
    {We use \jwst\ NIRSpec IFU and MIRI MRS observations of the Orion Bar from the \jwst\ Early Release Science Program PDRs4All (ID: 1288). We lever bisecting k-means clustering to generate highly detailed spatial maps of the spectral variability in the $3.2-3.6$, $5.95-6.6$, $7.25-8.95$, and $10.9-11.63$~\mum wavelength regions. We analyse and describe the variations in the cluster profiles and connect them to the conditions of the physical locations from which they arise. We interpret the origin of the observed variations with respect to the following key zones: the \HII\ region, the atomic PDR zone, and the layers of the molecular PDR zone stratified by the first, second, and third dissociation fronts (DF~1, DF~2, and DF~3, respectively).}
   % results heading (mandatory)
    {Observed PAH emission exhibits spectral variation that is highly dependent on spatial position in the PDR. We find the $8.6$~\mum band to behave differently than all other bands which vary systematically with one another. Notably, we find uniform variation in the $3.4-3.6$~\mum bands and $3.4/3.3$ intensity ratio. We attribute the carrier of the $3.4-3.6$~\mum bands to a single side group attached to very similarly sized PAHs. Further, cluster profiles reveal  a transition between characteristic profiles classes of the $11.2$~\mum feature from the atomic to the molecular PDR zones. We find the carriers of each of the profile classes to be independent, and reason the latter to be PAH clusters existing solely deep in the molecular PDR. Clustering also reveals a connection between the $11.2$ and $6.2$~\mum bands; and that clusters generated from variation in the $10.9-11.63$~\mum region can be used to recover those in the $5.95-6.6$~\mum region.}
   % conclusions heading (optional), leave it empty if necessary 
    {Clustering is a powerful and comprehensive tool for characterising PAH spectral variability on both spatial and spectral scales. For individual bands as well as global spectral behaviours, we find ultraviolet-processing to be the most important driver of the evolution of PAHs and their spectral signatures in the Orion Bar PDR.}


   \maketitle
%
%________________________________________________________________

\defcitealias{boersma2016}{B2016}
\defcitealias{sidhu2021}{S2021}

\section{Introduction}
\label{sec:intro}

  Infrared (IR) spectra are dominated by a rich set of emission features at wavelengths ranging from around 3 to 20 $\mu$m, referred to as Aromatic Infrared Bands (AIBs). The most prominent of these features have been identified predominantly at 3.3, 6.2, 7.7, 8.6, 11.2, 12.7, and 16.4 $\mu$m. A wealth of weaker features have been observed including at 3.4, 3.5, 5.25, 5.76, 6.0, 6.9, 10.5, 11.0, 12.0, 13.5, 14.2, 15.8, 17.0, 17.4, and 17.8 $\mu$m as well. These features are commonly attributed to polycyclic aromatic hydrocarbons (PAHs) or related species \citep{sellgren1984, leger1984, allamandola1985, allamandola1989}. For a detailed spectroscopic inventory in the Orion PDR, we refer the reader to \citet{Chown:23}. 
 PAHs are pervasive in the Universe and their signature broad emission features dominate the IR emission of almost all UV-visible illuminated interstellar environments. PAHs reside predominantly in photo-dissociation regions (PDRs) where the physics and chemistry of the gas are driven by far-ultraviolet (FUV; $6-13.6$~eV) photons \citep{verstraete1996sws, genzel1998powers, moutou1999upper, hony2001ch, geers2006c2d, armus2007observations}. Indeed, these emission bands arise from the vibrational emission cascade of PAHs upon absorption of such photons \citep{leger1984, allamandola1985, sellgren1983, hony2001ch, allamandola1985, allamandola1989}.\par 

Previous studies revealed the presence of distinct variations in the profiles of the main PAH bands, which led to their classification in classes A, B, C and D and transitions between them \citep{peeters2002, vanDiedenhoven2004, Sloan:07, Matsuura:14}. These studies also revealed that largely but not unequivocally classifications of one PAH band had predictive values for the classification of other PAH bands. \par 

\citet{peeters2002} identified classes A, B, and C for the $6.2$, $7.7$ and $8.6$~\mum bands largely based on their peak positions. Class A represent the profiles with the bluest peak position, class C those with the reddest peak position, and class B profiles with intermediate peak positions. 
Class A and B for the $7.7$~\mum complex also showed a dominant $7.6$~\mum component and a dominant $7.8$~\mum component, respectively. Those which do not show a typical $7.7$~\mum complex are assigned to class C and peak near 8.2~\mum. A fourth Class, D, shows broad emission peaking near 8~\mum \citep{Matsuura:14}.
\citet{vanDiedenhoven2004} extended the classification to the $3.3$ and $11.2$~\mum bands. Class $\mathrm{A_{3.3}}$ profiles are symmetric, with more blue peak positions whereas class $\mathrm{B_{3.3}}$ profiles are asymmetric and have peak positions towards the red. They identify class $\mathrm{A_{11.2}}$ profiles if they peak between $11.20-11.24$~\mum whereas those profiles belonging to the $\mathrm{B_{11.2}}$ category peak instead at $\sim11.24$~\mum. Both studies also found that in most cases, objects which belong to class $\mathrm{A}$ ($\mathrm{B}$ or $\mathrm{C}$) for one band, also belong to $\mathrm{A}$ ($\mathrm{B}$ or $\mathrm{C}$) for the other bands. Moreover, the profile classes also strongly depend on object type and ISM-type sources exhibit class A profiles \citep{peeters2002, vanDiedenhoven2004}.

PAH emission characteristics are coupled to the physical conditions of the regions in which they reside. Parameters describing the local physical conditions, such as the FUV radiation field strength, gas density, and gas temperature  determine the molecular attributes of the local PAH population \citep{bakes1994, galliano2008, pilleri2012evaporating, sidhu2022, knight2022}. In turn, these changes drive variability in the observed PAH emission signatures \citep{Joblin1996, berne2007analysis, pilleri2012evaporating, boersma2016, Candian2014, peeters2017, Bauschlicher2008, bauschlicher2009, Ricca2012, hony2001ch}. A photochemical evolution of the underlying population of emitting PAH species must give rise to this observed variability in PAH emission signatures. \par 

The effect of physical changes to the emitting PAH population on their observed spectral signatures are well realized, however our understanding of the mechanism for these changes is limited. This is due mainly to the highly blended nature of these spectral signatures, and to combinations of insufficient sensitivity, spatial resolution and spectral resolution of the IR facilities used to observe them historically. The \textit{James Webb Space Telescope} (\jwst) is able to resolve, with an unrivaled spatial resolution, the highly-blended spectral signatures of PAHs in the IR wavelength regime. Hence, $\jwst$ is ideally suited to improve our understanding of the mechanisms of the photochemical evolution of PAHs.\par 

As part of the PDRs4All \textit{JWST} Early Release Science (ERS) program $1288$, we use high spatial resolution spectroscopic data to directly probe the photochemical evolution of PAHs. We use IR observations of a nearby PDR, the Orion Bar, which is an ideal source for such a study. This well-studied, nearby star-forming region is known to exhibit strong PAH emission and is oriented in space in an edge-on fashion (see Fig.~\ref{orion_illustration}). We exploit the high spatial resolution of these \jwst\ observations to study the key physical zones within the PDR in great detail.\par

In addition to having access to this ideal \jwst\ dataset, we lever the powerful pattern recognition abilities of an unsupervised machine learning algorithm to expand our understanding of PAH emission variability in the Orion Bar. Unsupervised machine learning is a branch of machine learning which involves the use of algorithms that do not require labels on input data sets. That is to say that these algorithms work to extract underlying relationships between features they detect on their own. Due to the nature of such algorithms, unsupervised techniques are ideal for analyses in which the outcome or results are largely unknown. Unsupervised machine learning algorithms can therefore act as powerful tools which can, when employed thoughtfully, reveal underlying information in data sets such as PAH emission spectra which may not be extracted by classical analysis techniques alone. \par 

Machine learning methodologies in general have become increasingly popular among astronomers \citep[e.g.][]{rhea2021, gaetan2020, davies2019, tammour2016, meng2023}. For a detailed overview of machine learning techniques employed in astronomy, we refer the reader to \citet{baron2019}. While these methods are relatively new to studying PAH emission variations, some recent studies have explored using unsupervised machine learning as a technique to probe the characteristics of PAH variability through their spectral emission features \citep[e.g.][]{rapacioli2005, berne2007analysis, sidhu2021,sidhu2022,zhang2019, boersma2014}. 
In line with our own study, \citet{boersma2014} applied a traditional k-means clustering algorithm to PAH emission spectra in the PDR of NGC 7023. \citet{zhang2019} used clustering on the spatial structures of PAH species maps for the $7-9$~\mum emission region in NGC 2023 to understand the population carrying these bands. \citet{sidhu2021, sidhu2022} have used Principal Component Analysis (an unsupervised dimensionality reduction technique) to probe drivers in the variability of observed PAH bands in various reflection nebulae. \par 

We aim to expand on the goals of these studies and establish the promise of bisecting k-means clustering (and clustering algorithms in general) as a powerful tool for PAH spectral analyses. In Sect. 2, we describe the observations, data reduction and pre-processing steps, and we outline the bisecting k-means algorithm itself. In Sect. 3, we detail the goals and methodology of the analysis. In Sect. 4, we present the results from each of our clustering experiments in detail, and discuss their implications in Sect. 5. Finally, we summarize our results and highlight the performance of our unsupervised machine learning algorithm as a tool for PAH analysis in Section 6.\par

 \section{The Orion Bar}
 
The Orion Nebula is arguably the best-studied \HII\ region to date, illuminated by the brightest member of the Trapezium cluster, $\theta^1$ Ori C, an O6V type star with $T_{\rm{eff}}$=38 950K \citep{odell2008, odell2017}. Centered on the Trapezium cluster is a large ionized cavity, beyond which a molecular cloud is located that is composed of gas and dust that are bright at mid-infrared (MIR) wavelengths. Beyond this, along the outer boundary of the Orion Nebula, is a large, expanding shell of neutral gas which is being driven outwards by stellar winds emanating from the Trapezium cluster \citep{pabst2019}. Part of this PDR boundary is the Bar itself, which is a compressed shell, observed edge-on \citep{salgado2016}. The Bar is often referred to as the ``Bright Bar'' or ``Orion Bar'' \citep[e.g.][]{Elliott1974,Tielens93,ODell:20}. In the following, we name it the ``Bar''. Within the Bar, gas densities have been found to range from a few $10^4$ cm$^{-3}$ in the atomic PDR \citep{Tielens93, bernard-salas2012} to $0.5 - 1.0\times 10^5$ cm$^{-3}$ within the ambient molecular PDR \citep{TielensHollenbach1985, goicoechea2017,bernard-salas2012, Hoger95}. The maximum strength of the FUV radiation field (G$_0$) varies between $2.2$ and $7.1\times 10^4$ at the ionization front (IF), with a median value of $5.9\times 10^4$ \citep{Habart:im, Peeters:nirspec}.\par 

\begin{figure}
     \centering
     \includegraphics[width = \linewidth]{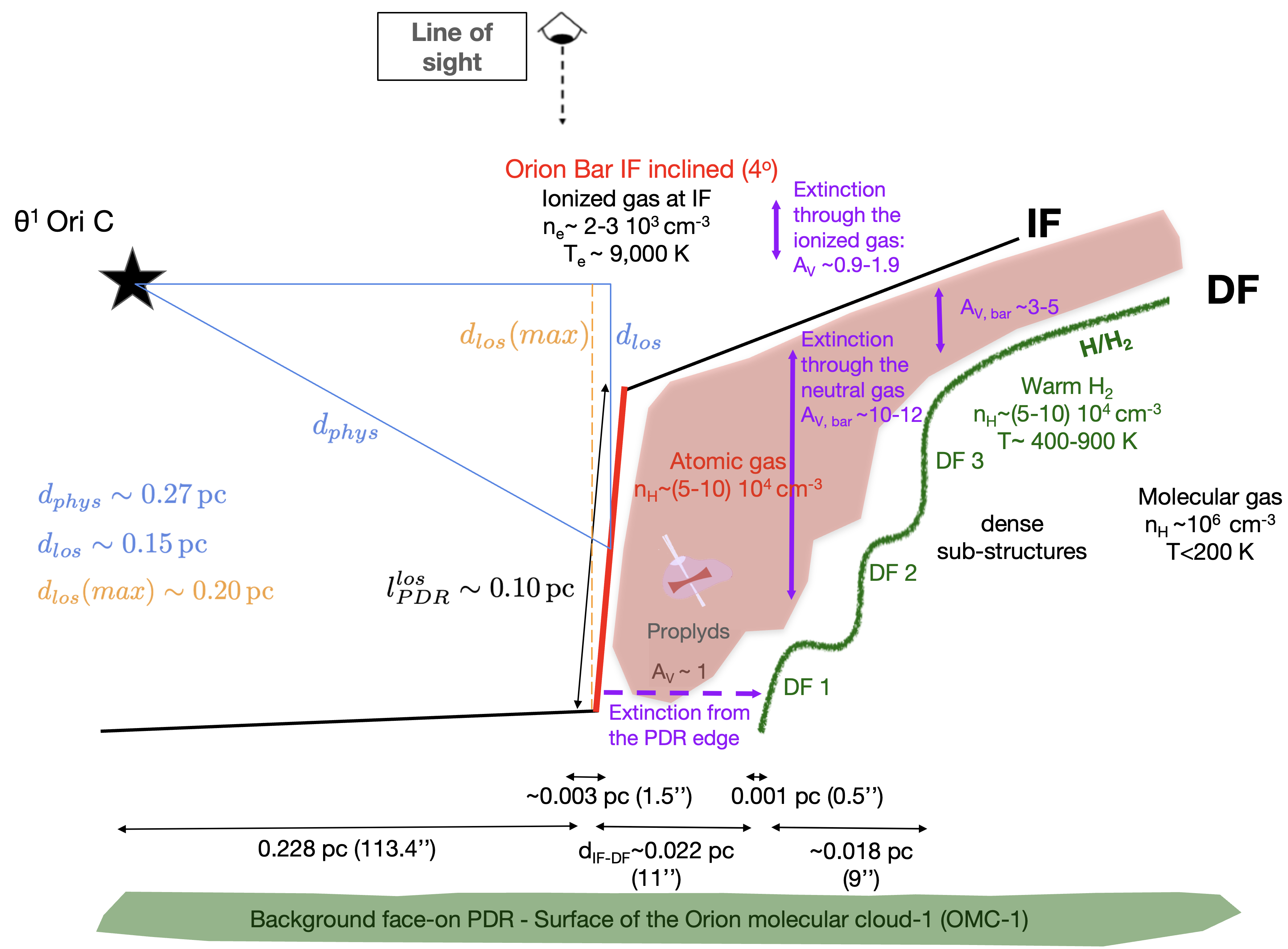}
     \caption{ Schematic sketch of the anatomy of the Bar. The dimensions perpendicular to the bar are not to scale but are given at the bottom of the figure. We also note that  foreground material is not depicted. This includes a layer of ionised gas \citep{ODell:20} and the Veil \citep[e.g.,][]{Rubin2011, Boersma2012, vanderWerf13, Pabst19, Pabst20}. Figure taken from \citet{Peeters:nirspec}. }
     \label{orion_illustration}
 \end{figure}

At a distance of 0.235~pc from the ionizing source, $\theta^1$ Ori C, the IF is very sharp (approximately 0.005~pc wide) and is well traced by the [FeII] 1.644~\mum and [OI] 6300 \AA \, emission \citep[e.g.][]{Tielens93, bernard-salas2012,Herrmann97, Ossenkopf13}. The IF marks the boundary between the ionized cavity (predominantly \HII\ gas) and the PDR. After this point, the gas is predominantly neutral, atomic H, and emission from low ionization potential energy species such as C, S, and Fe in the near-infrared (NIR) regime and from aromatic particles (such as PAHs) in the MIR is observed. A transition from predominantly neutral to molecular hydrogen takes place approximately 15$''$ ($\sim$0.03\,pc) from the IF due to attenuation of the FUV photon flux. We observe three consecutive ridges in the Bar which are the edges of this transition zone, the dissociation front, abbreviated as ``DF'' \citep{Habart:im, Peeters:nirspec}. These ridges can be likened to a terraced-field-like structure with three steps that we view edge-on. We refer to each of them as DF~1, DF~2, and DF~3 (according to the nomenclature established in \citet{Habart:im}) in order of increasing distance from the IF, respectively (see Fig.~\ref{orion_illustration}). Observations have not yet pinpointed the exact location for the transition zone from neutral to molecular hydrogen \citep[e.g.,][]{Tauber95,Wyrowski97,Cuadrado19,Salas19}, though \citet{Habart:im} and \citet{Peeters:nirspec} find this transition zone to coincide most closely with DF~2. The density of the gas in the molecular PDR is \mbox{$n_{\rm H}$\,$\simeq$\,10$^5$-10$^6$\,cm$^{-3}$} \citep{Habart:im, Peeters:nirspec}. This spatial stratification between ionized, atomic and neutral, and molecular gas has been extensively studied and confirmed by several series of IR and radio observations \citep[e.g.][]{habart2023,Peeters:nirspec, Tielens1993, goio2016} and are illustrated in Fig.~\ref{orion_illustration}.\par 

 The Bar is an ideal probe of the PDR environment as its edge-on orientation and close proximity allows for probing of the stratified PDR morphology in order to investigate the photoprocessing of the gas and dust in relation to spatial proximity to the ionizing source \citep[e.g.][]{cesarsky2000, goicoechea2015, knight2021}. The Bar is thus an ideal source for studying the photochemical evolution of PAHs. Further, as the Bar is so well-studied, it acts as a well-calibrated control upon which to derive new and reliable results using novel, high-quality technology (i.e. $\jwst$).\par

 \begin{figure*}
     \centering
     \includegraphics[width = \linewidth]{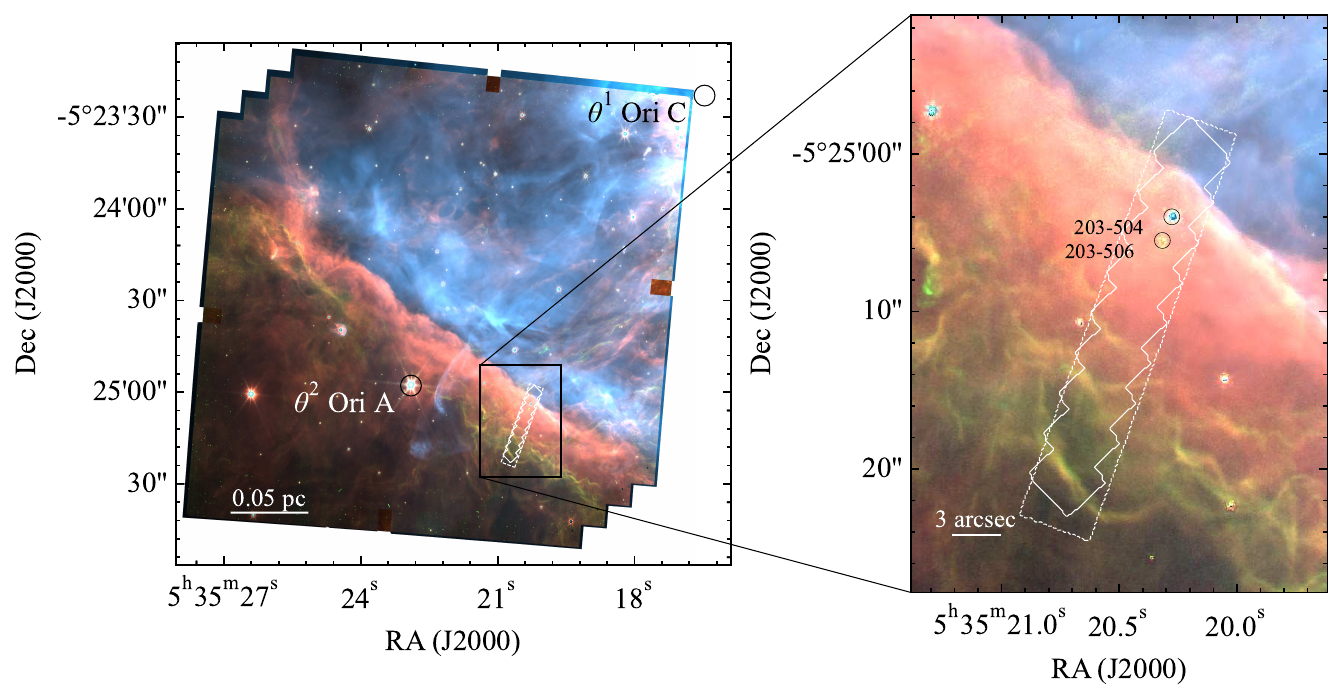}
     \caption{The PDRs4All FOVs for both NIRSpec 
  IFU and MIRI MRS datasets are illustrated with the solid and dashed white lines, respectively. The FOVs are overlaid on a composite NIRCam image of the Bar constructed from PDRs4All imaging data \citep{Habart:im}, where red, green, and blue are encoded as F335M ($3.3$~\mum aromatic infrared band emission), F470N-F444W (H$_2$ emission), and F187N (Paschen $\alpha$ emission), respectively. The location of the ionizing source, $\theta^1$ Ori C, is shown with the white circle in the top right corner of the left panel. The two proplyds, 203-504 and 203-506, are outlined with black circles. Figure adapted from \citet{Chown:23}.  }
 \label{fig:FOV} 
 \end{figure*}

 \section{Methodology}

 \subsection{The Data}

 We use the 0.97-28.3~\mum spectroscopic observations of the Bar taken with $\jwst$ as part of the PDRs4All Early Release Science (ERS) program \citep[ID 1288;][]{pdrs4all, Peeters:nirspec, Chown:23, lettergas}. The NIRSpec mosaic is a $9\times1$ mosaic taken with the integral field unit (IFU) mode at high spectral resolution ($R\approx2700$) from $0.9$ to $5.27\,\mu$m and centred on $\alpha$ = $05$:$35$:$20.4749$ and $\delta$ = $-05$:$25$:$10.45$ with a position angle of $43.74^\circ$ \citep{boker2022}. The MIRI observations were taken with the medium-resolution spectroscopy (MRS) observing mode to produce a $9\times1$ pointing mosaic with a spectral resolution $R$ ranging from $\sim 1700$ to $\sim 3700$. The MIRI MRS mosaic is positioned to overlap with that of the PDRs4All \jwst\ NIRSpec IFU observations \citep[][]{Peeters:nirspec}. The field of view (FOV) for both mosaics is shown in Fig.~\ref{fig:FOV}. 

 The NIRSpec data were taken from \citet{Peeters:nirspec}. The MRS data were re-reduced using version 1.11.1 of the JWST pipeline\footnote{\url{https://jwst-pipeline.readthedocs.io/en/latest/}}, and JWST Calibration Reference Data System\footnote{\url{https://jwst-crds.stsci.edu/}} (CRDS) context jwst\_1097.pmap. We refer the reader to \citet{Peeters:nirspec}, \citet{Chown:23}, and \citet{lettergas} for a detailed description of the data reduction, calibration and mosaic stitching techniques. 
 
 \subsection{Data Pre-Processing}

 As we aim to investigate PAH evolution, we isolate the PAH emission features we know to be associated with PAHs, from our data. We subtract the continuum emission component and remove spectral lines from each spectrum in the dataset. We use the same line list for removal of spectral lines not attributed to PAHs as was used in \citet{Chown:23}, and therefore refer the reader to \citet{Peeters:nirspec} and \citet{lettergas} for these complete inventories corresponding to the NIRSpec and MIRI MRS datasets, respectively. The continuum components of the spectra from both the NIRSpec and MIRI MRS datasets were the same as those used in \citet{Peeters:nirspec, Chown:23}, hence we refer readers to \citet{Peeters:nirspec, Chown:23} for a detailed description of how the continuum components were derived for both datasets. For the 7.7~\mum complex, we used an alternative continuum (a linear continuum instead of a spline continuum) to improve the continuum estimate in the \HII\ region.  Edge pixels in both NIRSpec and MIRI datatsets often have unreliable data, so we remove these. We also mask out all pixels coinciding with proplyds 203-504 and 203-506 (see Fig.~\ref{fig:FOV}) to focus our analysis on the PAH emission at the various depths in the Orion Bar PDR. We make use of the MIRI MRS data shortward of $13.2$~\mum and the NIRSpec data between $3.2$ and $3.6$~\mum.\par 

 The intensity of the AIB emission varies considerably with distance from the Trapezium cluster \citep{Habart:im, Peeters:nirspec, Chown:23}. We find that clustering results without normalization by total integrated intensity trace solely the variations in the strength of observed AIB emission. To prevent this effect from dominating the cluster assignment (see Section ~\ref{sec:the_algorithm} for a description of cluster assignment), we therefore normalize all continuum-subtracted spectra, cleared of lines, by their total integrated intensity (over the wavelength range considered) prior to clustering.  We will refer to these final spectra as cleaned, standardized spectra. \par 

 \subsection{The Clustering Algorithm}
 \label{sec:the_algorithm}

 We employ a bisecting k-means clustering algorithm for this study. We use the open-source Python implementation of the algorithm in the Scikit-Learn library \citep{scikit-learn}. This clustering scheme is hierarchical in nature as it combines the strategies of divisive and k-means clustering algorithms to perform cluster assignment. In this context, clusters are interpreted as groups of pixels in a given dataset whose spectra share similar properties. Going forwards, this algorithm may be referred to as ``the clustering algorithm''. \par 

 Bisecting k-means clustering applies a divisive hierarchical clustering scheme (described in detail in \citet{savaresi2001performance, steinbach2000comparison, wang1998content, zhao2005hierarchical}) to the cleaned, standardized spectra. Each spectrum in the dataset is assigned a label which designates it as a member of a given cluster. The labelling process is done in a top-down approach which begins by treating the entire sample of spectra as a single cluster before iteratively splitting the clusters further. In each iteration of splitting, the cluster which has the highest variance between members is split into two (``bisecting''). The splitting of a given cluster into two is done via k-means \citep[e.g.][]{forgy1965cluster} clustering, which involves an assignment of cluster members such that the total Euclidean distance between all points and the cluster centroid (the data point at the center of a cluster) are minimized. For this specific use case, distance between two spectra can be interpreted as Euclidean distance between two n-dimensional vectors where each spectrum is a vector; each wavelength position an n$^{\mathrm{th}}$ dimension and the intensity there the magnitude of that `vector' component. K-means clustering is also described in detail in \citet{boersma2014}. This process is repeated until either each observation is the only member of each cluster or a preset number of clusters, specified by the user, has been reached. The optimal number of clusters for a given dataset can be determined using a number of analytic methodologies, though whether the optimal number of clusters for a given dataset has been found is outside the scope of this study.\par

As we do not set out to characterize or classify the PAH emission profiles in the Bar, we employ heuristic methods, such as consulting distortion curves and silhouette scores, strictly to support our analyses. We consult these metrics to know the quality of our clustering applications only as it supports our understanding of the variation in the local PAH population which we probe. \par 

In order to quantify the quality of cluster assignments throughout our experiments, we fit our clustering algorithm for a range of total clusters and examine the relationship between the distortion within clusters as a function of the number of clusters. Distortion is calculated as the average sum of squared Euclidean distances between each cluster member and corresponding centroid. The inflection point on this curve (known as the ``elbow point'') is used to identify the optimal number of clusters detected by the clustering algorithm within the dataset. \par 

We also use a silhouette score ($S$) to quantify the quality of our clustering. The silhouette score is a value ranging from 0 to 1 that quantifies the goodness of cluster assignment and is given by Eq.~\ref{sil_score}, where $a$ is the average distance between each point within a cluster and $b$ is the average distance between all clusters.

\begin{equation}
 \centering
     S = \frac{b - a}{\mathrm{max}(a, b)}
     \label{sil_score}
 \end{equation}

The silhouette score for a set of clusters close to 0 indicates the clusters are indifferent from one another, while a score close to 1 indicates that clusters are well-distinguished from one another.  \par 

 Bisecting k-means is a popular algorithm for many unsupervised learning applications such as document retrieval, pattern recognition and image analysis due to its use of hierarchical binary taxonomy \citep{zhao2005hierarchical, steinbach2000comparison}. It is also ideal for this science application for multiple reasons. Its hierarchical nature enables us to probe increasingly specific groups of spectra based on both large-scale and more subtle variability. Bisecting k-means also offers improvements over basic k-means clustering in both computational power and the ability to recognize irregularly shaped (non-circular) clusters, especially in cases where the number of clusters is much smaller than the sample size \citep[e.g.][]{zhao2005hierarchical,steinbach2000comparison, bangoria2014enhanced, ristoski2015mining}. It is important to note, however, that bisecting k-means does not resolve a fundamental shortcoming of the k-means clustering algorithm, that is, the general convergence on a solution regardless of the dataset \citep[e.g.][]{selim1984k, banerjee2015empirical, di2018bisecting}.\par

 In order to focus cluster assignments on changes in spectral profiles of specific PAH features or regions, we only apply the clustering algorithm to selected wavelength ranges: $3.2-3.6$, $5.95-6.6$, $7.25-8.95$, and $10.9-11.63$ \mum to cover the main vibrational emission bands at 3.3, 6.2, 7.7, and 11.2 \mum, respectively. For each of these experiments pertaining to a single PAH emission band (the 3.3, 6.2, 7.7,  and 11.2 $\mathrm{\mu m}$ wavelength regions), a single round of clustering is applied to the input dataset. In the case of the experiment using all the MIRI MRS data shortward of 13.2 $\mathrm{\mu m}$, we apply an initial round of clustering to the data with a total of three clusters specified. Three clusters are chosen for this initial round of clustering to capture the \HII\, region, the atomic PDR, and the molecular PDR zones within the MIRI MRS footprint with the intention of focusing subsequent rounds of clustering on the atomic PDR zone. We isolate pixels identified by the clustering algorithm that belong to the atomic PDR where the PAH emission dominates. 
 We examine the resulting clusters and identify the cluster which coincides spatially most closely with the atomic PDR. We then mask out all pixels which do not belong to this cluster, and apply a secondary round of clustering to this subset of pixels only. We refer to this specific experiment as ``Atomic PDR-focused'' clustering.\par 

 For each experiment, once clustering has been applied, the global properties of the spectra belonging to each cluster are analysed by constructing the PAH emission profiles which belong to each cluster. This is done by averaging the intensity at each wavelength position across all cleaned, standardized spectra assigned to a given cluster. In the same way, the standard deviation at each wavelength position is calculated for all cleaned, standardized spectra belonging to the same cluster as a measure of the dispersion within the cluster spectra. Differences between these spectral profiles are then directly compared. We also examine the spatial locations of the pixels belonging to each cluster with respect to physical structures in the Bar such as the IF, and each of the three dissociation fronts.\par 

\begin{figure*}
    \centering
    \resizebox{.99\hsize}{!}{%
    \includegraphics[clip,trim =2.5cm 0cm 2.2cm 1.5cm, width = 6cm ]
    {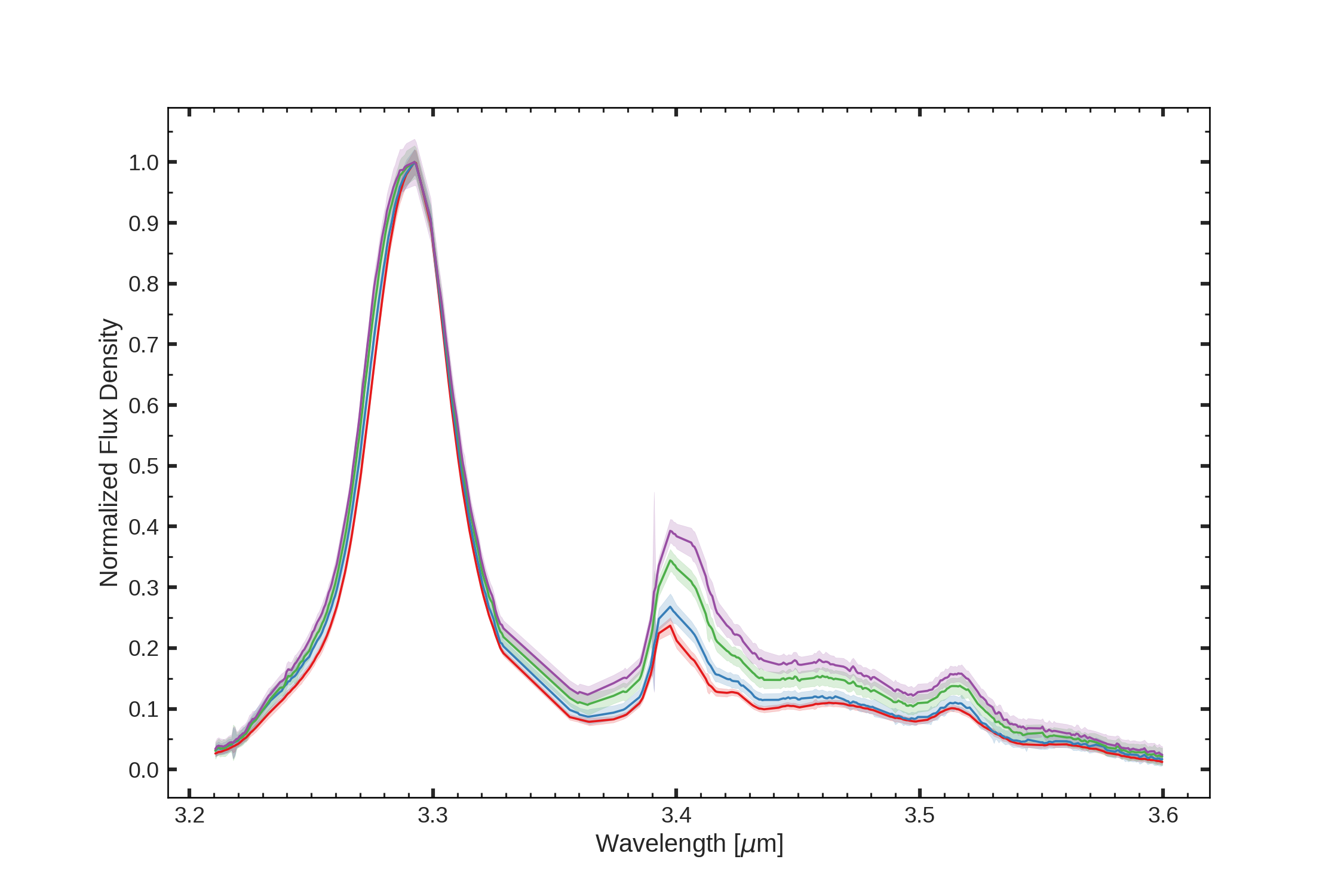}
    \includegraphics[width = 2.8cm]{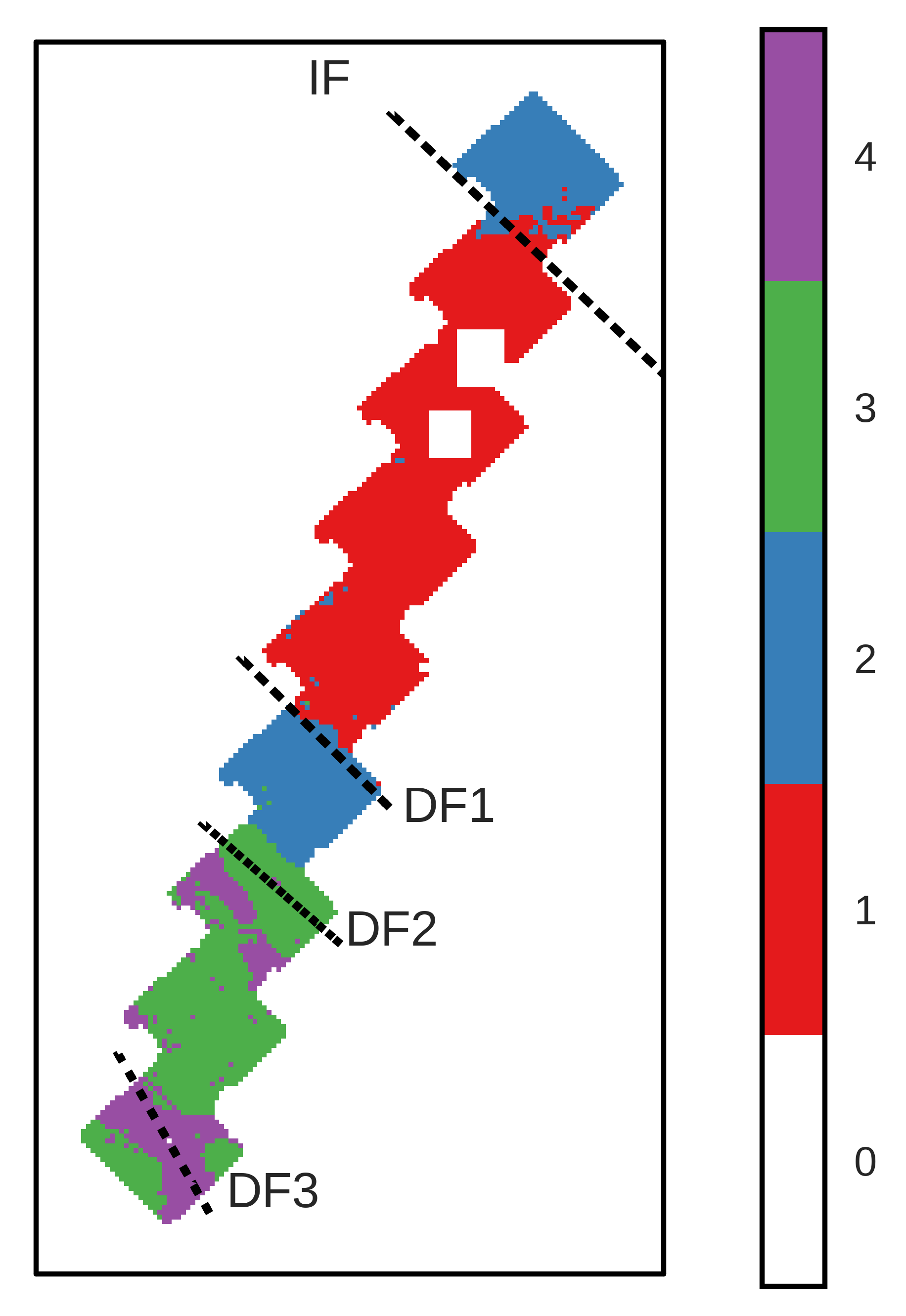}
 }
    \caption{The average spectral profile (left) and spatial footprint (right) for all clusters determined in the $3.2-3.6$~\mum region. Each cluster is labelled with an integer numbered 1 through 4 (in an arbitrary manner). Left panel: The shaded regions around each profile illustrate one standard deviation from the mean profile intensity at a given wavelength value. Each spectrum is normalized to its peak 3.3 $\mathrm{\mu} m$ intensity for visualization purposes. Right panel: The spatial footprint of the NIRSpec FOV color-coded by cluster assignment. Any masked pixels or those which have been otherwise masked out are labelled with 0. The black, dashed lines indicate the locations of the ionization front (IF) and the three dissociation fronts (DF~1, DF~2 and DF~3) as defined in \citet{Peeters:nirspec} using nomenclature established in \citet{habart2023}. }
    \label{fig:results_33}
\end{figure*}

\section{Results}
\label{sec:results}

We use clustering as a tool to probe the variation of observed PAH emission in both spatial and spectral dimensions. We emphasize that we do not set out to use the clustering algorithm to \textit{classify} PAH emission into any discrete number of classes. Instead, we study this variability to further our understanding of the (photochemical) evolution of the underlying PAH population. At the same time, we test the abilities of employing such an unsupervised machine learning technique for spectral analysis of PAH emission. We do so by examining the clustering results (Sect.~\ref{sec:results}) while considering those revealed by traditional spectral analysis methods (Sect.~\ref{discussion}). \par 

We report the results for each of the $3.3$, $11.2$, $6.2$ and $7-9$ ~\mum region cluster experiments, using a total of four clusters each. For clustering applied with four clusters, the average silhouette scores vary between $0.30$ and $0.52$ for the above wavelength regions. We found the elbow of the distortion curves for each wavelength range to occur at four clusters (except that of the $7-9$~\mum region which occurs at five clusters) and the corresponding silhouette scores to be acceptable for the purposes of our analysis. Therefore, we compare the results from each experiment with four clusters and give give the results from clustering with higher numbers of clusters in Appendix~\ref{app:more_clusters}. We also report the results from the atomic PDR-focused clustering experiment for the entire MIRI MRS wavelength range shortward of 13.2~\mum. For each wavelength range experiment, we show the elbow plots in Appendix~\ref{app:nr_clusters} and the silhouette scores and elbow points for each in Table \ref{scores_table}. \par

\begin{figure*}
    \centering
    \resizebox{.99\hsize}{!}{%
    \includegraphics[clip,trim =2.5cm 0cm 2.2cm 1.5cm, width = 6cm]{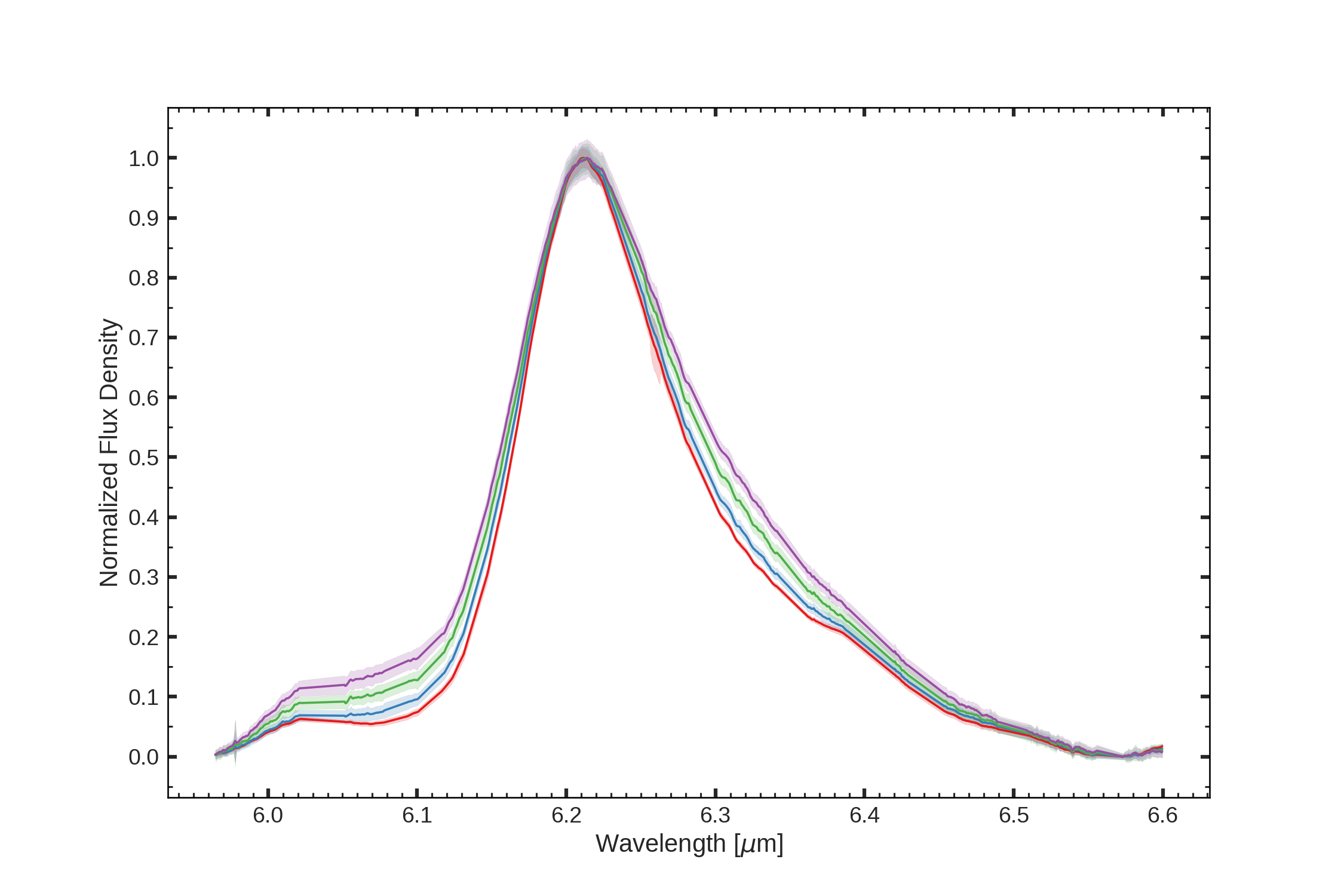}
    \includegraphics[width = 2.8cm]{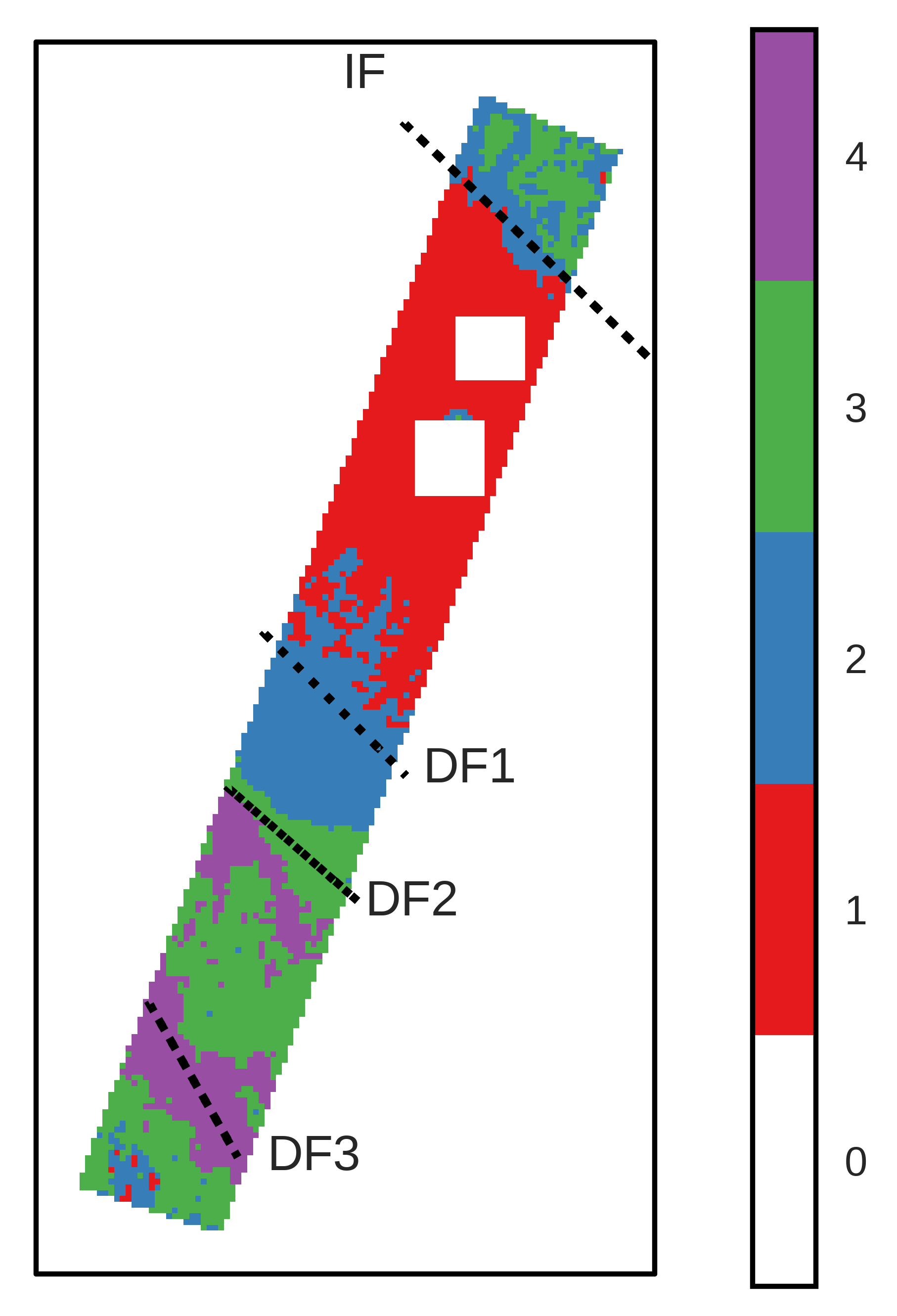}
 }
    \caption{The average spectral profile (left) and spatial footprint (right) for all clusters determined in the 5.95-6.6 ~\mum region. Each cluster is labelled with an integer numbered 1 through 4 (in an arbitrary manner). Left panel: The shaded regions around each profile illustrate one standard deviation from the mean profile intensity at a given wavelength value. Each spectrum is normalized to its peak 6.2 $\mathrm{\mu} m$ intensity for visualization purposes. Right panel: The spatial footprint of the MIRI MRS FOV is color-coded by cluster assignment. Any masked pixels or those which have been otherwise masked out are labelled with a 0. The black, dashed lines indicate the locations of the IF, DF~1, DF~2 and DF~3.}
    \label{fig:results_62}
\end{figure*}

\subsection{The $3.2-3.6~\mathrm{\mu} m$ region}

Many PAH emission bands arise in the $3.2 - 3.6$~\mum region, the strongest of which being the $3.29$~\mum band. Weaker bands arising from PAHs include the $3.25$ and $3.33$~\mum components. We note that the $3.25$ and $3.33$~\mum bands are not as clearly detected in the averaged spectra shown in Fig.~\ref{fig:results_33} as the others \citep{Peeters:nirspec}. The 3.40~\mum band is composed of three sub-components centred at $3.395$, $3.403$, and $3.424$~\mum \citep{Peeters:nirspec} and sit on top of a broad plateau emission component in this region \citep{geballe1989, Sloan:97}. Unlike the aromatic components which make up the $3.3$~\mum band, these components arise from aliphatic side groups on PAHs \citep{Joblin1996, Maltseva:18, bernstein1996, pla2020, buragohain2020}. \par 

The $3.29$~\mum feature dominates the emission profile in this wavelength region, the $3.4$~\mum band being weaker.  The $3.29$ and $3.4$~\mum bands are the strongest features in the averaged profiles for each cluster. As such, these features largely affect the cluster assignment as seen in the average spectral profiles for each cluster (Fig.~\ref{fig:results_33}). Indeed, the cluster profiles clearly trace variations in the 3.4/3.3 peak intensity ratio. In addition, the cluster profiles probe the relative intensity in roughly the $3.33-3.6$~\mum interval with respect to the 3.29~\mum PAH intensity and show subtle variation in the strength of the blue wing of the 3.29~\mum band. These variations are linked with each other:  enhanced 3.4/3.29 peak intensity ratio corresponds to enhanced overall $3.33-3.6$~\mum emission with respect to the 3.29~\mum PAH intensity and enhanced width of the 3.29~\mum profile. \par 

Indeed, the clustering assignment is determined by the emission in the $3.4-3.6$~\mum wavelength interval. We re-applied the clustering to only the $3.29$~\mum feature by selecting only the emission in the $3.2-3.36$~\mum wavelength interval. We found no significant difference between the profiles of the $3.29$~\mum feature from this round of clustering and the results from clustering on the entire $3.2-3.6$~\mum range. The cluster zone map from this round of clustering can be found in Appendix \ref{app:more_clusters} (see Fig.~\ref{fig:33_exclusive}). The \HII\ region remains in its own cluster, however a portion of the molecular PDR behind DF~1 now joins the cluster which covers the atomic PDR. The slight broadening of the $3.29$~\mum feature with increasing distance from the IF as well as the variation in the strength of the blue wing are also observed by \citet{Chown:23} and \citet{Peeters:nirspec}.\par 

The spatial zones within the PDR to which each cluster profile corresponds are shown in Fig.~\ref{fig:results_33}. The four clusters form very regions in the NIRSpec FOV that are parallel to the IF and thus to the stratification within PDRs. Specifically, the largest region encompasses the atomic PDR and represents cluster 1. This cluster exhibits the smallest 3.4/3.29 peak intensity ratio, the smallest ($3.33-3.6$)/3.29 emission, and the smallest full width at half maximum (FWHM) of the 3.29~\mum PAH.  Subsequently, cluster 2 exhibits a slightly enhanced 3.4/3.29 peak intensity ratio with respect to cluster 4. Cluster 2 encompasses the pixels in the \HII\ region, along DF~1, and the region between DF~1 and DF~2.  Compared to cluster 2, cluster 3 exhibits again a slightly enhanced 3.4/3.29 peak intensity ratio and originates from DF~2, most of the region between DF~2 and DF~3, and the region beyond DF~3. Finally, cluster 4, with the largest 3.4/3.29 peak intensity ratio, traces DF~3 as well as a small region just behind DF~2. \par

\begin{figure*}
    \centering
    \resizebox{.99\hsize}{!}{%
    \includegraphics[width = 6cm, height = 3.6cm]{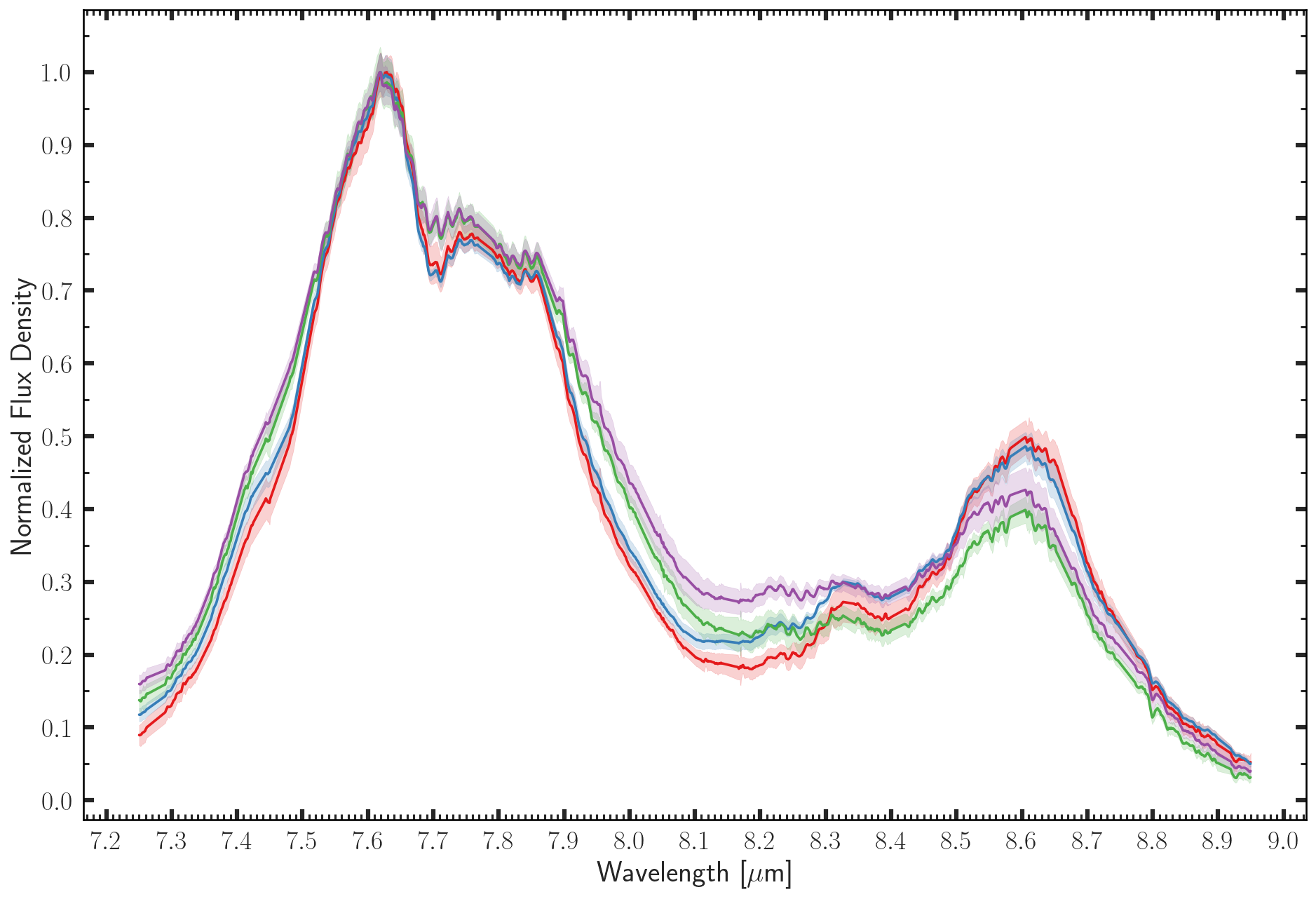}
    \includegraphics[width = 2.5cm]{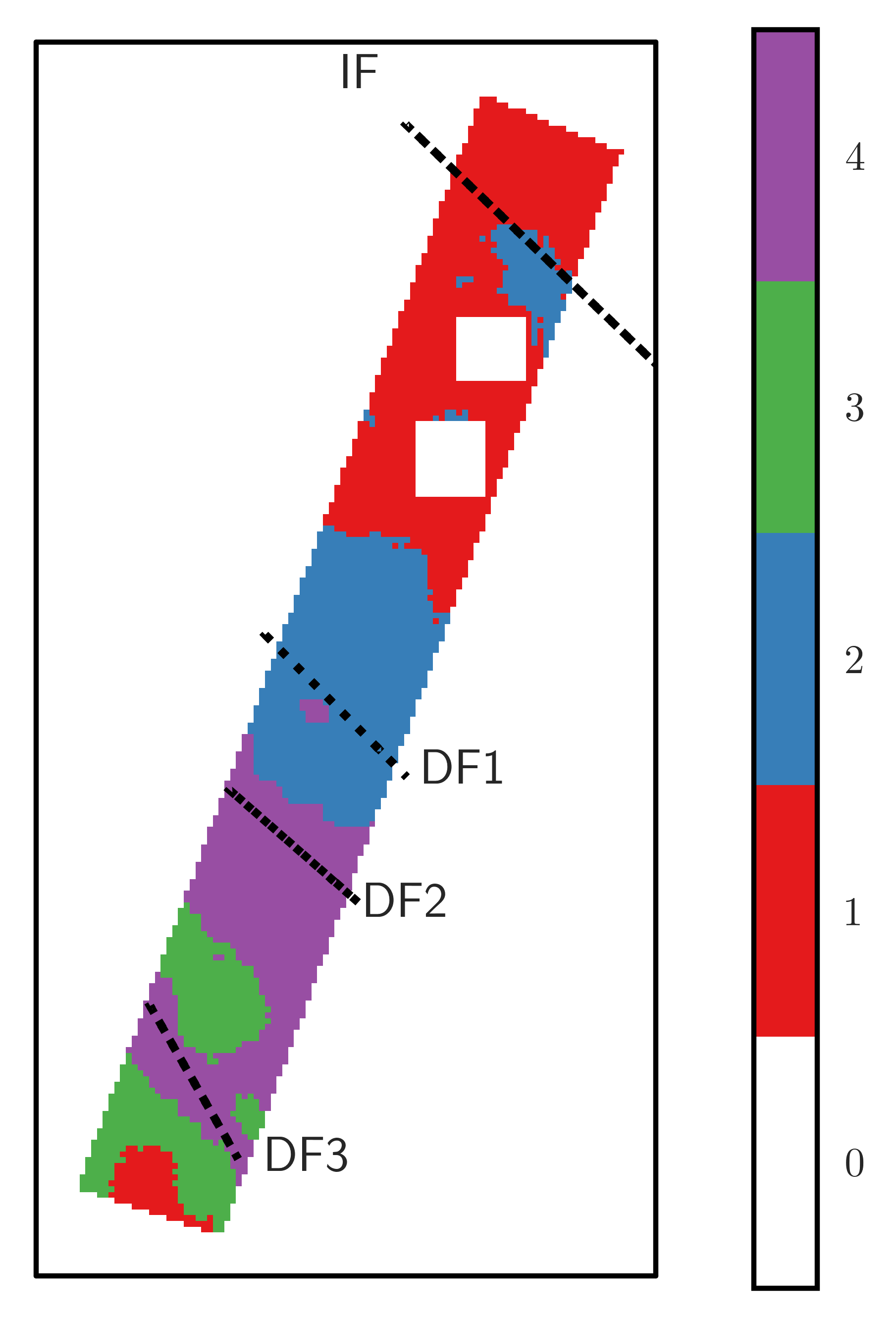}
 }
    \caption{The average spectral profile (left) and spatial footprint (right) for all clusters determined in the 7.25-8.95 ~\mum region. Each cluster is labelled with an integer numbered 1 through 4 (in an arbitrary manner). Left panel: The shaded regions around each profile illustrate one standard deviation from the mean profile intensity at a given wavelength value. Each spectrum is normalized to its peak 7.7 $\mathrm{\mu} m$ intensity for visualization purposes. Right panel: The spatial footprint of the MIRI MRS FOV is color-coded by cluster assignment. Any masked pixels or those which have been otherwise masked out are labelled with a 0. The black, dashed lines indicate the locations of the IF, DF~1, DF~2 and DF~3.}
    \label{fig:results_79}
\end{figure*}

\subsection{The $5.95-6.6 ~\mathrm{\mu} m$ region}

The 6.2 ~\mum band dominates PAH emission in the 5.95-6.6 ~\mum wavelength region, peaking at 6.212~\mum. Distinct weaker features also present themselves at 6.024 and 6.395 ~\mum \citep{Chown:23}, but are not strongly pronounced in the profiles of the mean cluster spectra presented in Fig.~\ref{fig:results_62}.\par 

The 6.2 ~\mum band very clearly governs the cluster assignment within this wavelength interval as it is the most prominent feature here. We do not observe any variation in the location of the peak intensity for the 6.2 ~\mum band between clusters, though variation in the width of this profile is present. Notably, the red wing of the 6.2 ~\mum band varies slightly more than what is observed for the blue wing. However, variation in the contribution from the 6.024~\mum component is observed across clusters. The 6.395 ~\mum component has a very subtle contribution to the 6.2 ~\mum profile which is present only in two of the mean cluster profiles. Once again, these spectral variations correlate with one another. As profiles decrease in width, the 6.024/6.2 intensity ratio also decreases but the prominence of the 6.395~\mum feature increases.\par 

We now consider each cluster in order of decreasing FWHM of the 6.2~\mum profile. First, cluster 4 has the widest profile with a barely noticeable 6.395~\mum component and highest 6.024/6.2 intensity ratio, and is found in DF~3 and a filament behind DF~2. Next, cluster 3 arises in part of the \HII\ region, in DF~2, in most of the region bounded by DF~2 and DF~3, and beyond DF~3.  Subsequently, cluster 2 which exhibits a noticeable 6.395~\mum component traces the IF and a zone which extends from a hard boundary just in front of DF~2, to a softer, more blended boundary with cluster 1 in front of DF~1. A few pixels at the farthest edge of the FOV from the ionizing source behind DF~3 as well as in the \HII\ region also belong to cluster 2. Lastly, cluster 1 has the narrowest 6.2 ~\mum band profile and most obvious presence of the 6.395 ~\mum feature. Cluster 1 covers most of the atomic PDR. The boundaries of the clusters assigned to this wavelength region are mostly parallel to the IF. However, more gradient-like transitions between zones are observed (see the transition between clusters 2 and 1 or between clusters 2 and 3 in the \HII\ region) for this wavelength region which are not observed for the 3.2-3.6 or 10.9-11.63 ~\mum regions.\par 

The pixelation of the boundaries between the clusters from the $6.2$~\mum wavelength band can be attributed to the fact that the changes probed by the clustering algorithm are (almost) solely the FWHM of the main $6.2$~\mum band and the variation in the (slight) broadening of this feature takes place very gradually. The pixelation may also be attributed to the nature of bisecting k-means clustering. The clustering algorithm operates under the assumption that all clusters are circular in shape, therefore those variations which deviate from this geometry will not result in well-defined cluster boundaries. If the variation between cluster zones is, in reality, not linearly separable, then the clustering algorithm in general will struggle to establish well-defined boundaries.\par

\subsection{The $7.25-8.95 ~\mathrm{\mu} m$ region}

There are multiple components to the 7.25-8.95 ~\mum PAH emission \citep{bregman1989, peeters2002, Cohen:southerniras:89}, the most prominent of which is the 7.7 ~\mum complex, followed by the 8.6 ~\mum band. The 7.626 ~\mum feature forms the main component of the 7.7 ~\mum complex, accompanied by other strong bands at 7.8 and 7.85 ~\mum. Weak features at 7.43,  8.223 and 8.330 ~\mum are also present, though subtle \citep{Chown:23}.\par 

These features are all observed in the mean spectral profiles for each of the clusters calculated in this wavelength region (Fig.~\ref{fig:results_79}). We observe four cluster profiles which vary most noticeably in the width of the 7.7 ~\mum profile and the 8.6/7.7 intensity ratio, as well as in the relative strength of the weaker 7.43 and  7.626 ~\mum components. As the width of the 7.7 ~\mum component decreases, the strength of the 8.6/7.7 intensity ratio increases. The presence of the 7.43~\mum component also varies between clusters, though independently from the width of the 7.7~\mum complex and the 7.8/7.626 intensity ratio.\par 

The spatial zones of each of these clusters are well-defined (Fig.~\ref{fig:results_79}). Cluster 4, which exhibits the widest 7.7~\mum profile, covers a large portion of the molecular PDR behind DF~2. This cluster zone is bounded by DF~2 and DF~3, with some pixels belonging to cluster 3 between these bounds. Cluster 4 exhibits the second weakest 8.6/7.7 band intensity ratio and the strongest 7.8/7.626 and 7.85/7.626 band intensity ratios. Cluster 4 also presents the most prominent 7.43/7.626 bump. 

Cluster 3 shows a $7.7$~\mum profile slightly narrower than that of cluster 4 and the weakest 8.6/7.7 and 8.33/7.7 intensity ratios. Cluster 3 is located beyond DF~3 and in two smaller, irregularly-shaped regions just in front of DF~3. 

The $8.223$~\mum component of cluster 3 coincides with that of cluster 2. Cluster 2 occupies most of the pixels in the molecular PDR behind DF~1 and in front of DF~2, as well as extends into the atomic PDR. The profile of cluster 2 has a more narrow $7.7$~\mum complex profile than that of cluster 3 but with an overlapping 8.223~\mum component. The 8.330/7.7 and 8.6/7.7 ratios, however, are distinctly enhanced. 

Cluster 1 coincides largely with the atomic PDR, with some pixels extending into the \HII\ region as well. The profile of cluster 1 shows the most narrow 7.7 ~\mum profile and highest 8.6/7.7 intensity ratio. Cluster 1 also shows the weakest 8.223/7.7 and second weakest 8.330/7.7 ~\mum intensity ratios. \par 

\begin{figure*}
    \centering
    \resizebox{.99\hsize}{!}{%
    \includegraphics[width = 6cm, height = 3.6cm]{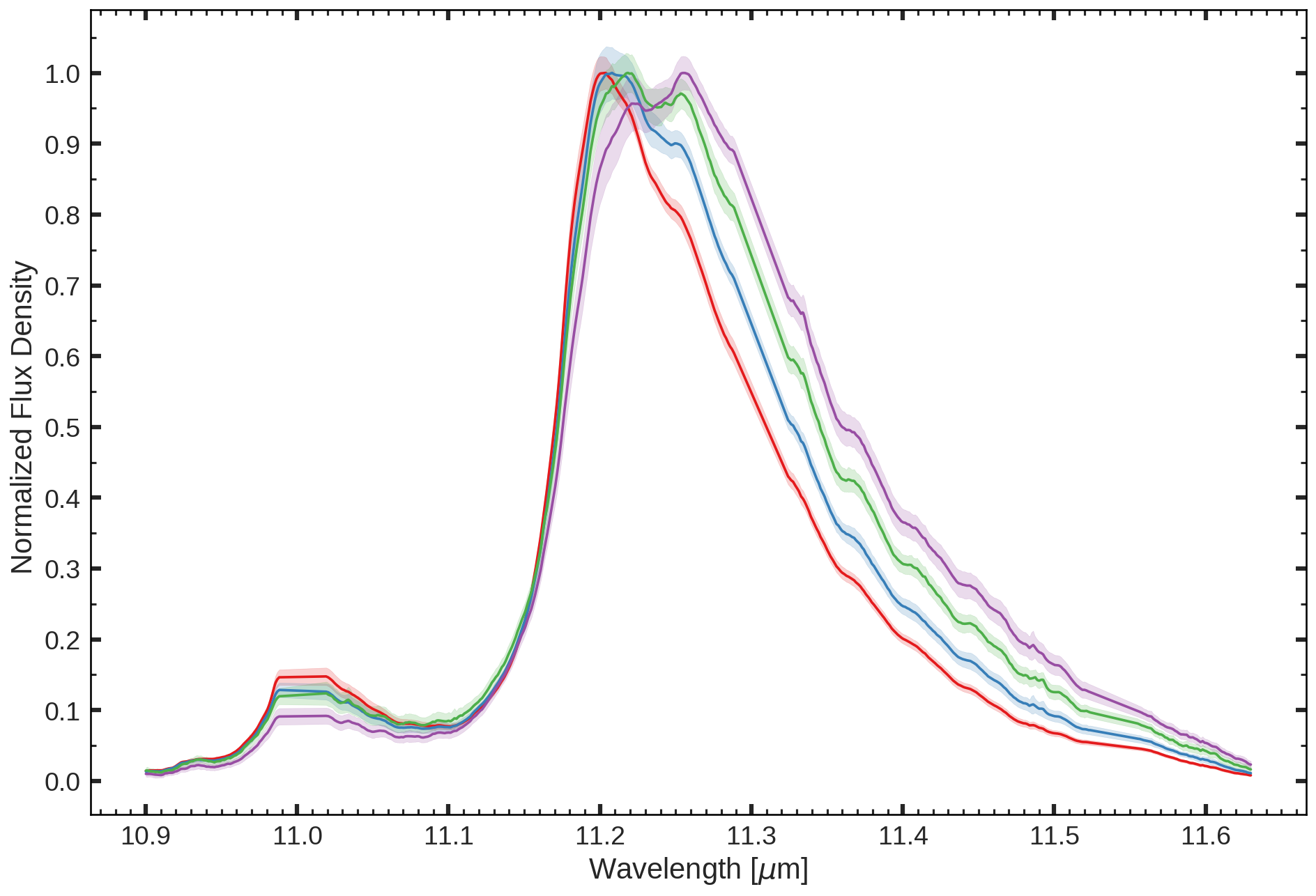}
    \includegraphics[width = 2.5cm]{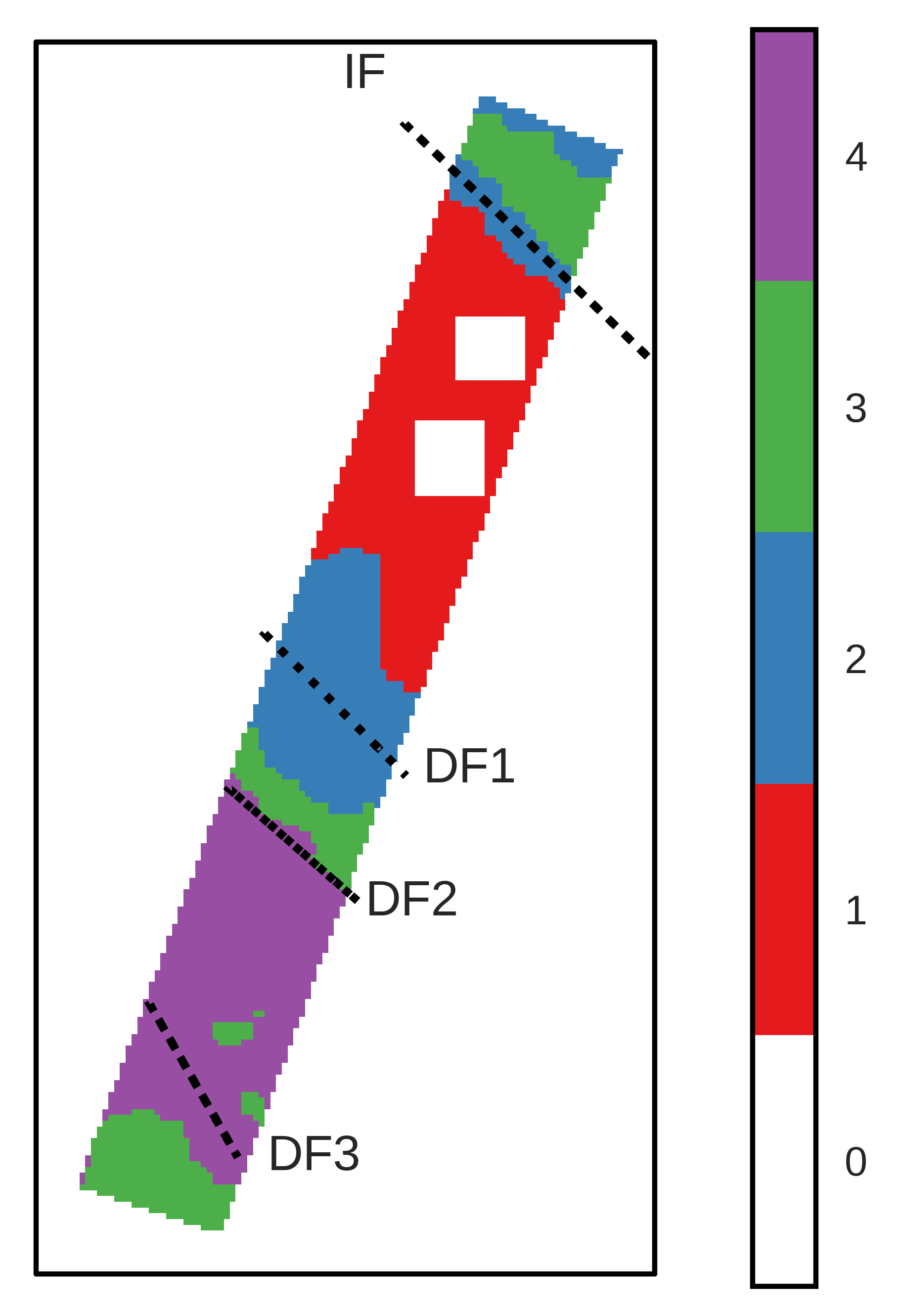}
 }
    \caption{The average spectral profile (left) and spatial footprint (right) for all clusters determined in the 10.9-11.63 ~\mum region. Each cluster is labelled with an integer numbered 1 through 4 (in an arbitrary manner). Left panel: The shaded regions around each profile illustrate one standard deviation from the mean profile intensity at a given wavelength value. Each spectrum is normalized to its peak 11.2 $\mathrm{\mu} m$ intensity for visualization purposes. Right panel: The spatial footprint of the MIRI MRS FOV is color-coded by cluster assignment. Any masked pixels or those which have been otherwise masked out are labelled with a 0. The black, dashed lines indicate the locations of the IF, DF~1, DF~2 and DF~3.}
    \label{fig:results_112}
\end{figure*}

\begin{figure}[t]
  \centering
     \resizebox{\columnwidth}{!}{%
    \includegraphics[height = 6.5cm]{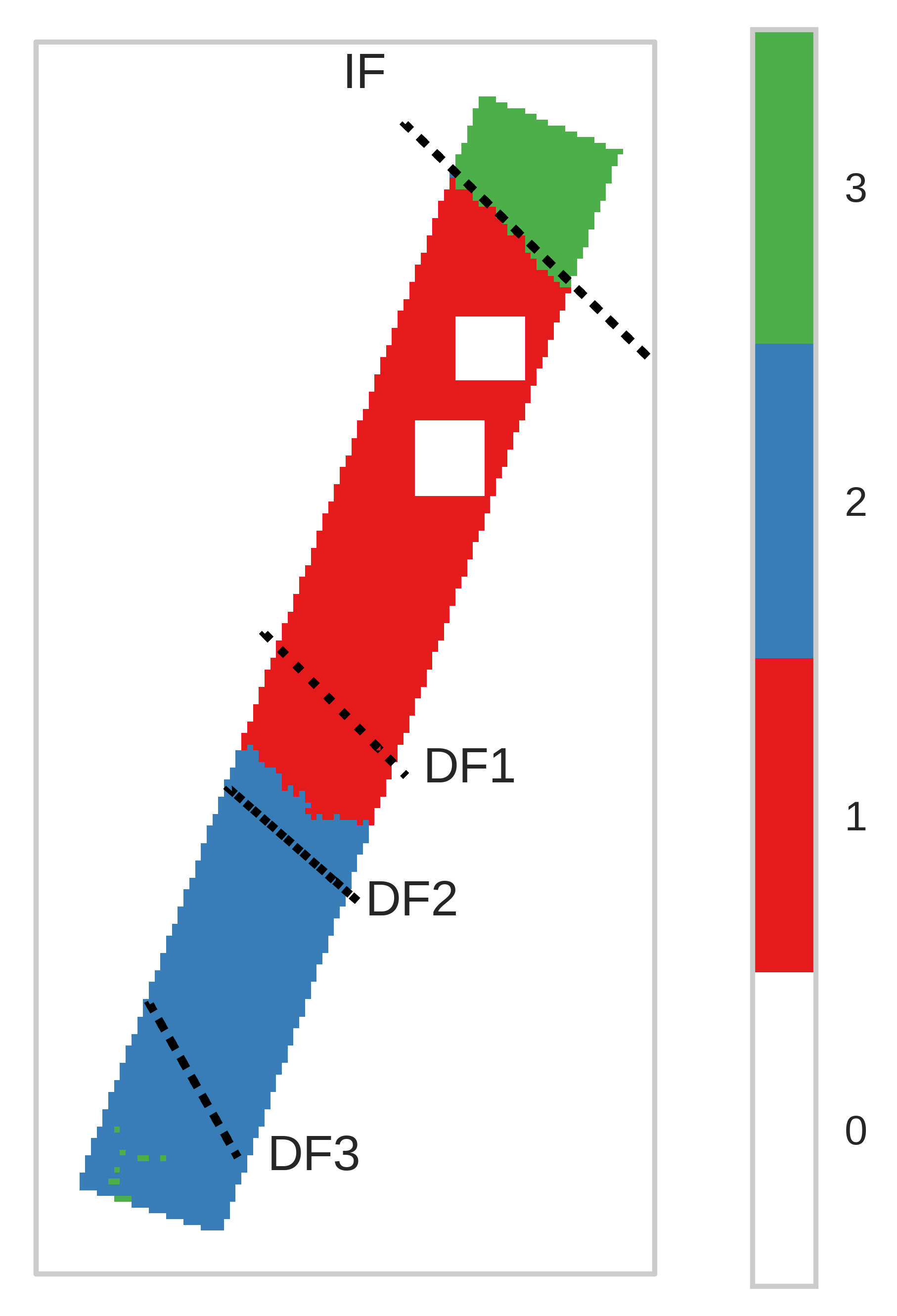}
    \includegraphics[height = 6.5cm]{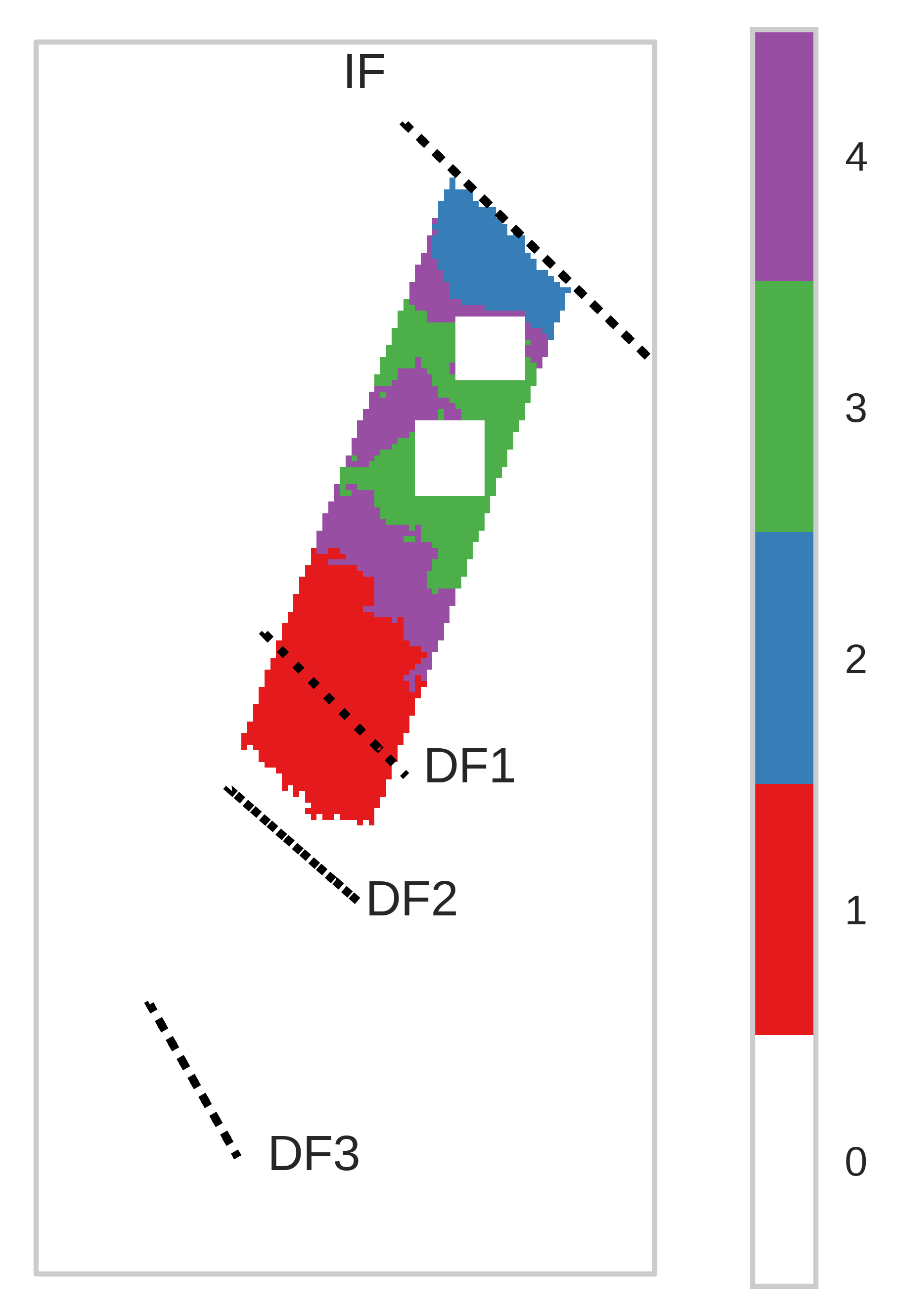}
 }
 
\caption{The spatial footprint of the MIRI MRS FOV is color-coded by cluster assignment calculated on an initial (left) and secondary (right) round of clustering on all spectral features in the MIRI MRS dataset shortward of 13.2 $\mu$m. The secondary round of clustering was applied only to those pixels belonging to cluster 3 in the initial round of clustering that largely coincide with the atomic PDR.  
Clusters are labeled 1 through 4 (in an arbitrary manner) and masked pixels are labeled 0. The black, dashed lines indicate the locations of the IF, DF~1, DF~2 and DF~3.}
\label{fig:double_clustering}
\end{figure}

\subsection{The $10.9-11.63 ~\mathrm{\mu} m$ region}

The $10.9-11.63$~\mum region is dominated by the 11.2 ~\mum emission band, with weaker bands at 10.95 and 11.005 ~\mum \citep{Chown:23}. The 11.2~\mum band is known to exhibit two components at 11.207 and 11.25~\mum \citep{Chown:23}. \par 

The 11.2 ~\mum band clearly dominates the PAH emission in the 10.9-11.63 ~\mum region, as such it plays a key role in cluster assignment here. This is observed in the significant 11.2 ~\mum profile variations between cluster spectra as shown in Fig.~\ref{fig:results_112}, specifically in the peak positions of the two components and the strength of the red wing for this prominent band. Additionally, more subtle variations in the 11.0/11.2 intensity are observed between the averaged spectral profiles. Profiles for the 11.2 ~\mum band which have a more prominent peak component at 11.207 ~\mum show the weakest red wing and strongest 11.0/11.2 intensity. As peak positions shift to the 11.25~\mum component, the 11.0/11.2 intensity decreases, and the 11.2 ~\mum red wing component widens. \par 

Fig.~\ref{fig:results_112} illustrates the spatial zones within the PDR to which the cluster assignment of pixels correspond. As was the case for the 3.2-3.6 ~\mum region, these cluster zones also exhibit well-defined boundaries roughly parallel to the IF, which correspond to the known physical stratification in PDRs. Cluster 1 arises from the most prominent region and covers most of the atomic PDR, stopping slightly short of DF~1. This cluster exhibits an 11.2 ~\mum profile dominated by the 11.207~\mum peak, with the smallest contribution from the 11.25 ~\mum component and weakest red wing. Cluster 1 also corresponds to the strongest 11.0/11.2 band intensity ratio. Cluster 2 traces the IF, part of the \HII\ region along the edge of the FOV closest to the ionizing source, and the region surrounding DF~1, stopping just short of DF~2. The profile for cluster 2 also shows a strong 11.2~\mum component peaking at 11.207~\mum, with a slightly stronger red wing than that of cluster 1. Accordingly, this profile has the second strongest 11.0/11.2 band intensity ratio. The 11.0/11.2 intensity ratio of cluster 2 is similar to that of cluster 3, however the 11.2~\mum profile appears doubly-peaked, with the strongest component at 11.207 ~\mum, though considerable strength in the second 11.25~\mum component. Cluster 3 is found in part of the \HII\ region (bounded by those in cluster 2 on both sides, parallel to the IF), in a stripe just in front of DF~2, past DF~3 in a region along the farthest edge of the footprint from the ionizing source, and in two small circular regions just in front of DF~3. Finally, cluster 4 is found in the region bounded by DF~2 and extends to the rear of DF~3. This profile has the strongest 11.2~\mum component peaking at 11.25~\mum, the widest red wing and the weakest 11.0/11.2 intensity ratio.\par 

\subsection{Atomic PDR, $4.9-13.2 ~\mathrm{\mu} m$ region}

We use the entire wavelength range from the MIRI MRS dataset shortward of 13.2 ~\mum in this atomic PDR-focused experiment. We report results from the initial round of clustering on the entire MIRI MRS footprint using three clusters, and using four clusters on the portion of the footprint found to coincide with the atomic PDR from the initial cluster assignment. We also generate increasingly granular stratification within this region using increasingly higher numbers of clusters during the secondary clustering application. We report these findings in Appendix ~\ref{app:more_clusters}. \par 

The initial round of clustering yielded three, well-defined zones (Fig.~\ref{fig:double_clustering}). The \HII\ region was assigned to cluster 3, with a hard boundary coincident with the IF (plus a few pixels in the molecular PDR past DF~3 near the farthest edge of the FOV from the ionizing source). The atomic PDR and a small portion of the first section of the molecular PDR just behind DF~1 belong to cluster 1. The remaining pixels past the midpoint between DF~1 and DF~2, extending to the far edge of the FOV, are assigned to cluster 2. \par 

We apply a subsequent clustering round to those pixels belonging to cluster 1, exclusively. The cluster zones thus generated are shown in the right-hand panel of Fig.~\ref{fig:double_clustering}. This clustering application returns very well defined clusters with boundaries parallel to the IF, that is, it picks out increasingly granular variations in the local physical conditions. Starting closest to the IF, cluster 2 is bounded by the IF and a stripe of pixels in front of proplyd 203-504. This boundary marks the beginning of cluster 4, covering three main stripes of pixels in the atomic PDR, between which are the pixels belonging to cluster 3. Finally, cluster 1 is centered on DF~1, extending just in front of DF~1 into the deepest layers of the atomic PDR and out into the first layers of the molecular PDR past DF~1 (ending in front of DF~2). \par

\section{Discussion}
\label{discussion}

We begin our discussion by noting that trends in the spectral characteristics of the PAH emission captured by the clustering algorithm align very closely with those reported by \citet{Peeters:nirspec} and \citet{Chown:23}. In both studies, template spectra were selected from five regions probing the key zones in the Bar. In addition, \citet{Peeters:nirspec} reported on the $3.2-3.7$~\mum PAH emission variability across the NIRSpec mosaic. \citet{Chown:23} established that spectral variations in this region are linked to spatially distinct regions (as opposed to a slowly-changing behaviour). Our study serves to expand on these results by applying unsupervised machine learning to more spatial data \citep{Chown:23} and spectral data \citep{Peeters:nirspec}. We report an elaborate map of the variation in the PAH emission throughout the Bar based on the study of the entire NIRSpec and MIRI MRS FOVs. The consistency between our results and those of these authors therefore highlights the excellent range in spectral variability captured by the five template spectra in these previous studies.\par 

\subsection{Connecting the clustering results on different PAH emission regions.}
\label{subsec:connecting_results}

We applied the clustering algorithm to several individual, disjoint wavelength regions which exhibit strong PAH emission bands. While the spatial locations of each of the cluster zones for each wavelength region do not necessarily correspond one-to-one, we find a relationship between the changes in the mean spectral profiles for each of the cluster assignments. That is to say, the variations observed in the PAH bands, traced by the clustering results, are connected to one another.

In general, those regions which exhibit a strong $3.4$/$3.3$ intensity ratio will also demonstrate broader $6.2$ and $7.7$~\mum bands, a weaker $8.6$/$7.7$ band intensity ratio, and likely a so-called ``class B'' profile for the $11.2$~\mum feature. The same trend follows for many of the weaker PAH bands as well. For example, these same cluster profiles will also show stronger $3.424$~\mum component and $3.52$~\mum components in the $3.2-3.7$~\mum complex, and a weaker $11.005$~\mum component in the $11$~\mum region. In contrast, those clusters which select spectra for their weaker $3.4$/$3.3$ intensity ratio will correspond to those profiles which also show more narrow $6.2$ and $7.7$~\mum profiles, stronger $8.6$/$7.7$ band intensity ratios, and will have a class A $11.2$~\mum profile peaking closer to $11.207$~\mum. At the same time, the $3.424$~\mum and $3.52$~\mum components will weaken and the $11.005$~\mum component will strengthen.\\

It is also interesting to note that the clustering results can be used, in some cases, to probe/predict the cluster assignments in other wavelength ranges. Indeed, we find that clustering results calculated on the $11.2$~\mum wavelength region recovers strikingly similar clustering results for the $6.2$~\mum region (see Fig.~\ref{fig:discussion_112_62}). The only major difference between the mean cluster spectra, noticeable by eye, concerns the relative sizes of the $1-\mathrm{\sigma}$ deviations of the cluster profiles. The variances in the $6.2$~\mum profiles calculated using the $11.2$~\mum cluster assignments are slightly larger than those which were calculated using the $6.2$~\mum profile itself. This is to be expected, given that the clusters were not assigned based on this information. Overall, the cluster zones themselves also overlap with one-another quite well. The clusters which coincide with the major zones in the PDR such as the atomic PDR, molecular PDR, IF, and \HII\ region all agree for the majority of their pixels. Deviations between the zones are most noticeable in the boundaries of the clusters themselves, as well as especially between the zones which lie between DF~2 and DF~3. The cluster zones originating from the $11.2$~\mum region largely assign the region of pixels bounded by DF~2 and DF~3 to a single cluster, except for a few smaller groups of pixels. In the case of the clusters assigned based on the $6.2$~\mum band, this same cluster largely just traces DF~3 and the filamentary structure of pixels just behind DF~2.
This result suggests an intrinsic connection between the $11.2$ and $6.2$~\mum bands. The carriers of the $11.2$ and $6.2$~\mum bands are known to be largely neutral and cationic, respectively \citep[e.g.][]{allamandola1999}. Conditions at the origin of the variation in the 11.2~\mum band are thus also at the origin of the variations seen in the 6.2~\mum band. We discuss possible carriers in relation to these variations in Sect. ~\ref{subsec:astro_implications}.\par

Such a \textit{very} strong connection between clustering results was not found for other combinations of the other wavelength regions explored in this study. We illustrate comparisons between the results generated from clustering on the $7.25-8.95$ and $3.2-3.6$~\mum regions with those derived from the $10.9-11.63$~\mum region in Appendix ~\ref{app:comparisons}. As for the $6.2$~\mum results, the average spectral profiles in the $7.25-8.95$ and $3.2-3.6$~\mum regions derived from the clustering results of the $10.9-11.63$~\mum region (see Figs. ~\ref{fig:discussion_112_79} and ~\ref{fig:discussion_112_33}, respectively) are less well defined (i.e. have enhanced variance). Moreover, they also probe a slightly smaller range in spectral variability. Nevertheless, these results showcase the power of this clustering algorithm as a predictive tool in spectral analysis of PAHs. Should, for example, only a limited wavelength range be available for study, the variations in profiles outside of this wavelength range can be extrapolated, though this has yet to be confirmed by other studies for other PDRs. This could be done using results from a clustering application to the data available and knowledge of the connection between the bands observed and those which are connected to the observed bands, but not observed themselves.\\

Our application of clustering on the $11.2$~\mum wavelength region reveals the detailed, gradual transition between the class A and B profiles for the $11.2$~\mum band. Indeed, our cluster zones, coupled with the corresponding mean spectral profiles, act as a detailed map for this transition. Upon inspection of clustering results for increasingly higher numbers of clusters than $4$, we trace this transition in greater detail.

In Fig.~\ref{fig:results_112_all}, for 7 clusters (the highest number of clusters explored in this study), we see the transition between class $\mathrm{A_{11.2}}$ and class $\mathrm{B_{11.2}}$ profiles given by the cluster zones in great detail. The transition from class A to class B can be described spectroscopically by a slightly decreasing and redshifting $11.207$~\mum component combined with an increasing and slightly redshifting $11.25$~\mum component. Combined, this results in a less steep blue wing and enhanced red wing for the class B profiles with respect to class A profiles.

Those profiles which correspond to class $\mathrm{A_{11.2}}$ originate in the atomic PDR, and transition to class $\mathrm{B_{11.2}}$ profiles with distance from the ionizing source. The purple, green, red and blue clusters show the strongest class $\mathrm{A_{11.2}}$ profiles, in decreasing order. The green zone occupies a group of pixels in one of the outermost layers of the atomic PDR, slightly in front of DF~1, as well as a thin stripe of pixels just behind the IF and through the middle of the atomic PDR zone. The red zone lies just in front of DF~1 and as well consists of a small band of pixels just behind the IF. The blue profile originates largely between DF~1 and DF~2 but also with some pixels in the \HII\ region. We are also able to observe, for this number of clusters, more nuance to the intermediate class $\mathrm{A(B)_{11.2}}$ profiles. These orange and brown profiles belong largely to the region after the mid-point between DF~1 and DF~2 , excluding DF~3, with some clusters in the \HII\ region as well. The strongest class $\mathrm{B_{11.2}}$ profile, belonging to the yellow cluster, solely traces DF~3. \par 

Our clustering results suggest that the carriers of the class $\mathrm{B_{11.2}}$ profile originate largely in DF~3. The carriers of the class $\mathrm{A_{11.2}}$ profiles are located throughout the Bar, primarily in the atomic PDR. However, the cluster zones indicate a transition to the intermediate class profile when you travel both away from and in front of the IF (we note that in the latter case the PAH emission arises from the background PDR). When moving towards the molecular PDR, we also see the broadening of most PAH emission profiles and an intensifying $3.4$/$3.3$ band ratio. The carriers of the class $\mathrm{B_{11.2}}$ profiles thus originate in the much colder conditions of the molecular PDR, especially within DF~3 itself. The regions in the cluster zone map given in Fig.~\ref{fig:results_112_all} (along DF~3 and a filament just behind DF~2, seen in numbers of clusters greater than $4$) for $11.2$~\mum which align with the most pronounced class B profile are also picked out for the $3.3$ and $6.2$~\mum wavelength regions. These structures are not as well defined in the case of the $7-9$~\mum region where they begin to appear with $6$ clusters (see Fig.~\ref{fig:results_79_all}).

\subsection{Clustering probes the changing physical conditions.}
\label{subsec:phys-conditions}

We emphasize again that we removed the effects of total intensity from each of the PAH emission regions prior to any of our clustering experiments (by normalizing the input spectra). Nevertheless, the clusters pick up spectral variation corresponding to different regions in the PDR structure/anatomy consistently for each of the wavelength bands used to generate the clusters as illustrated by the spatial morphology of various cluster zones (Figs.~\ref{fig:results_33} to~\ref{fig:results_112}). 

We find the atomic PDR, for example, to be largely uniform for the clusters calculated on the main PAH bands (the $3.3$, $6.2$, $7-9$ and $11.2$~\mum wavelength regions). In the case of the $3.3$~\mum region, the atomic PDR is covered entirely by a single cluster, even through seven clusters (see Fig.~\ref{fig:results_33_all}) while for the other wavelengths just two clusters are found for the atomic PDR (with one of them covering most of this zone). Similarly, our clustering results cover the \HII\ region quite consistently between experiments. In all cases, the \HII\ region is mostly covered by a single cluster, with some variance in the shape of this cluster zone between those from the $3.3$ and $7-9$~\mum region results and those from the $6.2$ and $11.2$~\mum regions. The outer zone of the molecular PDR is identified in a similar fashion for all clustering results based on the $3.3$, $6.2$ and $7-9$~\mum wavelength regions. This cluster zone, in each experiment, is well bounded by DF~1 along the edge closest to the IF, and bounded by pixels varying in position from along DF~2 in the case of the $3.3$~\mum band results, slightly in front of DF~2 in the case of the $6.2$~\mum band and $7-9$~\mum region results. The clustering results based on the $11.2$~\mum feature instead group pixels centered on DF~1, cutting into both the atomic PDR and the outer region of the molecular PDR. This cluster zone also corresponds to a stripe of pixels along the IF and at the edge of the MIRI FOV closest to the ionizing source. Finally, the deepest zone of the molecular PDR, behind DF~2, shows the most structure in the clustering results of any of the zones. This structure is captured by the clusters corresponding to all four of the major wavelength regions explored and becomes increasingly detailed when greater numbers of clusters are used (see Figs.~\ref{fig:results_33_all} to~\ref{fig:results_112_all}). Most notably, the clustering identifies two filamentary patterns, the first of which coincides with DF~3, the second of which lies just behind of and at a slight angle with DF~2. \par 

Overall, we find the molecular zone is split largely into two distinct clusters. One of which corresponds to the outermost region of the molecular PDR, overlapping DF~1, and the second corresponding to the deeper regions of the molcular PDR, coinciding with DF~2 and DF~3. In the case of the cluster zones derived from the $11.2$ ~\mum region, this second cluster zone is quite uniform. In the cases of the cluster zones derived from the $3.3$, $6.2$, and $7-9$~\mum regions, however, this zone is seen to be split further between a cluster which traces DF~3 and a filament slightly behind DF~2 (the filament behind DF~2 is not apparent in the zone map for the $7-9$~\mum region). For all cases, though, we observe that the spectral characteristics of the cluster corresponding to DF~1 behave more like to those arising from clusters covering the atomic PDR and \HII\ region than those arising from the secondary molecular PDR cluster coincident with DF~2 or DF~3 (see Figs.~\ref{fig:results_33} to~\ref{fig:results_112}). This observation is consistent with the result reported by \citet{Peeters:nirspec} and \citet{Chown:23} that the spectra originating in DF~1 often behave more like those originating in the atomic PDR and \HII\ region than those originating in DF~2 or DF~3. \citet{Habart:im} and \citet{Peeters:nirspec} reported that DF~1 is likely located at a greater distance from us than are DF~2 and DF~3. Hence, a larger column of the atomic PDR is present in this line of sight (Fig.~\ref{orion_illustration}) which may be driving the characteristics of the PAH emission along it \citep{Peeters:nirspec}. \par 

Local physical conditions determine the characteristics of the PAH emission bands we observe. In other words, our clustering results, based on varying \textit{spectral} characteristics of the PAH emission, trace the physical conditions which give rise to these properties.

\subsection{Comparison with previous observations}
\label{subsec:literature}

\subsubsection{Relative intensity variation}

The clustering highlights several variations in relative intensity ratios. The observed dependence of the 3.4/3.3 ratio with distance from the IF and thus with decreasing intensity of the FUV radiation field is well established and attributed to the lower stability of the 3.4~\mum band carrier \citep[e.g.][]{geballe1989, Joblin1996, Sloan:97, Mori:14,  pilleri2015}. The clustering extends this behaviour to the 3.46, 3.51, and 3.56~\mum bands (relative to the 3.3~\mum band). 
We furthermore observe an increased importance of the 6.0~\mum band (relative to the 6.2~\mum band) with increasing depth into the PDR. Such a behaviour is consistent with the morphology of both bands in reflection nebulae \citep{peeters2017, Knight:1333:22}. In contrast, the 11.0/11.2 ratio decreases with depth into the PDR due to a decreasing ionization fraction of the PAH population \citep[e.g.][]{Rosenberg:11, peeters2012, boersma2013, peeters2017, knight2022, Knight:1333:22}. Finally, relative intensity variations between the 7.6, 7.8, and 8.6~\mum bands are commonly observed. The 7.6/7.8 ratio decreases with decreasing strength of the radiation field \citep[e.g.][]{boersma2014, peeters2017, Stock:17} and the 8.6/7.7 ratio increases with increasing strength of the radiation field \citep{peeters2017, knight2021, Knight:1333:22}.

\subsubsection{Profile Variation}

Using a set of template spectra of the key zones in the Bar,  \citet{Chown:23} reported class $\mathrm{A}$ profiles for the 3.3, 6.2, 7.7, and 8.6~\mum bands, consistent with our results.
These authors also found the profiles of the $11.2$~\mum band to belong to class A in the atomic PDR and to class B deep in the molecular PDR (DF~2, DF~3). This paper further extends these results to the entire mosaic. While a spatial evolution of the $11.2$~\mum band profile with distance to the ionizing source had previously  been reported for reflection nebulae \citep{boersma2013, boersma2014, shannon2016}, the PDRs4All observations of the Bar report, for the first time, a class $\mathrm{B_{11.2}}$ profile in the ISM and show the transition between class $\mathrm{A_{11.2}}$ and $\mathrm{B_{11.2}}$ \citep[this paper and][]{Chown:23}. Our clustering results as well as \citet{Chown:23} for the five template spectra also capture a broadening of the $7.7$~\mum band in the direction away from the ionizing source. A broadening of the $7.7$~\mum profile has previously been observed \citep[e.g.][]{BregmanTemi2005, boersma2014, Stock:17}. To our knowledge, a broadening of the  $\mathrm{A_{3.3}}$ or $\mathrm{A_{6.2}}$ has not been previously reported within the ISM.
\subsubsection{Basis set of PAH components}
\label{subsubsec:BSS}

A powerful analysis approach of PAH observations is the mathematical decomposition of PAH emission spectra as a linear combinations of elementary spectra using blind signal separation (BSS) methods \citep[e.g.][]{Boissel:01, Rapacioli:05, berne2007analysis}. The obtained basis set of elementary spectra details spatially different components of the observation, thus each component of the basis set has its unique morphology. Generally, three components have been derived which have been associated with neutral PAHs, cationic PAHs, and evaporating very small grains (eVSGs) or PAH clusters \citep{berne2007analysis, pilleri2012evaporating}. The strength of the $7.7$/$11.2$ ratio was strongest in the cationic PAH component spectrum and weakest for the neutral PAH component spectrum. The eVSGs component spectrum is quite distinct and exhibits slightly redshifted bands, significantly broadened $7.7$ and $11.2$~\mum profiles and no $8.6$~\mum feature \citep{berne2007analysis, Foschino:19}. These studies found that the eVSGs and PAH clusters components exist in deeper regions of the PDR, far from the ionizing source. Indeed, a transition, driven by photoprocessing, from clusters of PAHs to neutral PAHs and eventually to PAH cations resulting from the increasing strength of the UV radiation field occurs as distance to the ionizing source decreases \citep{cesarsky2000, Boissel:01, Rapacioli:05, berne2007analysis, pilleri2012evaporating}. This is consistent with our clustering results which also trace a layered stratification across the Bar which corresponds to varying spectral characteristics as given in our cluster zone maps. \citet{pilleri2012evaporating} reported a strong negative correlation between the fraction of eVSGs or PAH clusters and the strength of the local UV radiation field for several PDRs. These observations serve as evidence for the existence of a photodestruction process that breaks eVSGs or PAH clusters down into PAH molecules. Evidence for eVSGs as carriers for the class B profiles identified in \citet{peeters2002} has also been found \citep{Foschino:19}. These studies substantiate evidence for PAH clusters as carriers for ``component 2" of the $11.2$~\mum profile we detect far from the ionizing source (see Section ~\ref{subsec:astro_implications}). \par

\subsection{Astrophysical implications}
\label{subsec:astro_implications}

\begin{figure*}
    \centering
    \resizebox{.99\hsize}{!}{%
    \includegraphics[clip,trim =2.5cm 0cm 2.2cm 1.5cm, width = 9cm, height = 6.08cm]{figures/62_avg_specs_n_4.png}
    \includegraphics[width = 4cm]{figures/62_cluster_zones_n_4.png}
 }
    \resizebox{.99\hsize}{!}{%
    \includegraphics[width = 3.7cm, height = 2.5cm]{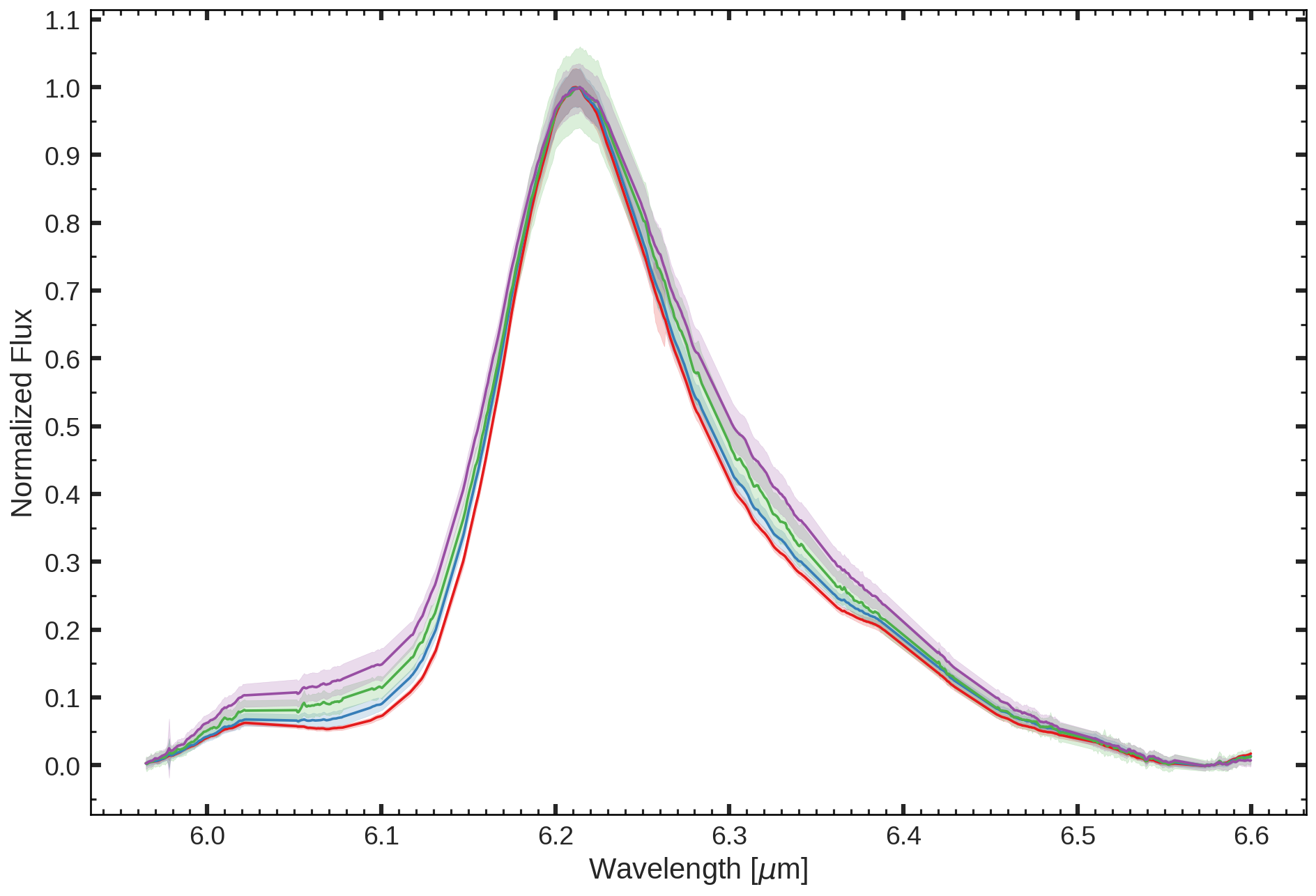}
    \includegraphics[width = 1.736cm]{figures/112_cluster_zones_n_4.png}

}
 
    \caption{The average spectral profiles (left) for the $6.2$~\mum band obtained from clustering assignment based on the $5.95-6.6$~\mum region and the corresponding cluster zones (right) are given in the top two figures. The average spectral profiles (left) for the $6.2$ ~ \mum band obtained from clustering assignment based on the $10.9-11.63$~\mum region and the corresponding cluster zones (right) are given in the bottom two figures. Left panels: The shaded regions around each profile illustrate one standard deviation from the mean profile intensity at a given wavelength value. Each spectrum is normalized to the peak 6.2 $\mathrm{\mu} m$ intensity for visualization purposes. Right panels: The spatial footprint of the MIRI MRS FOV color-coded by cluster assignment. Any masked pixels or those which have been otherwise masked out are labelled as 0. The black, dashed lines illustrate the locations of the ionization front (IF) and the three dissociation fronts (DF~1, DF~2 and DF~3) as defined in \citet{Peeters:nirspec} using nomenclature established in \citet{habart2023}.}
    \label{fig:discussion_112_62}
\end{figure*}

\begin{figure*}
    \centering
    \resizebox{.99\hsize}{!}{%
    \includegraphics[width = 4cm, height = 3cm]{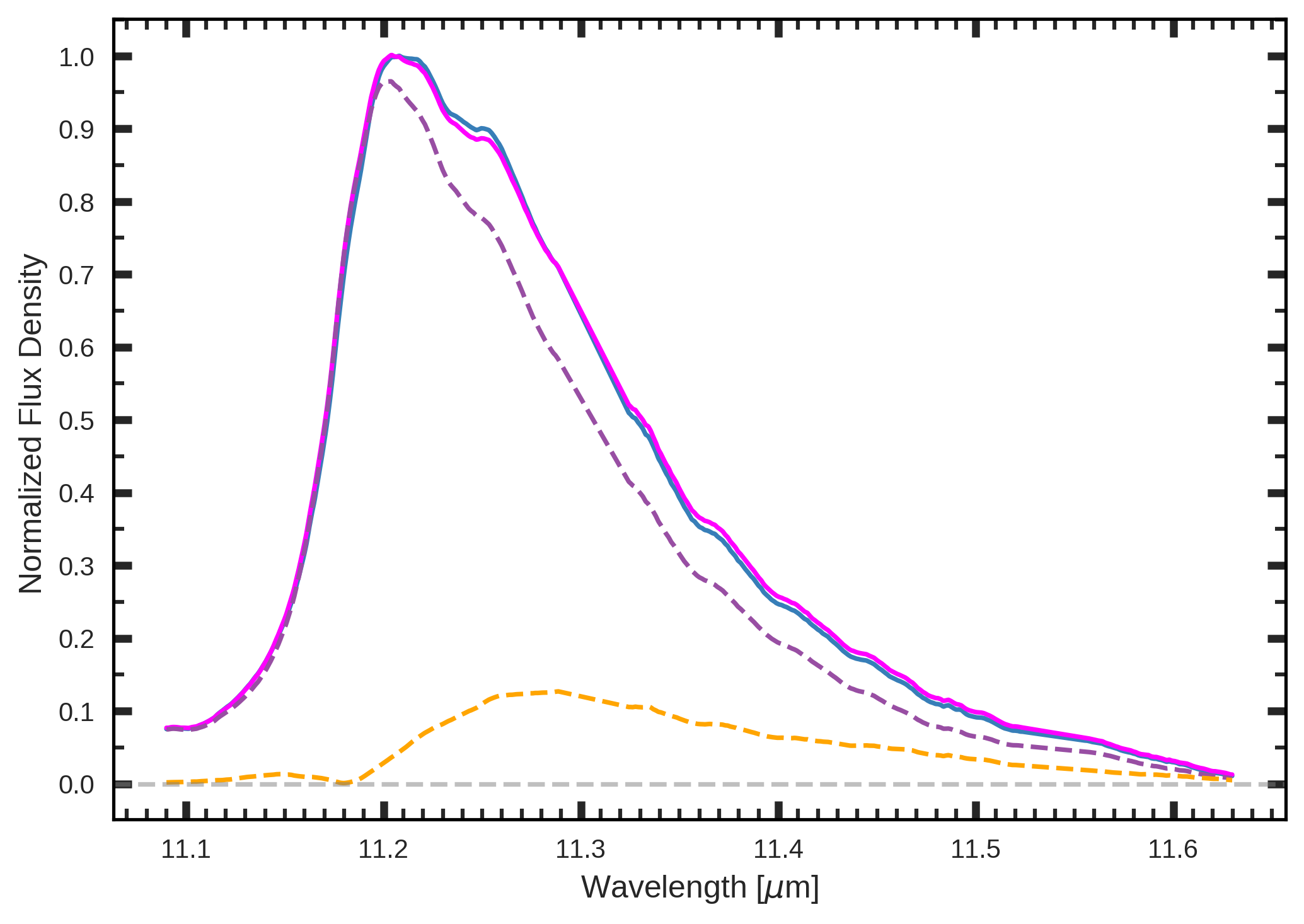}
    \includegraphics[width = 4cm, height = 3cm]{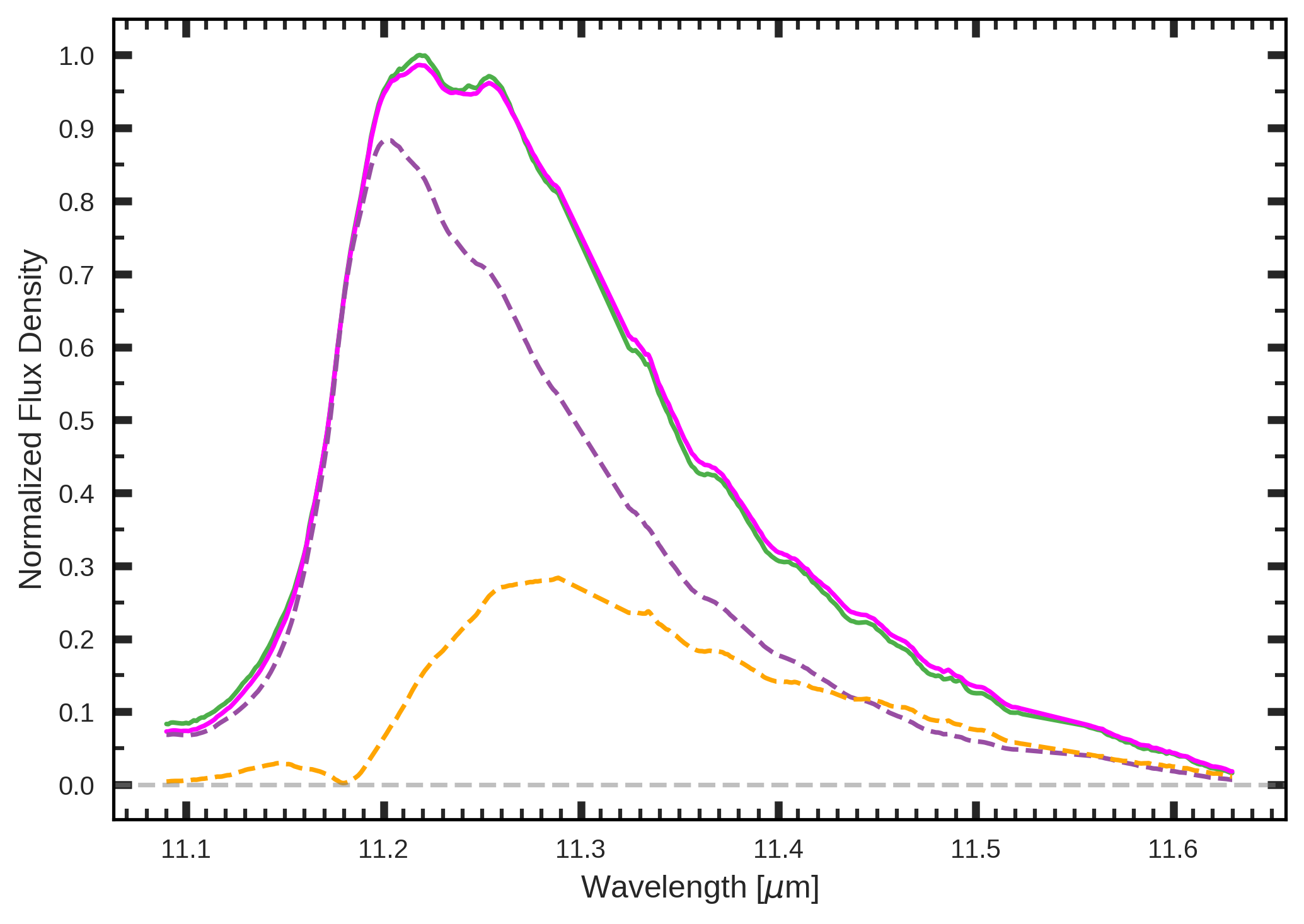}
 }
    \caption{The linear combinations of the class $\mathrm{A_{11.2}}$ profile and a ``secondary component''  which recover the ``intermediate" $11.2$~\mum profiles of clusters 3 (left) and 2 (right) from the $10.9 - 11.63$~\mum wavelength region (see Sect.~\ref{subsec:astro_implications} for details). The cluster profile is shown in blue and green in the left and right panel respectively. The dashed purple line represents the scaled class $\mathrm{A_{11.2}}$ profile given by cluster 4, the dashed orange line is the secondary component, and the bright magenta line represents the linear combination of these two components. We note that the small undulations in the red wing of the profiles are residual fringes present due to incomplete fringe removal.}
    \label{fig:112_lin_combs}
\end{figure*}

\subsubsection{The $3.2-3.6$~\mum region}

While the 3.3~\mum profile has been attributed to the aromatic C-H stretching mode \citep[e.g.][]{ATB, Leger:review:89}, various candidates have been put forward for the carrier of the components in the $3.4-3.6$~\mum region. These include the aliphatic C-H stretching mode in methyl ($-$CH$_3$) and methylene ($-$CH$_2$) side groups of PAHs \citep{ATB, Joblin1996, Maltseva:18, buragohain2020, steglich2013abundances} and in super-hydrogenated PAHs \citep{Bernstein:96, Wagner:00, Sadjadi:15, Maltseva:18, Sundararajan:19, buragohain2020, pla2020, Yang:20}, overtones of the aromatic C-H stretching mode \citep{Barker:anharm:87}, combination bands \citep{ATB}, and a combination thereof. Several studies reported that the relative importance of the aliphatic bands with respect to the aromatic bands decreases with increasing intensity of the FUV radiation field  reflecting that aliphatic bonds are less stable than aromatic ones \citep{geballe1989, Joblin1996, Sloan:97, Mori:14,  pilleri2015}. The clustering results clearly indicate a concerted behaviour of the bands in the $3.4-3.6$~\mum region:  the 3.4 complex, 3.46, 3.51, and 3.56~\mum bands all vary in the same fashion, peaking in DF~3 and the region slightly behind DF~2. This behaviour is confirmed by the morphology of these band intensities individually \citep{Peeters:nirspec}. This concerted behaviour of the $3.4-3.6$~\mum emission implies that the carriers of each of these components must all be most abundant in DF~3, and either get destroyed or evolve in the same way into carriers of other bands in the other PDR zones. We note that the distinct behaviour of the intensities of the components making up the 3.4~\mum complex (3.395, 3.403, and 3.424~\mum components), as reported by \citet{Peeters:nirspec}, is also picked up by the clustering algorithm. Indeed, the morphological similarity of the 3.395~\mum band intensity to the 3.29~\mum band intensity instead of the 3.403~\mum intensity is recovered in the 3.4~\mum band profile of the mean cluster profiles (Fig.~\ref{fig:results_33}). We are compelled by the similarity of the spatial variation of the $3.4-3.6$~\mum emission bands to propose that their carriers originate from the same PAH side group attached to similar-sized PAHs. Their concerted spatial behaviour requires that their carriers have similar sensitivity to photolysis. Different bonds within aliphatic subgroups attached to PAHs have distinct dissociation rates \citep{Joblin1996, tielens2008}. Our clustering results therefore point to a single sidegroup that is responsible for the observed emission. Furthermore, the dissociation rate for a given bond within an aliphatic subgroup depends on the size of the PAH to which it is attached \citep{Joblin1996, tielens2008}. Combined with the dependence of the PAH excitation on size \citep{schutte1993}, our results further suggest that this sidegroup is attached to similar-sized PAHs. Further, we eliminate resonances between CC and CH in-plane bending modes from the possible causes of this uniform variation of features in this wavelength region. We do not observe what would be consequential shifts in the peak position of the $3.3$~\mum feature or blending on the red wing as traced by our clustering results \citep{mackie2018, mackie2022}. Finally, we exclude superhydrogenated PAHs as the carriers of the 3.4~\mum band in the Orion Bar based on the fact that superhydrogenated PAHs only reside in very benign environments and cannot be the carriers of the 3.4~\mum band in the reflection nebula NGC~7023 \citep{Andrews:16}. 

\subsubsection{The $11.2$ Profile Classes}

The clustering results for the $11.2$~\mum band suggest the presence of at least two components/populations, one carrying the $11.207$ component and the other one carrying the $11.25$~\mum component. The former component dominates the class $\mathrm{A_{11.2}}$ profile while the latter dominate the class $\mathrm{B_{11.2}}$ profile. The transition between both classes can then be interpreted as a changing relative importance of these two components/populations. To investigate this hypothesis, we take the cluster profiles for the $10.9-11.63$~\mum region, normalized to peak intensity, and decompose the class $\mathrm{B_{11.2}}$ profile from our clustering results (given by cluster 4; Fig.~\ref{fig:results_112}) into a linear combination of the class $\mathrm{A_{11.2}}$ profile (given by cluster 1) and the difference between the class $\mathrm{B_{11.2}}$ profile and a ``secondary component''. We determine this secondary component (referred to as component 2) by subtracting the scaled class $\mathrm{A_{11.2}}$ profile ($77\%$) from the class $\mathrm{B_{11.2}}$ profile. We found $0.77$ to be the suitable coefficient by which to multiply this secondary component by manual experimentation with the aim to scale the class $\mathrm{A_{11.2}}$ profile to fit entirely inside the $\mathrm{B_{11.2}}$ profile. We can successfully re-create the other normalized cluster profiles using linear combinations of component 2 and the class $\mathrm{A_{11.2}}$ profile (Fig.~\ref{fig:112_lin_combs})\footnote{To recover the intermediate $11.2$~\mum band profile of cluster 2, we combine $96.5\%$ of the class $\mathrm{A_{11.2}}$ profile and $30\%$ of the component 2. To recover the intermediate profile of cluster 3, we combine $88.3\%$ of the class $\mathrm{A_{11.2}}$ profile and $67\%$ of the component 2.}. Certainly, the carriers of each of the class $\mathrm{A_{11.2}}$ and $\mathrm{B_{11.2}}$ components are independent. \par 

\begin{figure}
    \centering
    \resizebox{\columnwidth}{!}{ \includegraphics[height = 6cm]{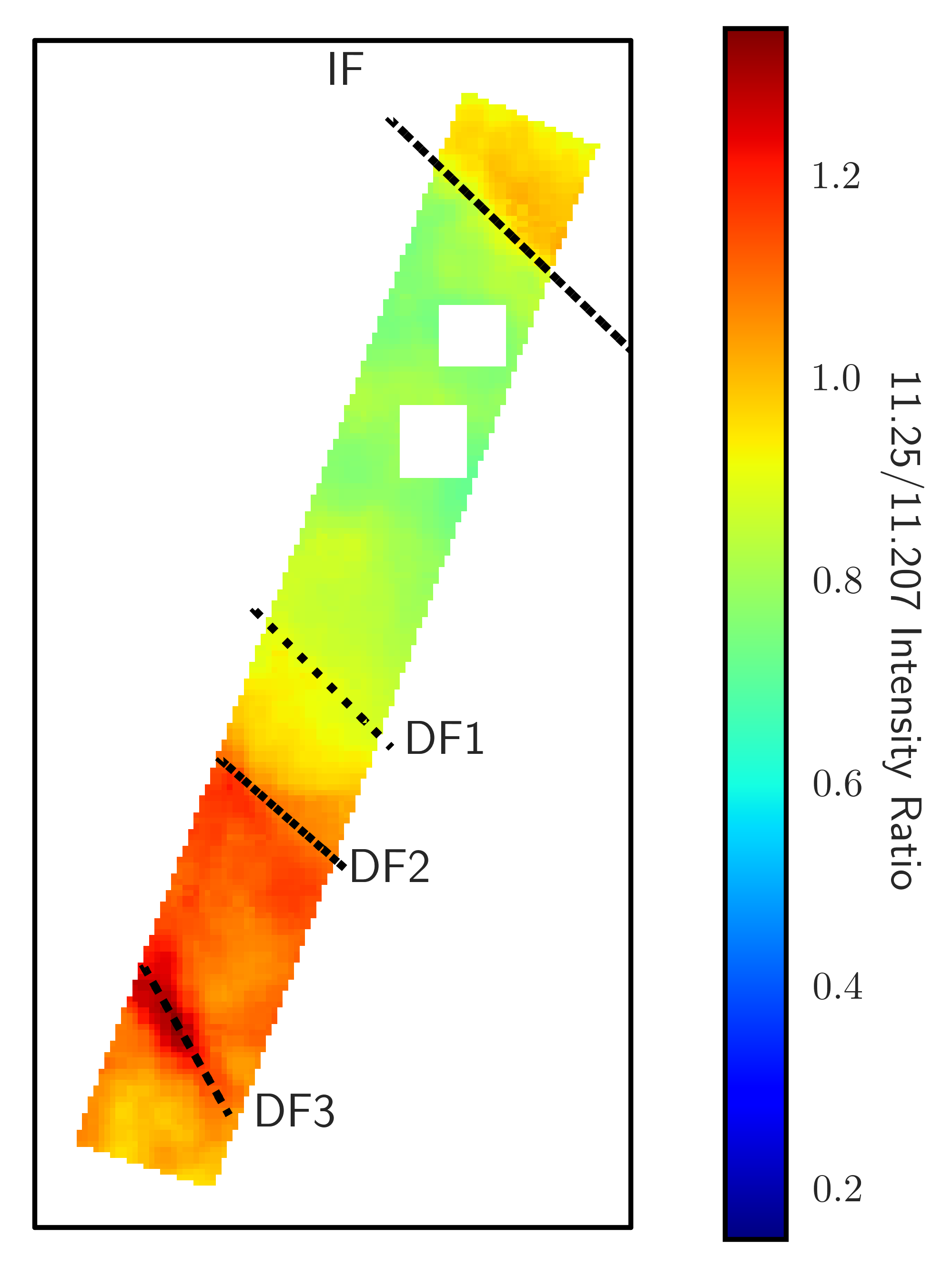}
 }
 \caption{The spatial morphology of the $11.25$/$11.207$ intensity ratio (i.e. the ratio of the surface brightness at 11.25~\mum to the surface brightness at 11.207~\mum  after continuum subtraction). The $11.25$~\mum component dominates over the $11.207$~\mum component in the deeper regions of the PDR, past DF~1, especially in DF~3.}
 \label{fig:1125_1120_heatmap}

 \end{figure}

We propose that PAH clusters or VSGs carry this $11.25$~\mum component. These PAH clusters (or VSGs) must exist predominantly in DF~3, deep into the molecular PDR, as visualized by the cluster zones in Fig. ~\ref{fig:results_112} as well as by the ratio of the surface brightness at 11.25~\mum to the surface brightness at 11.207~\mum (after continuum subtraction) illustrated in Fig. ~\ref{fig:1125_1120_heatmap}. This intensity ratio map shows the 11.25~\mum component peaking over the 11.207~\mum component in DF~3. The carriers of the $11.25$~\mum component must also be independent of the carriers of the $11.207$~\mum component, as demonstrated in Fig.~\ref{fig:112_lin_combs}, and thus exist solely in the deeper regions of the PDR, furthest from the IF. This can be attributed to the fact that the binding energy of PAH clusters is low \citep[$\sim$50~meV/C-atom, e.g. the binding energy for a coronene cluster is $\sim$2.5~eV;][]{Tielens:21}. PAH clusters or VSGs can easily be photolysed due to the increasing FUV radiation field in the direction of the PDR surface at the IF.

Though we do see variation in the other bands studied here which also vary as function of increasing distance from the IF (for example, the intensifying of the $3.4$/$3.3$ band ratio and the broadening of the band profiles), we do not see any other emission band component which appears or disappears in this fashion for any of the other zones in the PDR. Hence, there is no evidence in our results that point to a ``new'' contribution in any of the shorter-wavelength bands. The carriers of the $11.25$~\mum component must then undergo a molecular transition in the photolysis process such that they do not affect the other bands emitting shorter wavelengths. These carriers, the PAH clusters, are molecular predecessors to the carriers of the bands located at shorter wavelengths. 

However, it is important to note that we do indeed observe broadening in several shorter wavelength bands in DF~3 compared to those in the atomic PDR \citep[this paper,][]{Peeters:nirspec, Chown:23}. In addition, while we attribute part of the broadening of the 11.2~\mum band to the relative importance of 2 individual components ($\mathrm{A_{11.2}}$ profile and component 2; Fig.~\ref{fig:112_lin_combs}), component 2 has a broader profile than the $\mathrm{A_{11.2}}$ profile. Spectral broadening is generally attributed to more highly excited species and \citet{Peeters:nirspec} and \citet{Chown:23} concluded that smaller, labile PAH species are present in DF~3 and not in the atomic PDR.
The fact that the clustering results calculated on the $11.2$~\mum wavelength region recovers strikingly similar clustering results for the $6.2$~\mum region (see Fig.~\ref{fig:discussion_112_62}) then implies that these smaller, labile PAH species and PAH clusters are equally sensitive to photolysis and both are destroyed as the FUV radiation field increases in the direction towards the PDR surface. Moreover, a similar effect on these smaller species can then be expected for the 11.2~\mum emission as for the shorter wavelength band profiles in DF~3. 
This then implies that the width of the blue component in our linear decomposition ($\mathrm{A_{11.2}}$ profile) slightly increases from the atomic PDR towards DF~3 as observed for the shorter wavelength bands. This broadening of the blue component of the 11.2~\mum profile due to the smaller-sized PAH population however does not dominate the observed changes in the 11.2~\mum profile which is driven by the relative contribution of the PAH clusters. In addition, the larger width of component 2 with respect to that of the blue component then reflects different emission characteristics of PAHs and PAH clusters. A broader 11.2~\mum emission band is also observed in the template spectrum assigned to PAH clusters obtained with BSS \citep[Sect.~\ref{subsubsec:BSS};][]{berne2007analysis, pilleri2012evaporating, Foschino:19, Rosenberg:11}. This BSS template spectrum is dominated by a broad emission band peaking near 8.0~\mum, has additional emission bands near 6.2 and 3.3~\mum but lacks a well-defined emission band near 3.4~\mum. Hence, this template spectrum associated with PAH clusters differs from the emission attributed to PAH clusters in this paper. The clustering results indicate that the population in DF~3 consists of smaller, labile PAHs (with respect to those in the atomic PDR) and PAH cluster that are equally sensitive to photolysis. As such, the BSS method may be less sensitive to the distinction of both populations as it relies on spatial variation. However, we note that the BSS method was applied to a mosaic of low spectral resolution data \citep{berne2007analysis, pilleri2012evaporating} as well as integrated medium spectral resolution data \citep[without spatial information;][]{Foschino:19}. Hence, a firm conclusion about this difference awaits the results of the BSS method applied to the same data set as used here (Schroetter et al., in prep.). The BSS results also indicated that the contribution of PAH clusters increases with decreasing strength of the FUV radiation field \citep{pilleri2012evaporating}. This is consistent with our findings that the relative contribution of component 2 increases with distance from the IF (Figs.~\ref{fig:results_112}, ~\ref{fig:112_lin_combs}, and ~\ref{fig:results_112_all}) suggesting a dependence on the strength of the local FUV radiation field. This then extends the reported dependence of the central wavelength of the 11.2~\mum bands in circumstellar environments on the effective temperature of the central star \citep[which is a indication of the strength of the radiation field impinging on the PAH population;][]{Sloan:07, Keller:08} to resolved PDRs.  

 \section{Conclusions}

We performed a variability analysis on the PAH emission in the Orion Bar using bisecting k-means clustering, an unsupervised machine learning algorithm. We applied this clustering algorithm to spectra from both NIRSpec-IFU and MIRI-MRS onboard \textit{JWST} for the $3.2-3.6$~\mum region in the case of the NIRSpec data, and the $5.95-6.6$, $7.25-8.95$ and $10.9-11.63$~\mum regions for the MIRI-MRS dataset. These regions were selected as they correspond to major vibrational emission bands of PAHs at $3.3$, $6.2$, $7.7$, and $11.2$ ~\mum, respectively. Specifically, we generated mean spectral profiles for each cluster calculated from the PAH emission in each of these wavelength regions, as well as detailed spatial maps associated with each cluster throughout the Bar. This wealth of spatial and spectral data has allowed us to connect variations between bands with one another as well as to derived a nuanced understanding of the variation in the underlying PAH populations across the Bar's varying physical conditions. \par 

Applying the bisecting k-means clustering algorithm to multiple prominent PAH emission profiles results in well-defined cluster zones that correlate with the spatial stratification of physical conditions within the Bar across the ionized to the molecular, UV-shielded zone. We find that clustering captures subtle differences in profile changes driven by the strength  of the local FUV radiation field.
In addition, our clustering application on the $10.9-11.63$~\mum wavelength region reveals the transition between class A and class B $11.2$~\mum band profiles as defined in \citet{vanDiedenhoven2004}, consistent with the results of \citet{Chown:23}. We also find that the clustering application calculated on the $10.9-11.63$~\mum region reproduces the cluster assignments for the $5.95-6.6$~\mum region. \par 

Our clustering results capture a united behaviour in the bands which make up the $3.4-3.6$~\mum region. They are found to exist most abundantly in the Bar in DF~3, and evolve or are destroyed with increasing FUV radiation in the direction towards the IF. We deduce, from the similarity in the spatial variation in these bands, that they  are carried by the same PAH side group attached to similarly-sized PAHs. We also reveal that the carriers of the $11.207$ and $11.25$~\mum components of the $11.2$~\mum profile, which dominate the class $\mathrm{A_{11.2}}$ and $\mathrm{B_{11.2}}$ profiles, respectively, are independent. We have extracted the two components of the $11.2$~\mum feature which may help in the identification of their carriers. We propose PAH clusters or VSGs to be the carriers of the $11.25$~\mum component, as they are found to exist solely in the deeper regions of the PDR and their low binding energies dictate their destruction in the direction towards the IF. Finally, we find a dependence on the characteristics of the $11.2$~\mum profile and the strength of the FUV radiation field.\par 

Our clustering results shed light on the morphology of the physical zones within the Bar in great detail. Clustering on PAH emission, which is coupled to local physical conditions, therefore also identifies clusters of spatial regions which share similar physical conditions. These clusters, coupled with their corresponding spectral information, outline the specific physical conditions for each zone as well as give context for relative differences or similarities between those of the other clusters. Though, at this point in time, the exact physical and/or chemical processes which control the evolution of the interstellar PAH family (and therefore the observed spectral variations) are unknown. Future observations with \jwst\ and \textit{ALMA}, for example, may shed light on this. Such observations may, when coupled with PAH spectra and a similar clustering analysis, provide astronomers a tool for pinpointing these varying physical conditions in galaxies far away.   
Cluster zone maps therefore offer detailed maps which highlight transitions between physical zones limited only by spatial resolution. In this way, a clustering application on PAH emission is a much more powerful tool for identifying variations in physical conditions than what may be the case from studying intensity variations between lines alone. \par 

Overall, we find extraordinary potential for unsupervised machine learning techniques, namely clustering algorithms, as PAH spectral analysis tools. The consistency of our results with those of existing studies which employ traditional spectral analysis methods \citep[e.g.][]{Peeters:nirspec, Chown:23} validates the use of such a technique in this space. \par 

Further focused studies on PAH emission in the $3.2-3.6$ and $11.2$~\mum emission regions may uncover greater insight into the carriers of the $3.4-3.6$~\mum emission and the class $\mathrm{B_{11.2}}$ profiles. Future studies regarding the use of unsupervised machine learning in astrochemistry will lead to an enhanced understanding of the abilities of such algorithms within the context of such spectral analyses. We have only just begun to highlight the potential that such clustering algorithms can have for PAH variability studies. 

\begin{acknowledgements}

This work is based on observations made with the NASA/ESA/CSA James Webb Space Telescope. The data were obtained from the Mikulski Archive for Space Telescopes at the Space Telescope Science Institute, which is operated by the Association of Universities for Research in Astronomy, Inc., under NASA contract NAS 5-03127 for JWST. These observations are associated with program \#1288.
Support for program \#1288 was provided by NASA through a grant from the Space Telescope Science Institute, which is operated by the Association of Universities for Research in Astronomy, Inc., under NASA contract NAS 5-03127.

EP and JC acknowledge support from the University of Western Ontario, the Institute for Earth and Space Exploration, the Canadian Space Agency (CSA, 22JWGO1-16), and the Natural Sciences and Engineering Research Council of Canada. 
Studies of interstellar PAHs at Leiden Observatory (AT) are supported by a Spinoza premie from the Dutch Science Agency, NWO. CB is grateful for an appointment at NASA Ames Research Center through the San Jos\'e State University Research Foundation (80NSSC22M0107).
TO acknowledges support from JSPS Bilateral Program, Grant Number 120219939.

\end{acknowledgements}

\bibliographystyle{aa}
\bibliography{bibliography}

\begin{appendix}

\section{Elbow Plots}\label{app:nr_clusters}

\begin{figure*}
    \centering
    \resizebox{.99\hsize}{!}{%
      \includegraphics[width = 8cm, height = 5.5cm]{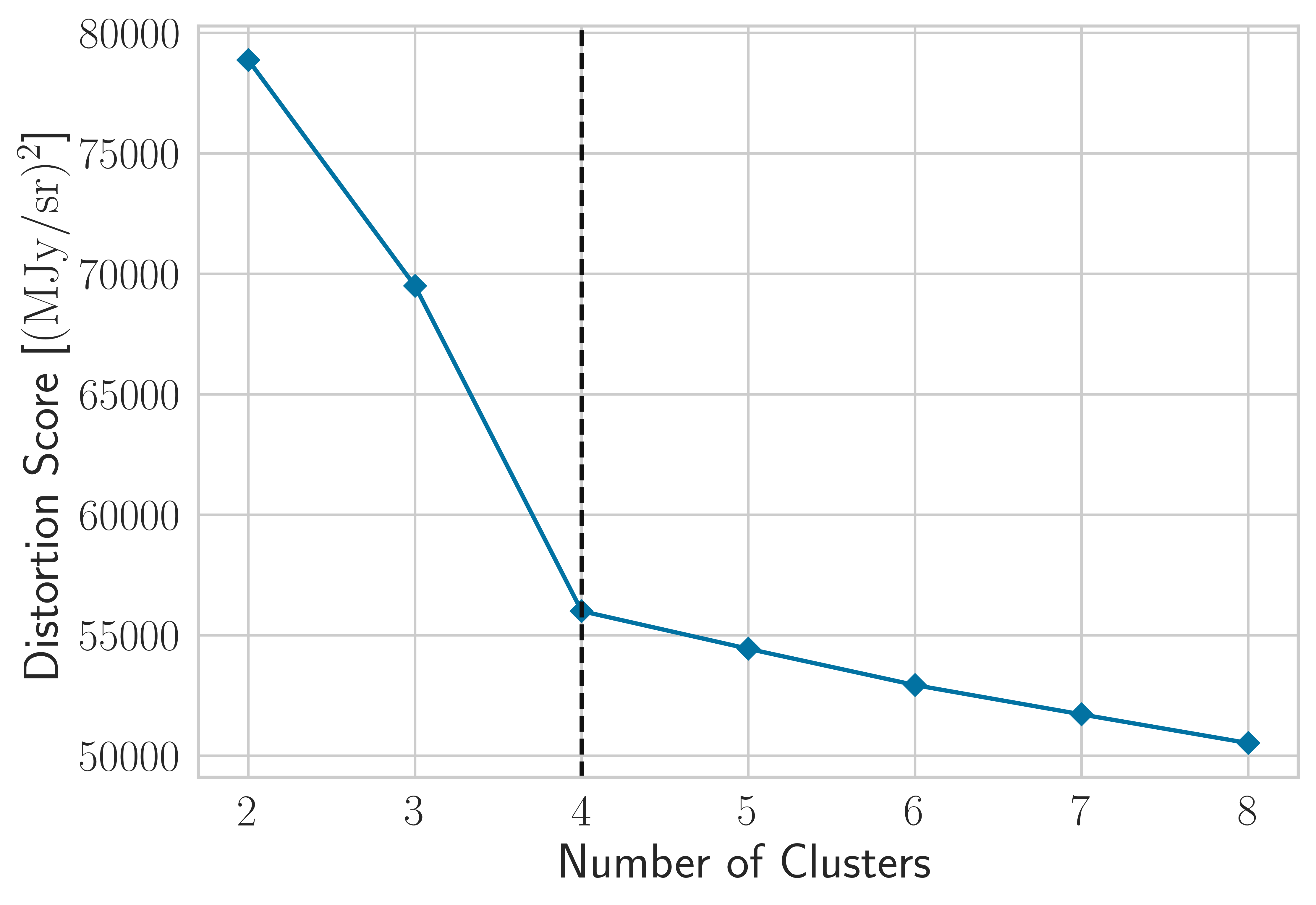}  \includegraphics[width = 8cm, height = 5.5cm]{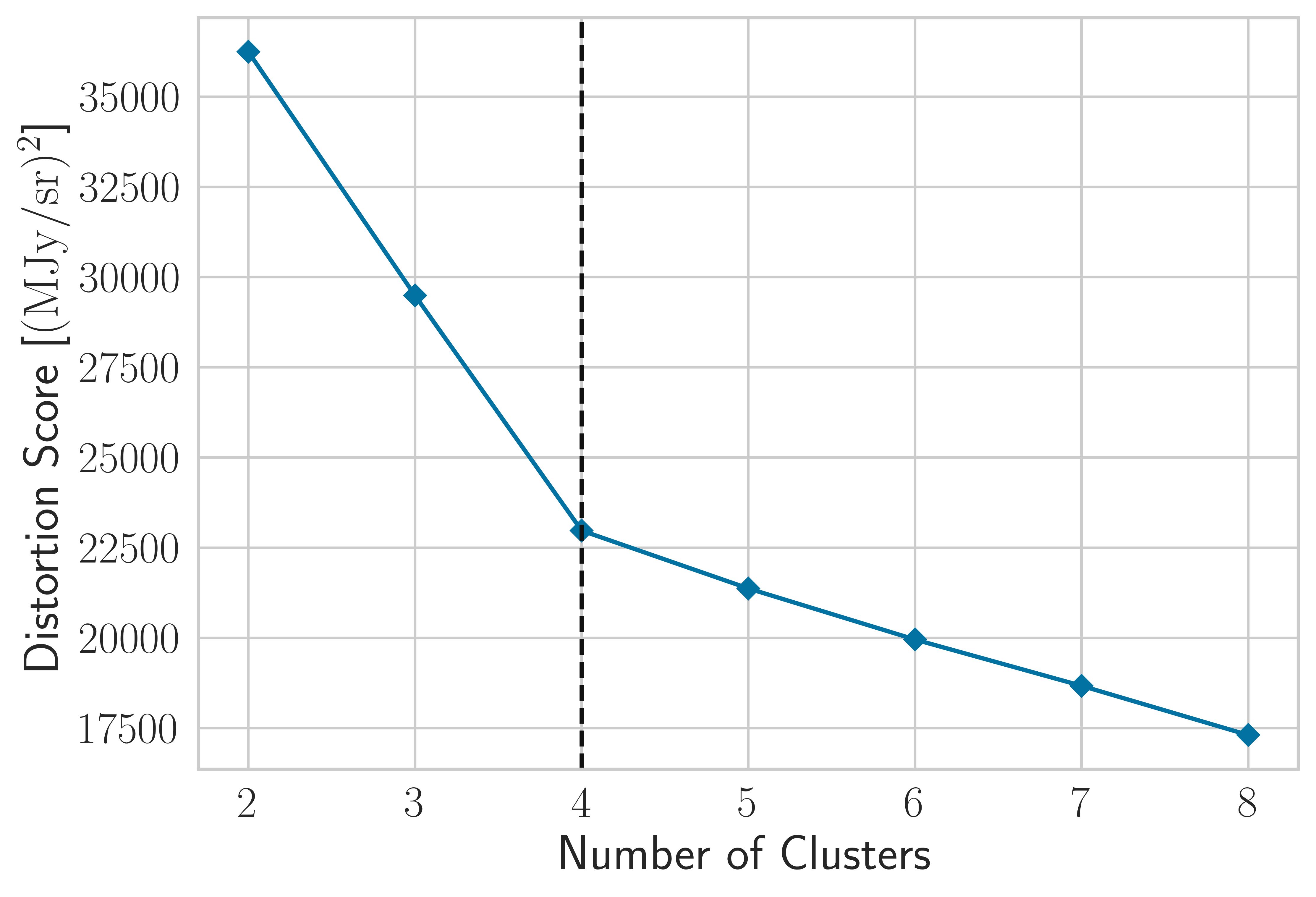}
}
   \resizebox{.99\hsize}{!}{%

   \includegraphics[width = 8cm, height = 5.5cm]{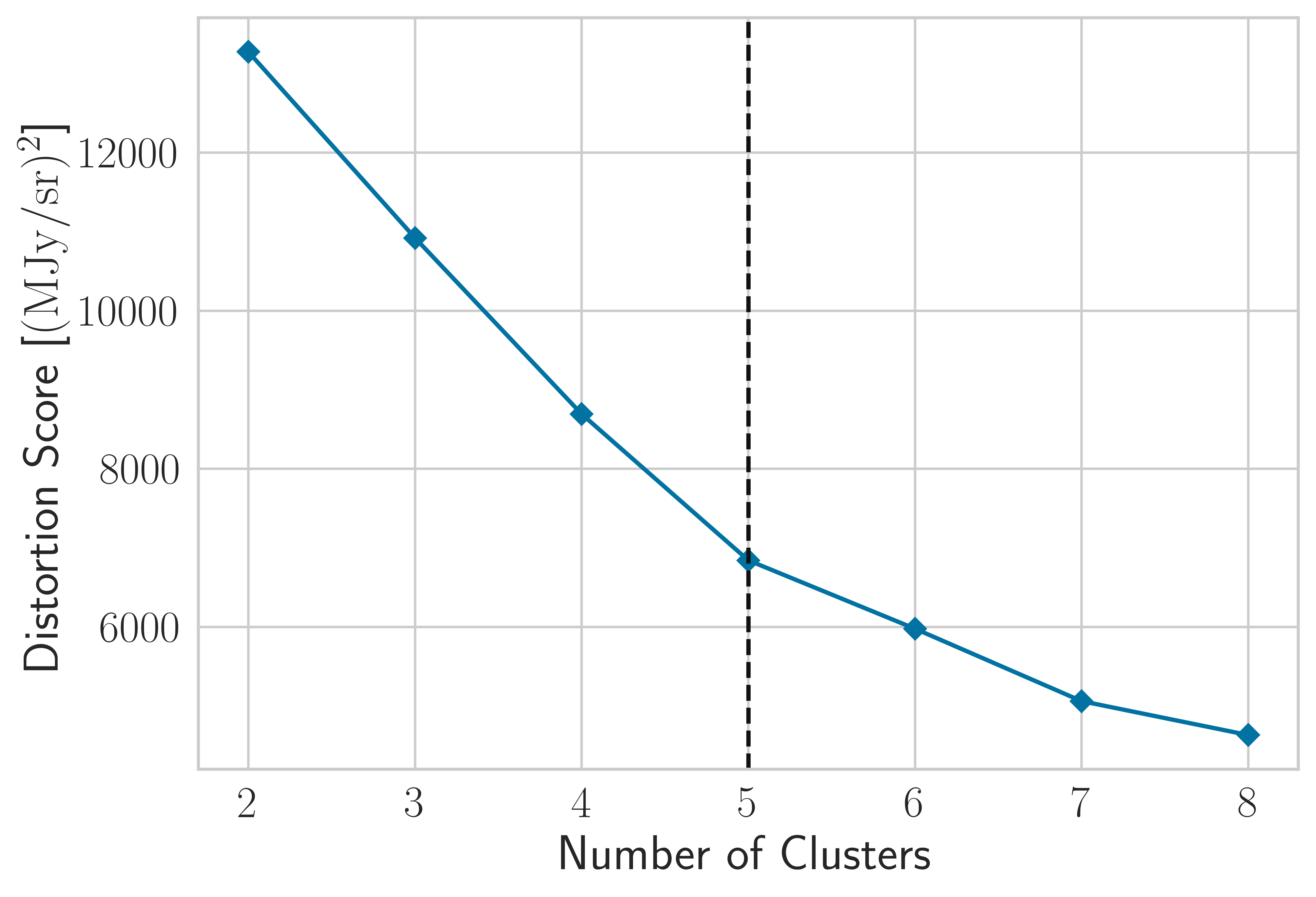}
    \includegraphics[width = 8cm, height = 5.5cm]{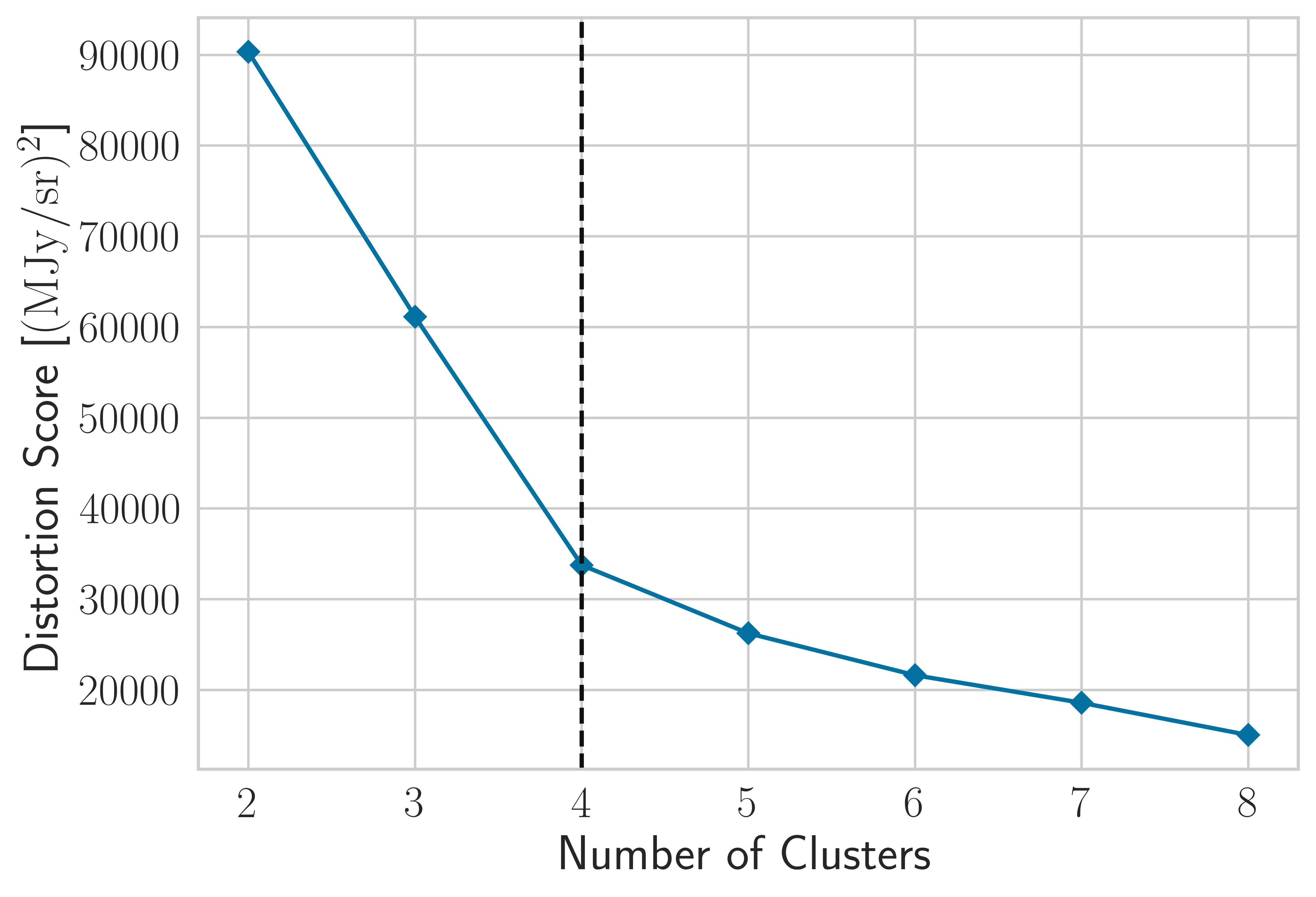}
 }

    \caption{The distortion scores for clustering applied in the 3.2-3.6, 5.95-6.6, 7.25-8.95, and 10.9-11.63 ~\mum region plotted as a function of the number of clusters used. The inflection point, or ``elbow'', in the distortion curve is highlighted with the dashed black line. As is shown, four clusters are optimal for each wavelength regime.}
    \label{fig:elbow_all}
\end{figure*}

\begin{figure*}
    \centering
    \resizebox{.99\hsize}{!}{%
    \includegraphics[width = 6cm, height = 4cm]{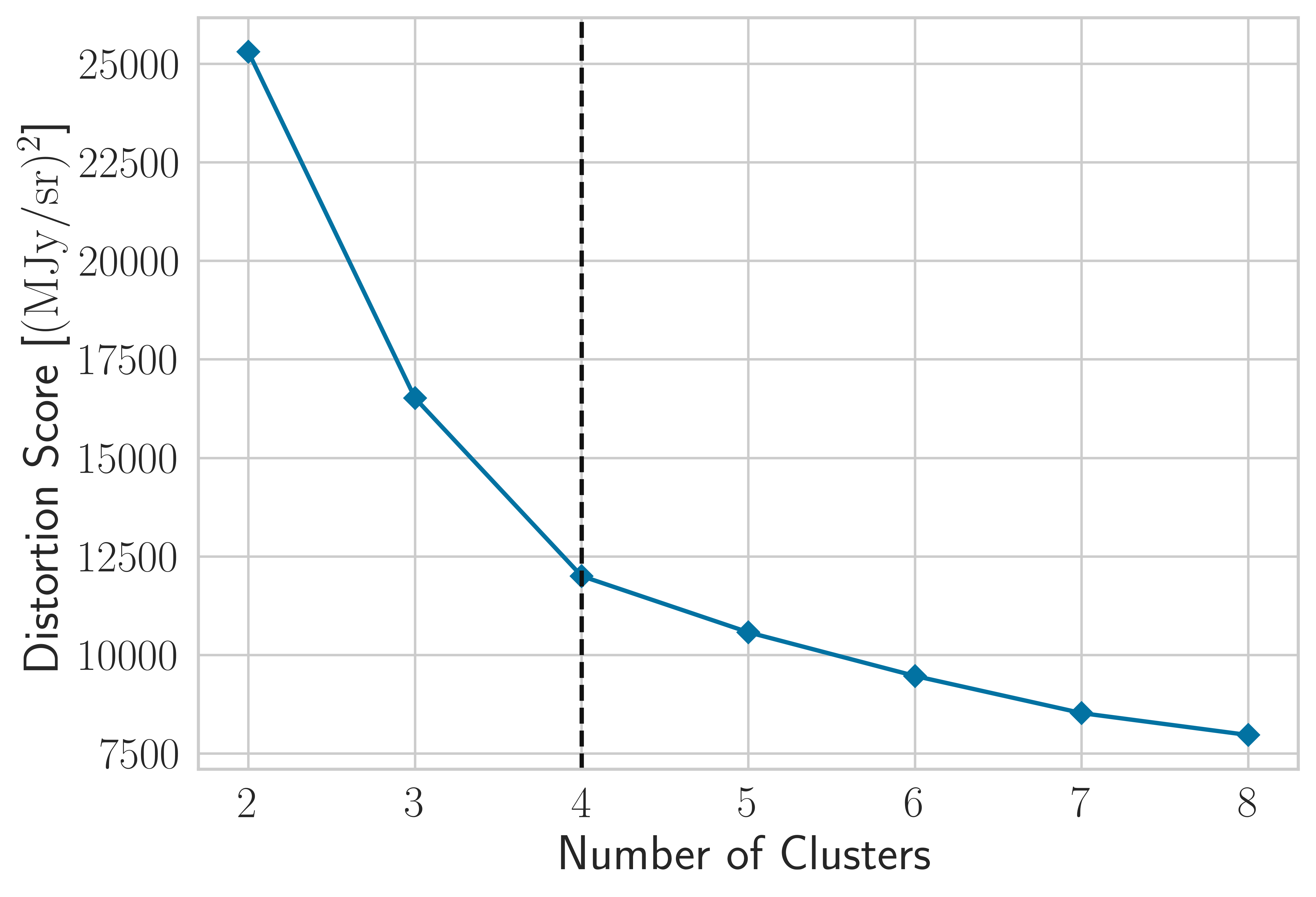}
    \includegraphics[width = 6cm, height = 4cm]{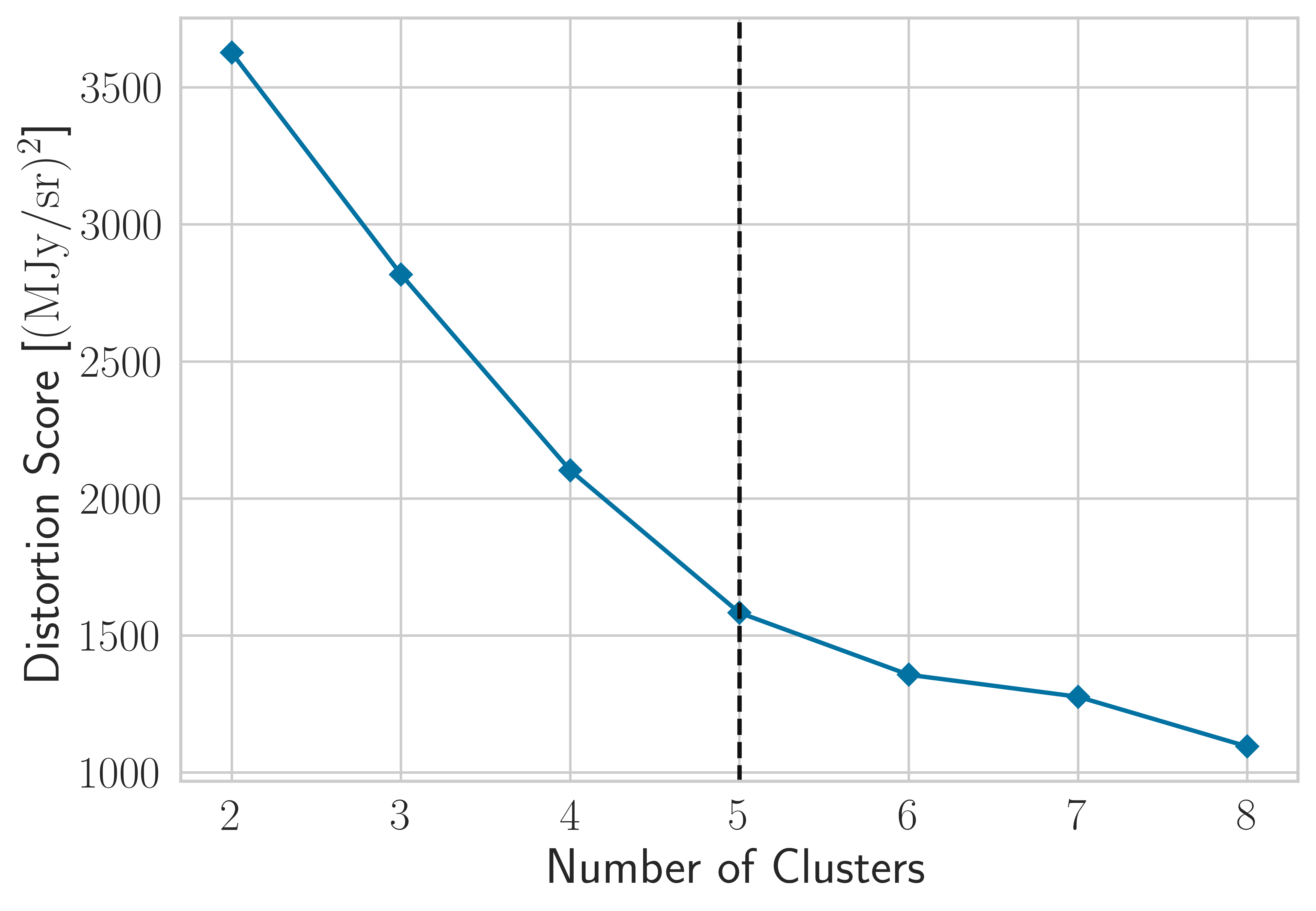}

 }
    \caption{The distortion scores for clustering on the first (left) and second (right) round of clustering for the 4.9-13.2 ~\mum region in the MIRI MRS dataset as a function of the number of clusters used. The inflection point, or ``elbow'', in the distortion curve is highlighted with the dashed black line. In the case of the first round of clustering across the entire, cleaned MIRI MRS footprint, four clusters are optimal. In the case of the clustering focused to pixels belonging to the cluster coincident with the atomic PDR, five clusters are optimal.}
    \label{fig:elbow_132}
\end{figure*}

The elbow plots generated for each of the 3.2-3.6, 5.95-6.6, 7.25-8.95, and 10.9-11.63 ~\mum wavelength regions are given in Fig.~\ref{fig:elbow_all}. The elbow plots generated prior to the initial and secondary rounds of clustering involved in the Atomic-PDR-focused clustering experiments are given in Fig.~\ref{fig:elbow_132}. We also report the silhouette scores and elbow points for each experiment in Table \ref{scores_table}.\par

\begin{table}
    
    \caption{\label{scores_table} The average silhouette scores and elbow points for each of the clustering experiments are given. }
    \begin{center}
     \begin{tabular}{cccc}
    \hline\hline\\[-10pt]
    \textbf{$\Delta \lambda$} & \textbf{\# } & \textbf{Silhouette} & \textbf{Elbow}
    \\ \textbf{[$\mathrm{\mu} m$]} & \textbf{Clusters} & \textbf{Score} & \textbf{Point}
    \\[5pt]
    \hline \\[-10pt]
    \\ 3.2-3.6 & 2& 0.57 & 4
    \\ 3.2-3.6 & 3& 0.457& 4
    \\ 3.2-3.6 & 4& 0.328 & 4
    \\ 3.2-3.6 & 5& 0.286 & 4
    \\ 3.2-3.6 & 6& 0.268& 4
    \\ 3.2-3.6 & 7& 0.271& 4
    \\ 
    \hline 
    5.95-6.60 & 2& 0.521& 4
    \\ 5.95-6.60  & 3& 0.404& 4
    \\ 5.95-6.60  & 4& 0.306& 4
    \\ 5.95-6.60  & 5& 0.273& 4
    \\ 5.95-6.60  & 6& 0.26& 4
    \\ 5.95-6.60  & 7& 0.212& 4
    \\
    \hline
    7.25-8.95 & 2& 0.403& 5
    \\ 7.25-8.95 & 3& 0.416 & 5
    \\ 7.25-8.95 & 4& 0.420& 5
    \\ 7.25-8.95 & 5& 0.423& 5
    \\ 7.25-8.95 & 6& 0.306& 5
    \\ 7.25-8.95 & 7& 0.228& 5
    \\ 
    \hline
    10.90-11.63 & 2& 0.612& 4
    \\ 10.90-11.63 & 3& 0.476& 4
    \\ 10.90-11.63 & 4& 0.469& 4
    \\ 10.90-11.63 & 5& 0.456& 4
    \\ 10.90-11.63 & 6& 0.405& 4
    \\ 10.90-11.63 & 7& 0.397& 4
    \\ 
    \hline

    \hline
    <13.2\tablefootmark{a} & 2& 0.471& 5
    \\<13.2 & 3& 0.493& 5
    \\<13.2 & 4& 0.349& 5
    \\<13.2 & 5& 0.336& 5
    \\<13.2 & 6& 0.33& 5
    \\<13.2 & 7& 0.33& 5
    \\
\hline \\[-5pt]    
    \end{tabular}
    \tablefoot{
    \tablefoottext{a}{Here are silhouette scores for the secondary round of clustering applied to the pixels belonging to the cluster from the initial round of clustering that coincides most closely to the atomic PDR.}}
           
\end{center}
\end{table}

\section{Further Clustering Applications}\label{app:more_clusters}

\begin{figure*}
    \centering
    \resizebox{.9\hsize}{!}{%
    \includegraphics[width = 3cm, height = 1.5cm]{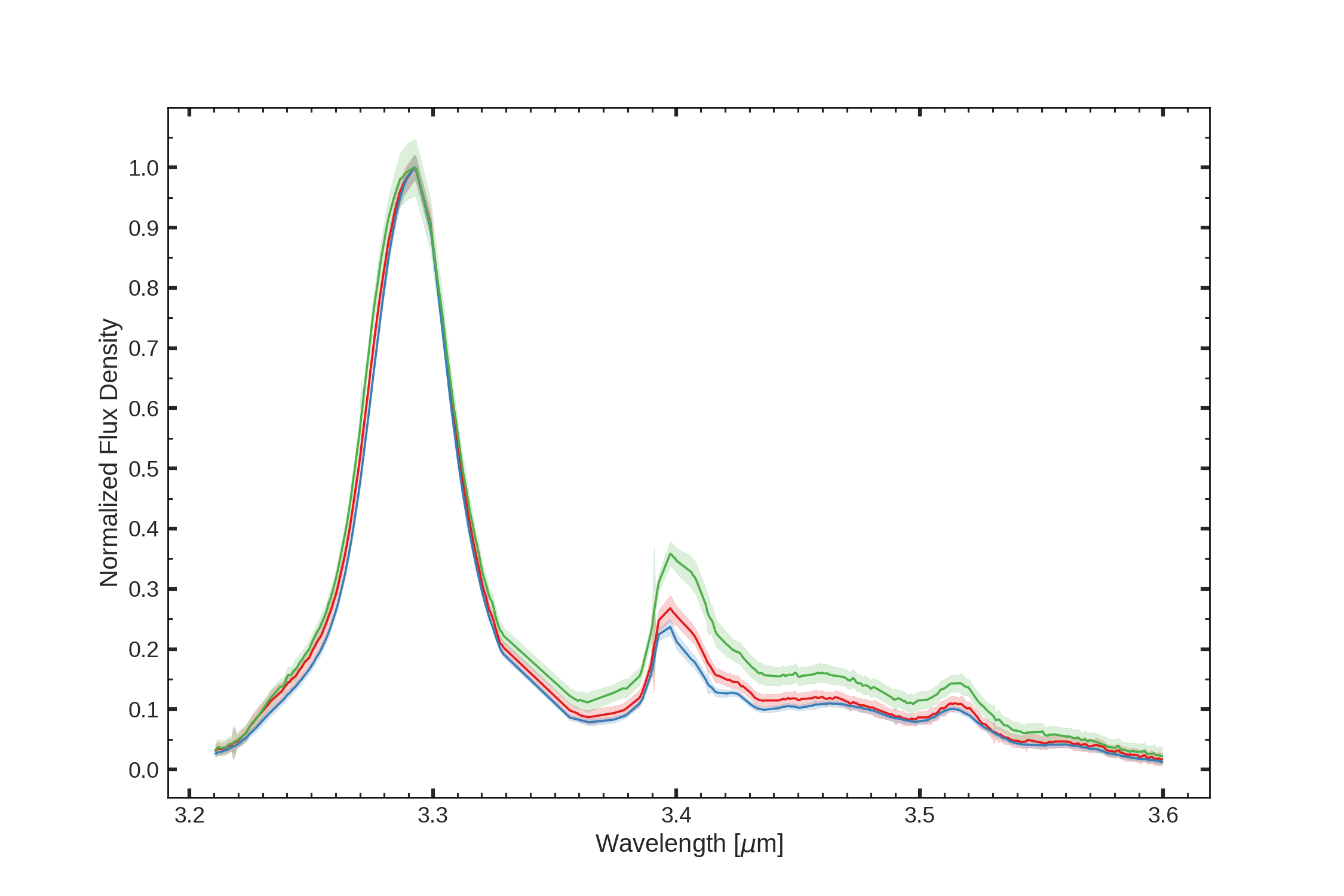}
    \includegraphics[width = 1cm, height = 1.5cm]{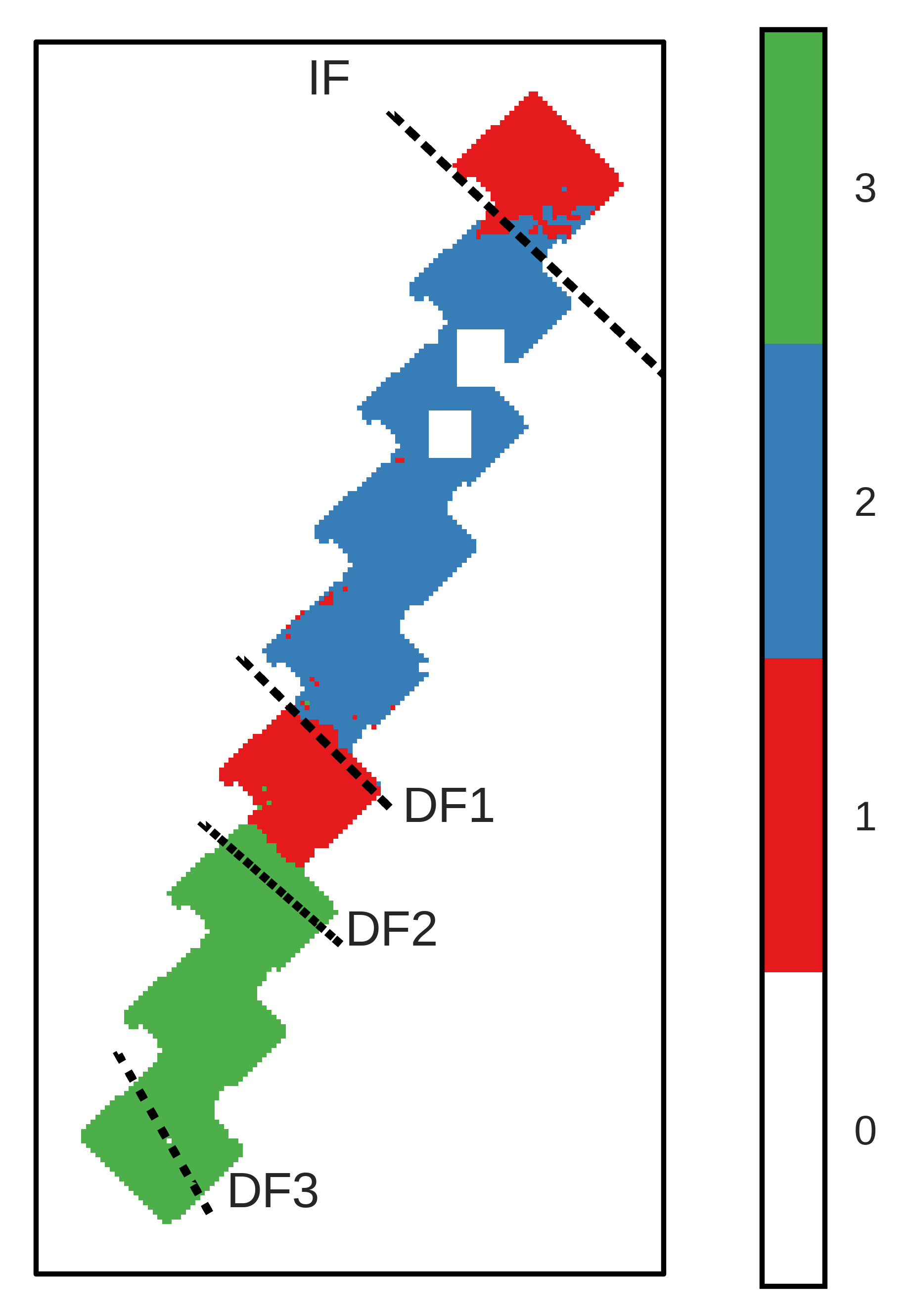}
 }
    \resizebox{.9\hsize}{!}{%
    \includegraphics[width = 3cm, height = 1.5cm]{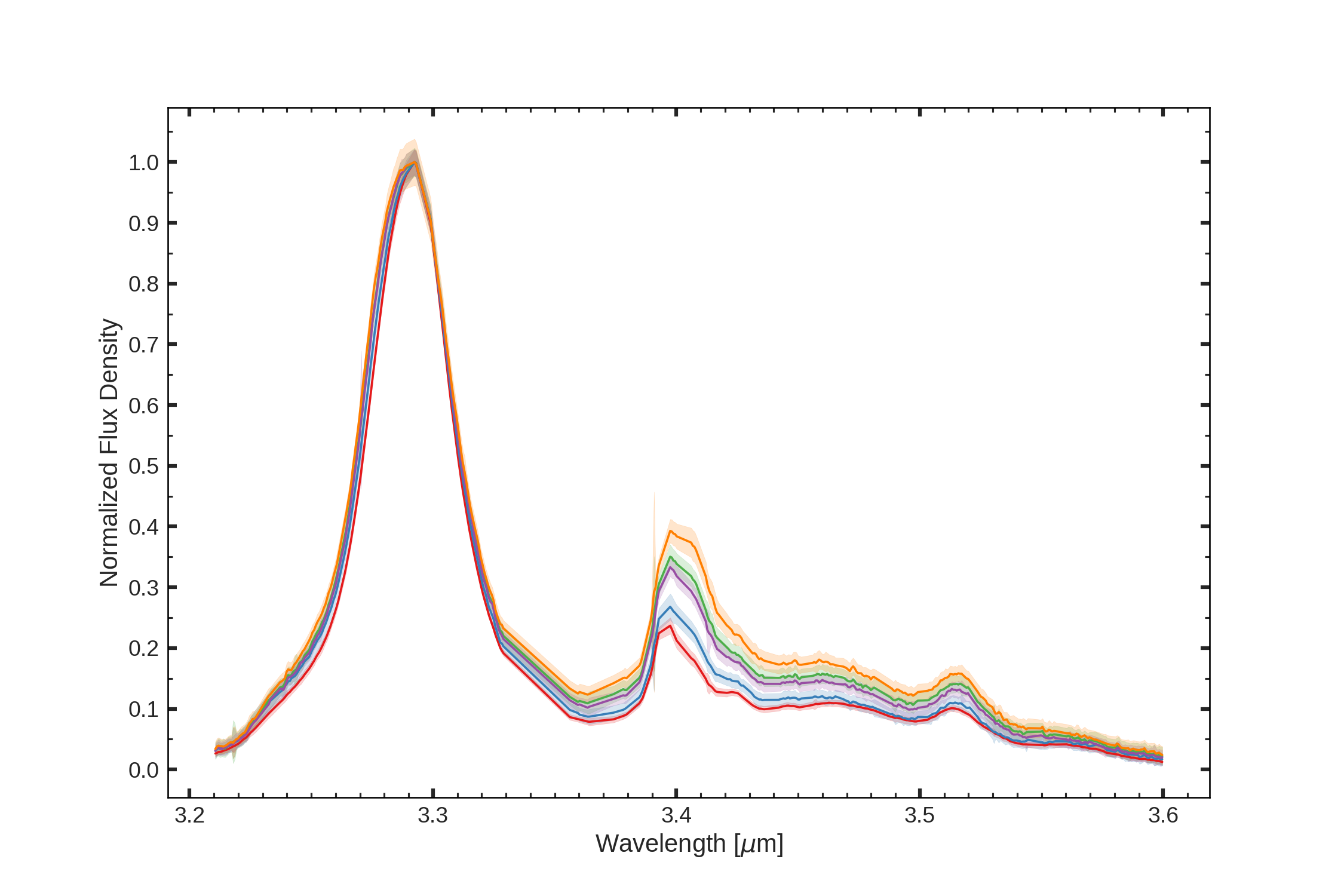}
    \includegraphics[width = 1cm, height = 1.5cm]{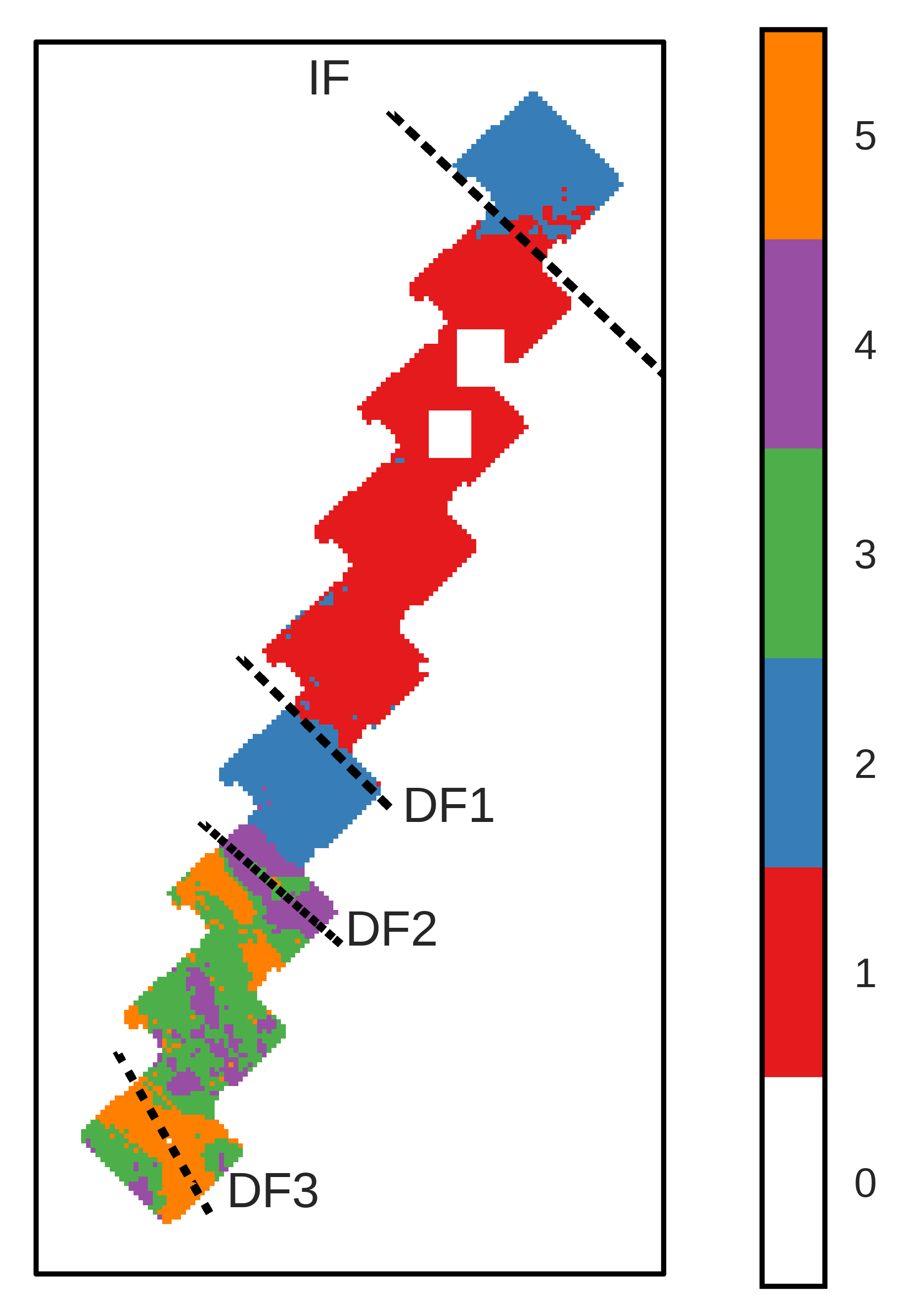}
 }
    \resizebox{.9\hsize}{!}{%
    \includegraphics[width = 3cm, height = 1.5cm]{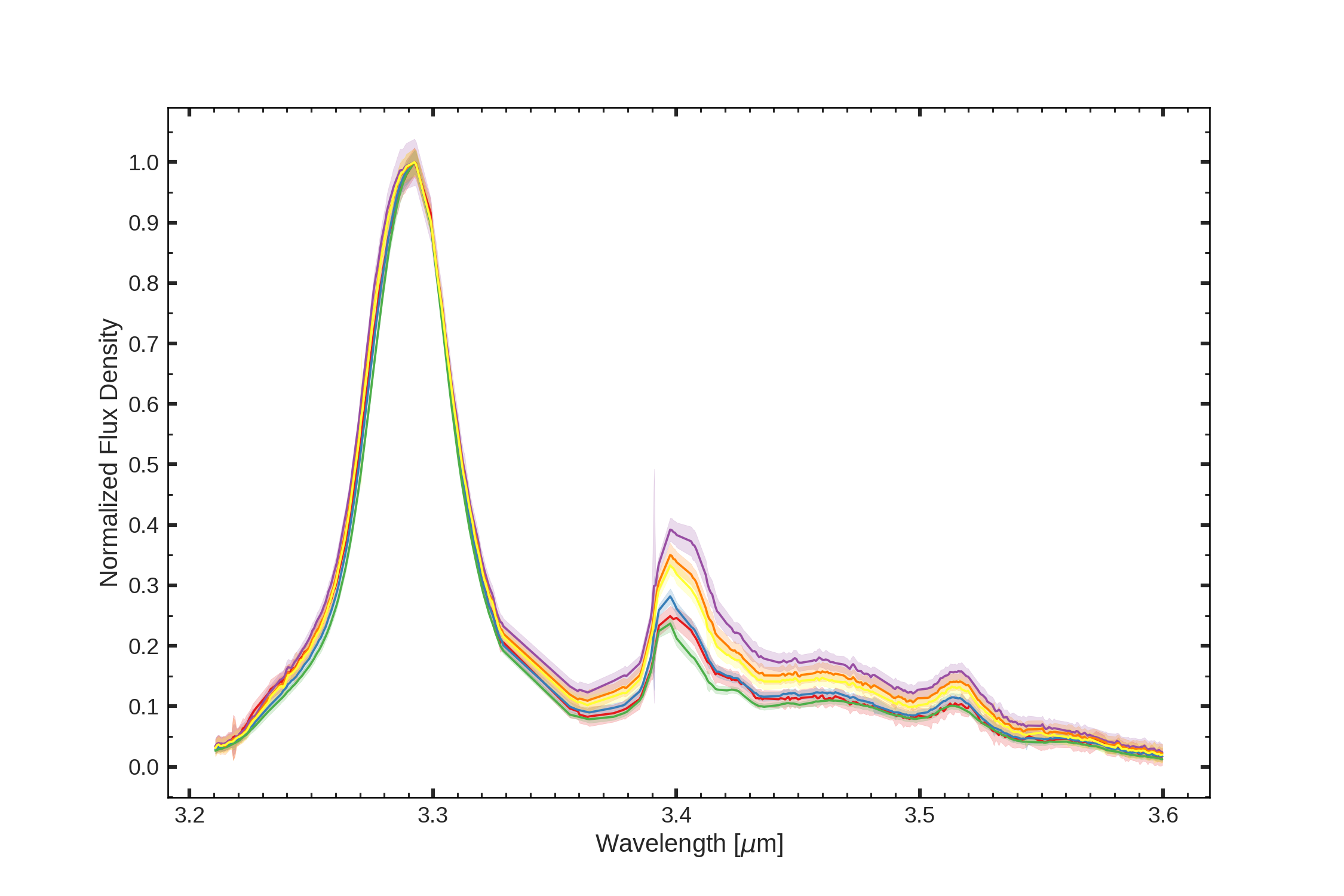}
    \includegraphics[width = 1cm, height = 1.5cm]{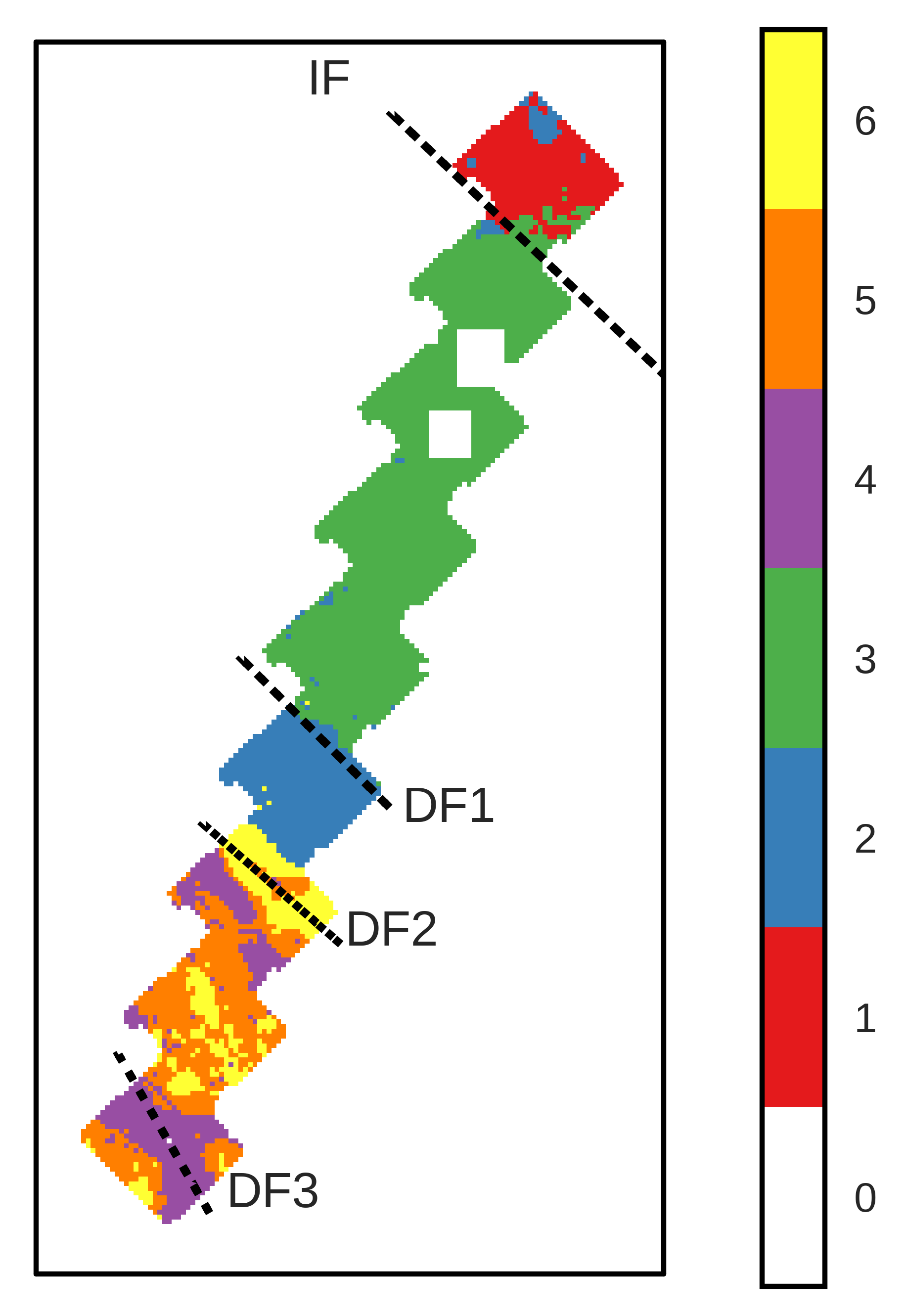}
 }
    \resizebox{.9\hsize}{!}{%
    \includegraphics[width = 3cm, height = 1.5cm]{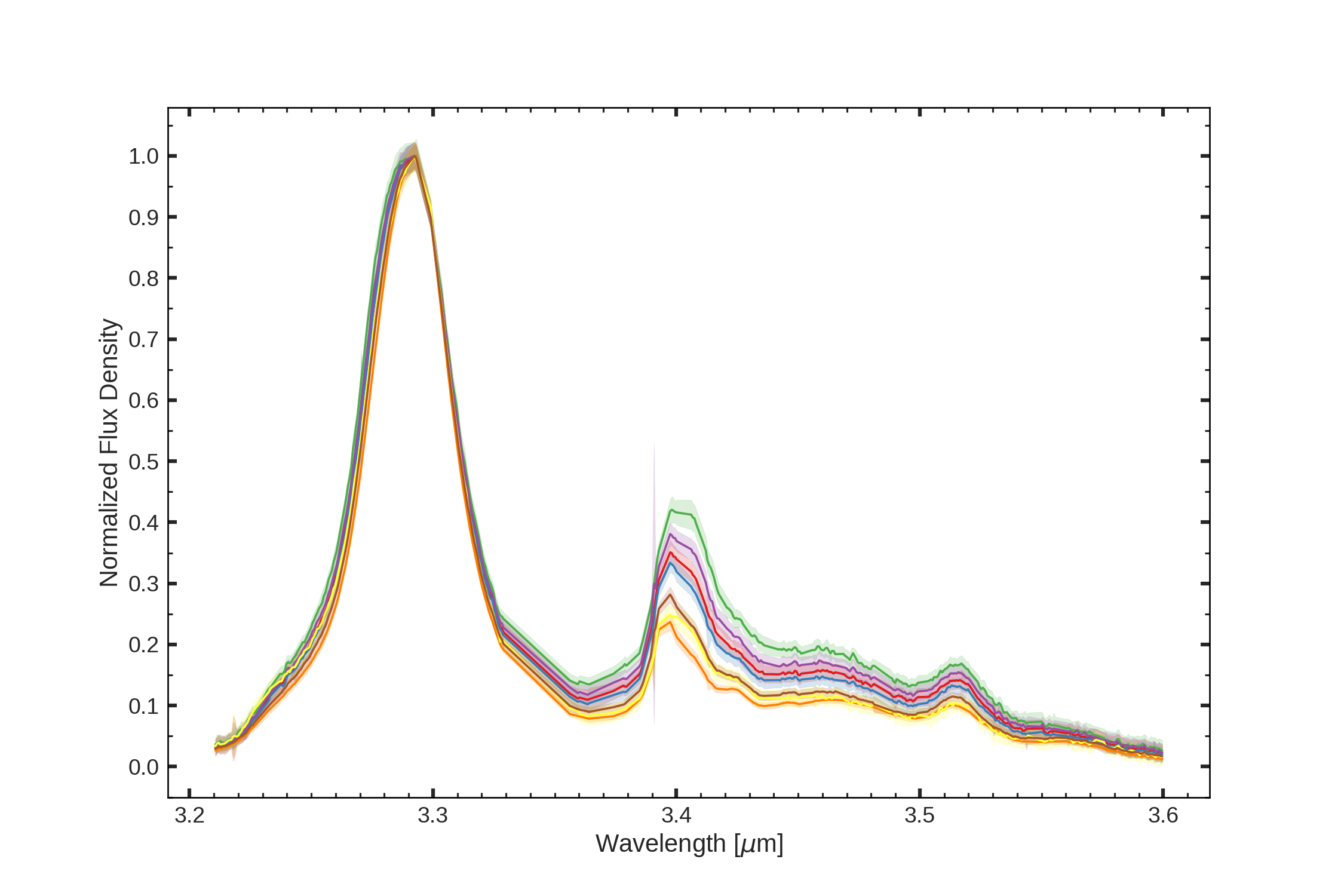}
    \includegraphics[width = 1cm, height = 1.5cm]{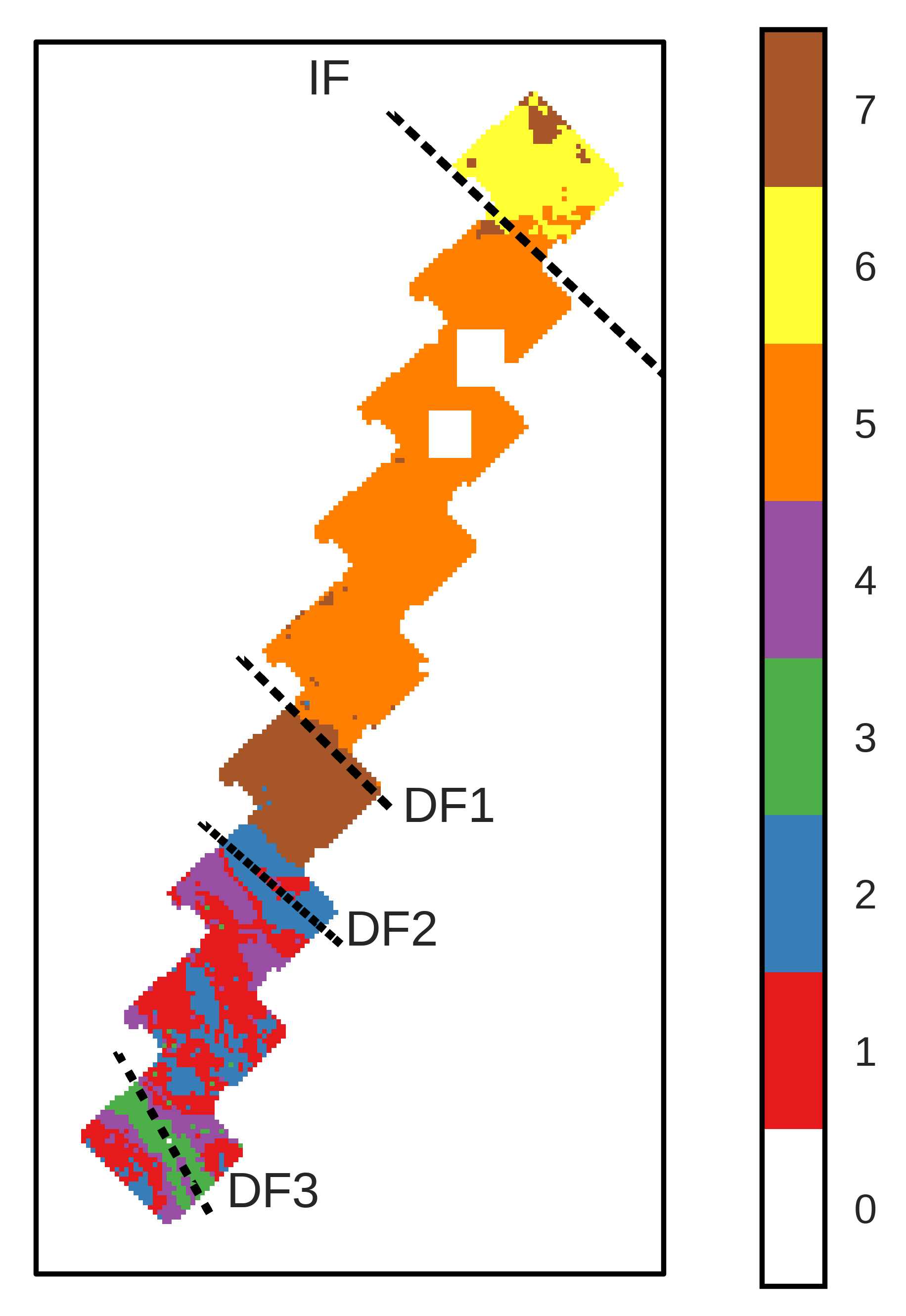}
 }

    \caption{The average spectral profile (left) and spatial footprint (right) for clusters determined in the $3.2-3.6$~\mum region. Each cluster is labelled with a number (in an arbitrary manner). Shading and  normalization (left), and labels and masked pixels (right) are the same as in Fig.~\ref{fig:results_33}.}
    \label{fig:results_33_all}
\end{figure*}

\begin{figure*}
    \centering
    \resizebox{.9\hsize}{!}{%
    \includegraphics[width = 3cm, height = 1.5cm]{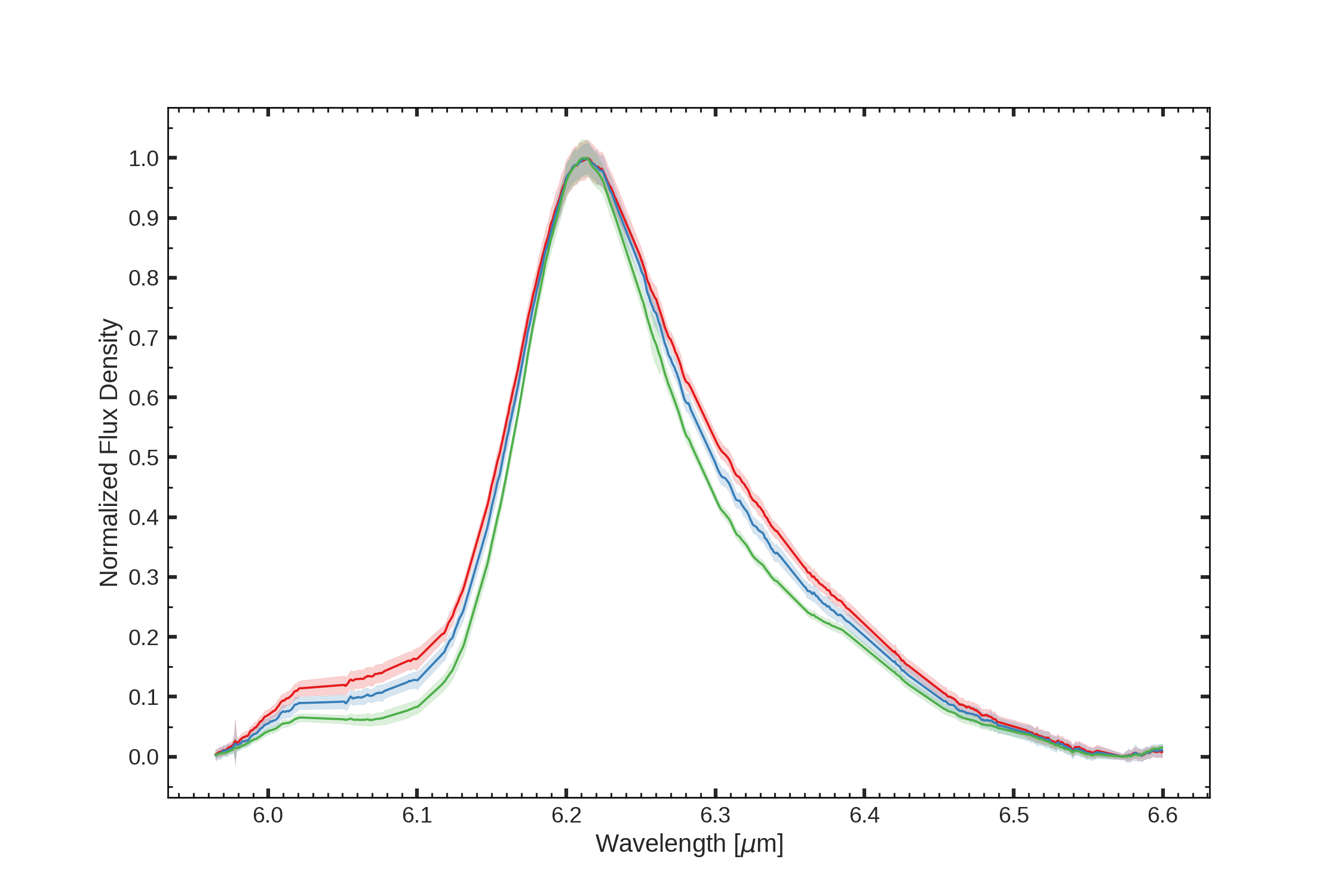}
    \includegraphics[width = 1cm, height = 1.5cm]{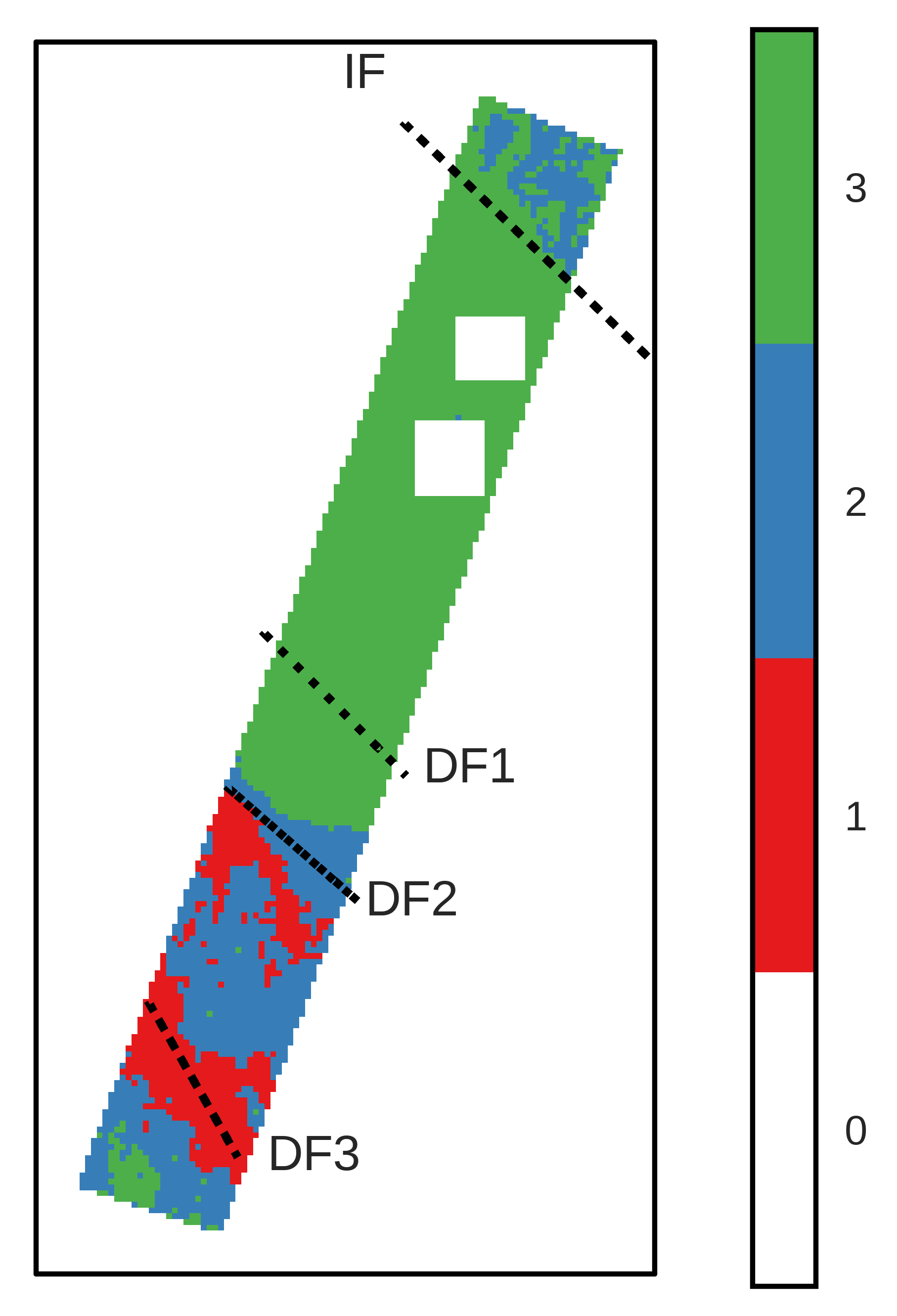}
 }
    \resizebox{.9\hsize}{!}{%
    \includegraphics[width = 3cm, height = 1.5cm]{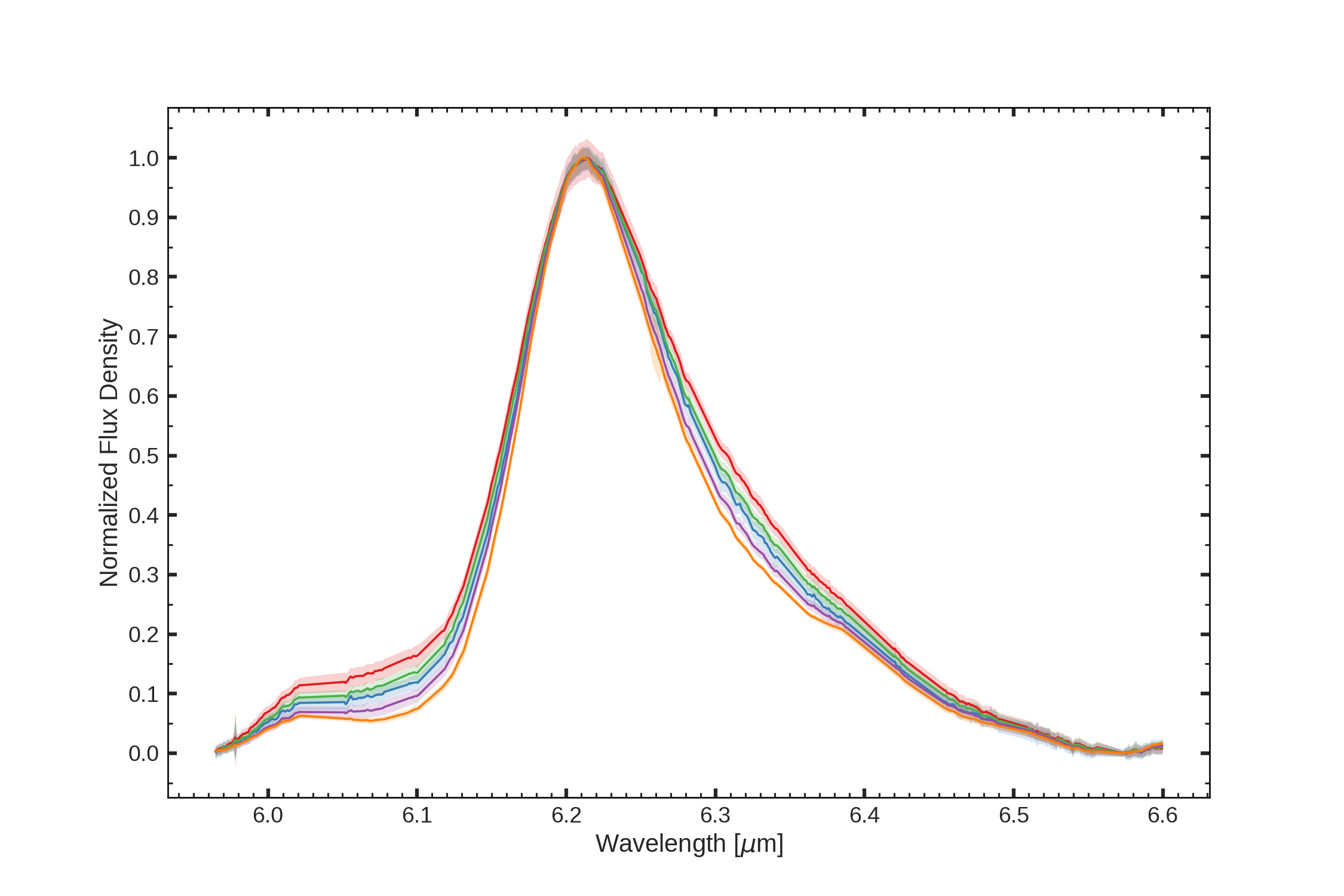}
    \includegraphics[width = 1cm, height = 1.5cm]{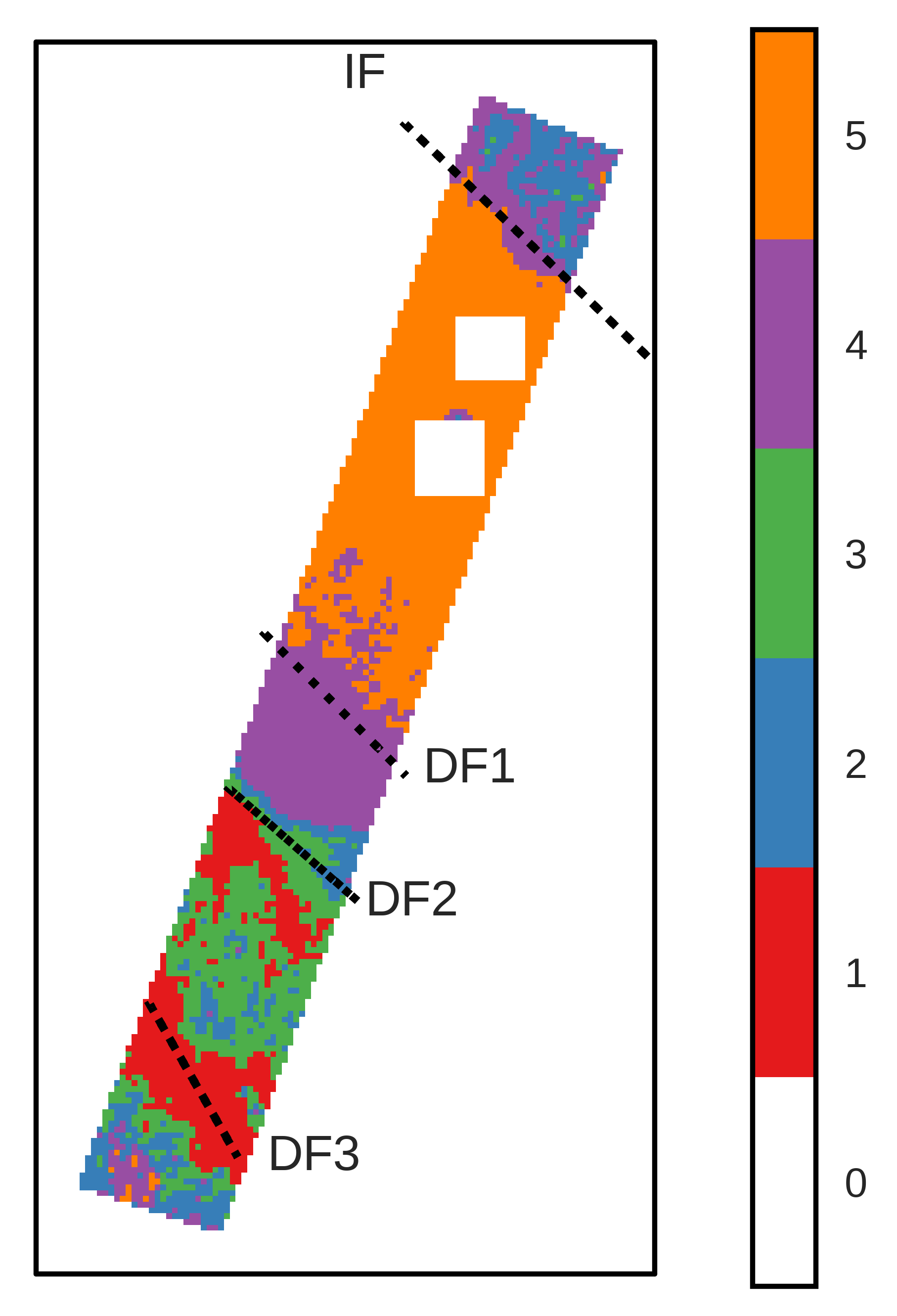}
 }
    \resizebox{.9\hsize}{!}{%
    \includegraphics[width = 3cm, height = 1.5cm]{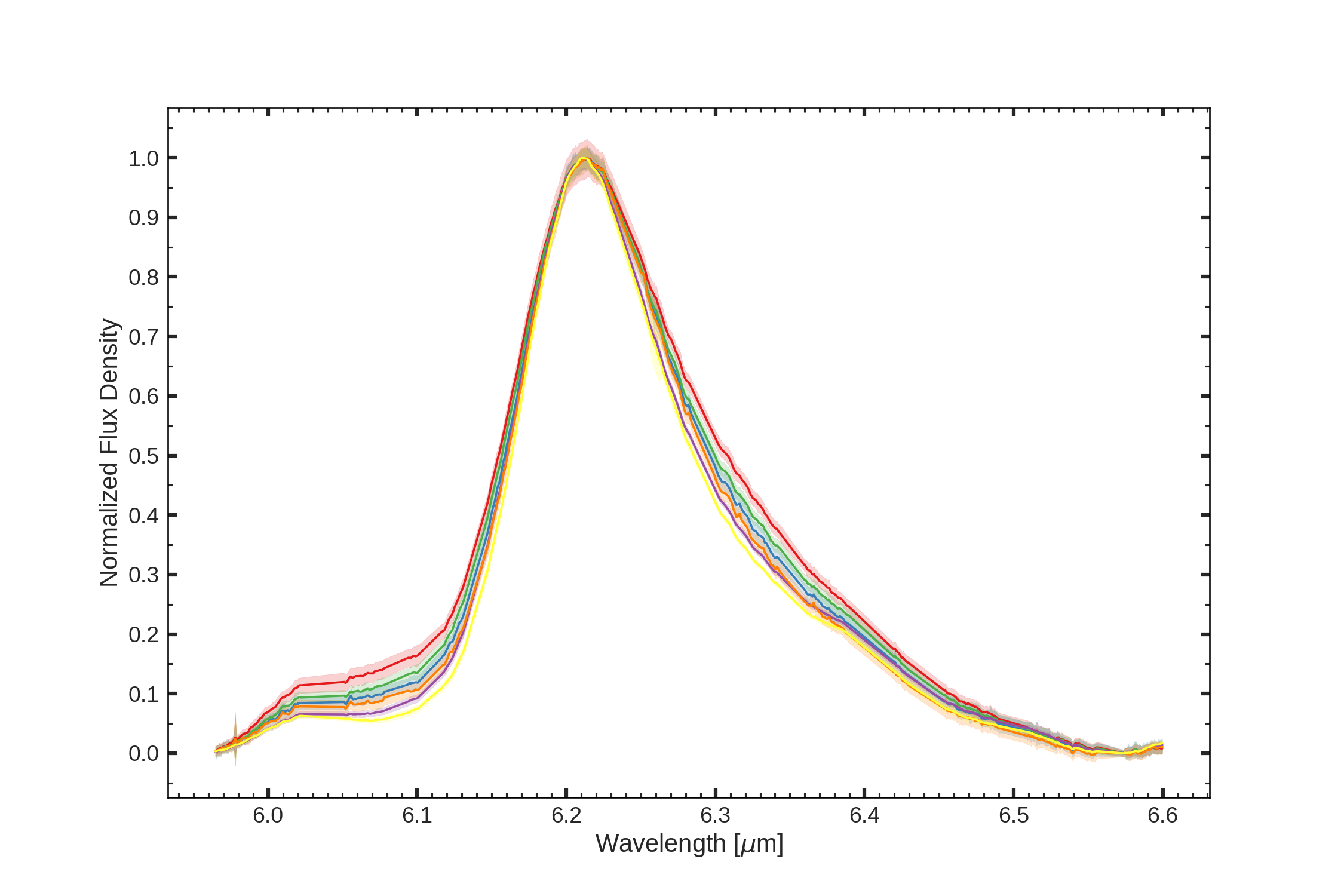}
    \includegraphics[width = 1cm, height = 1.5cm]{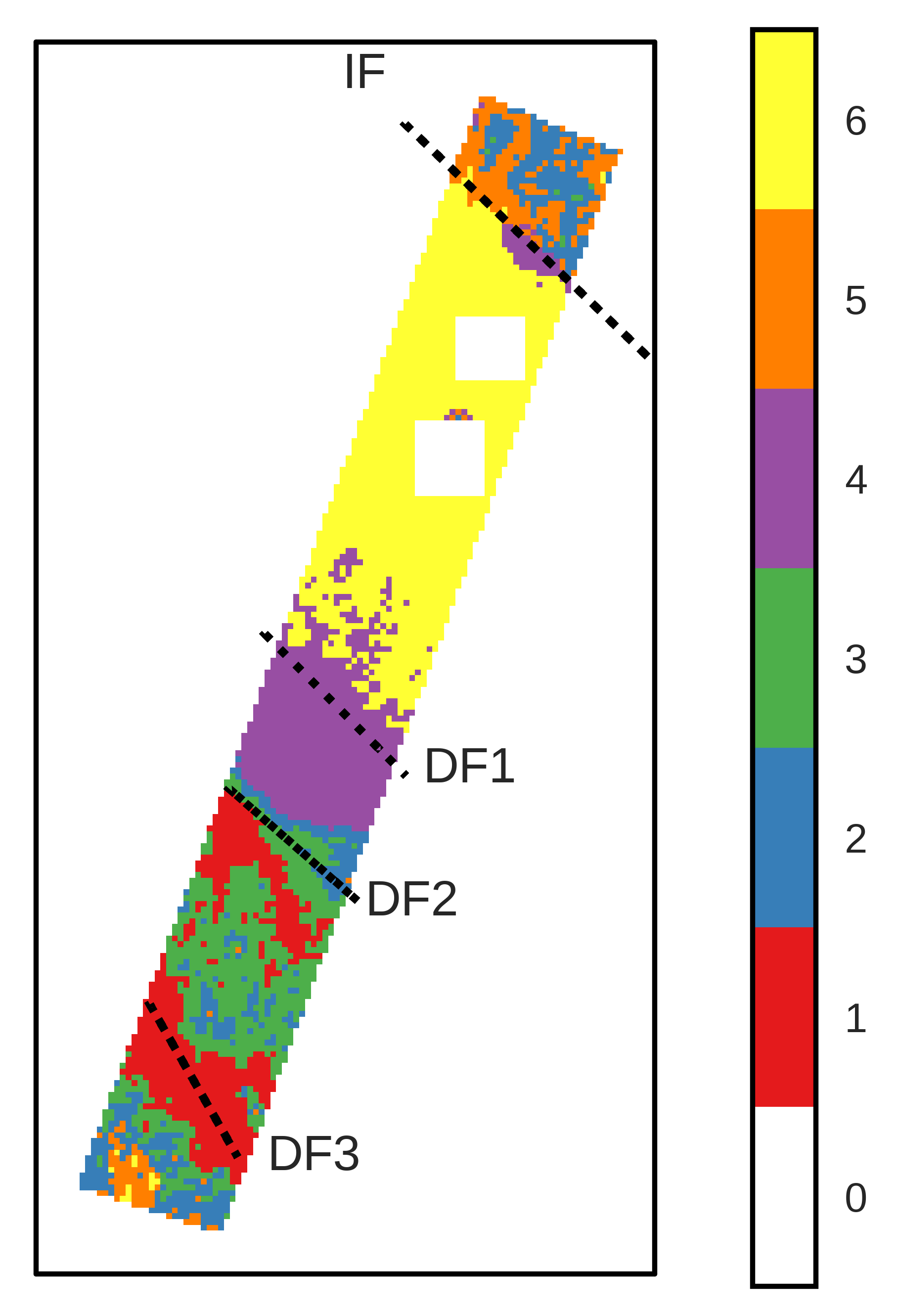}
 }
    \resizebox{.9\hsize}{!}{%
    \includegraphics[width = 3cm, height = 1.5cm]{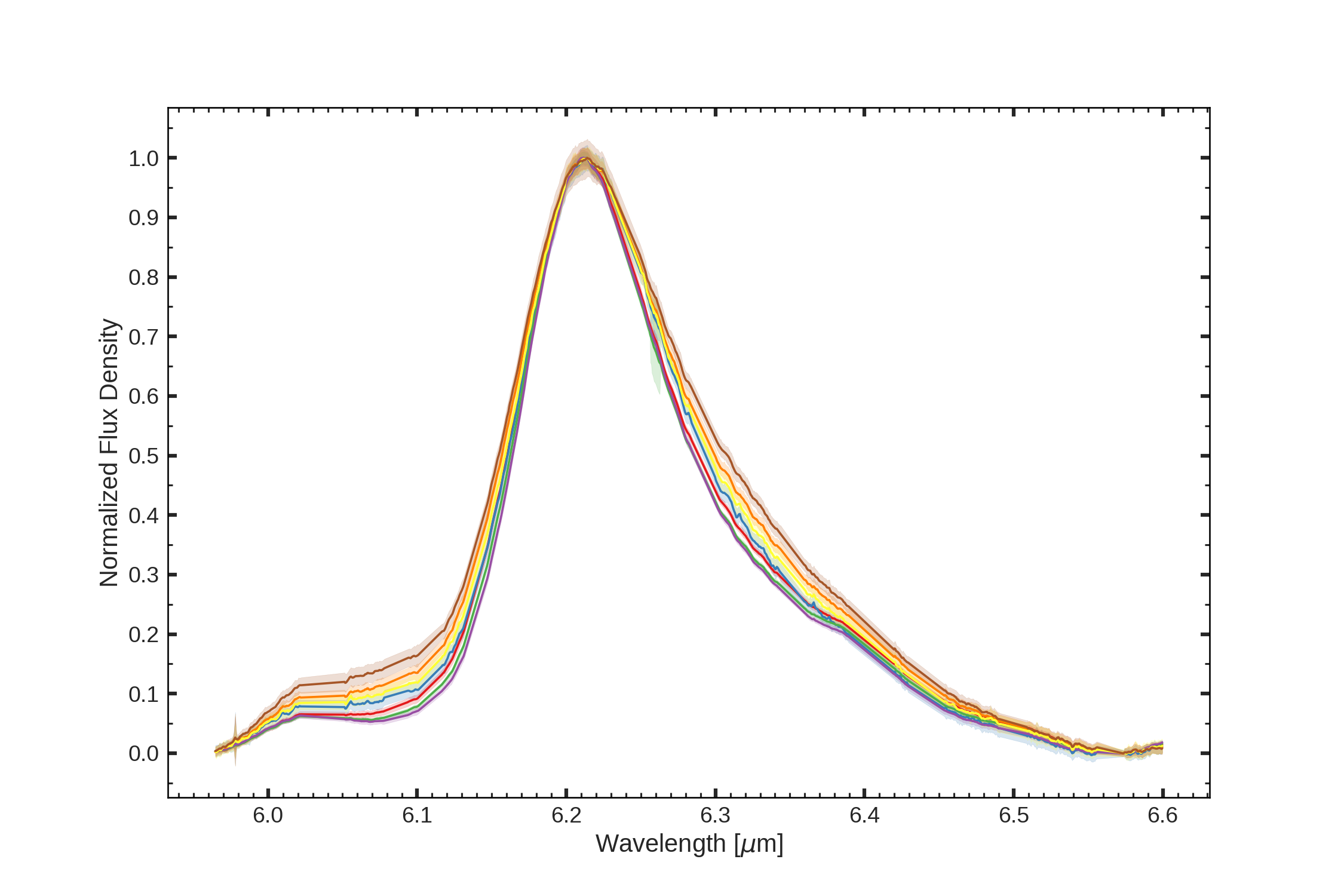}
    \includegraphics[width = 1cm, height = 1.5cm]{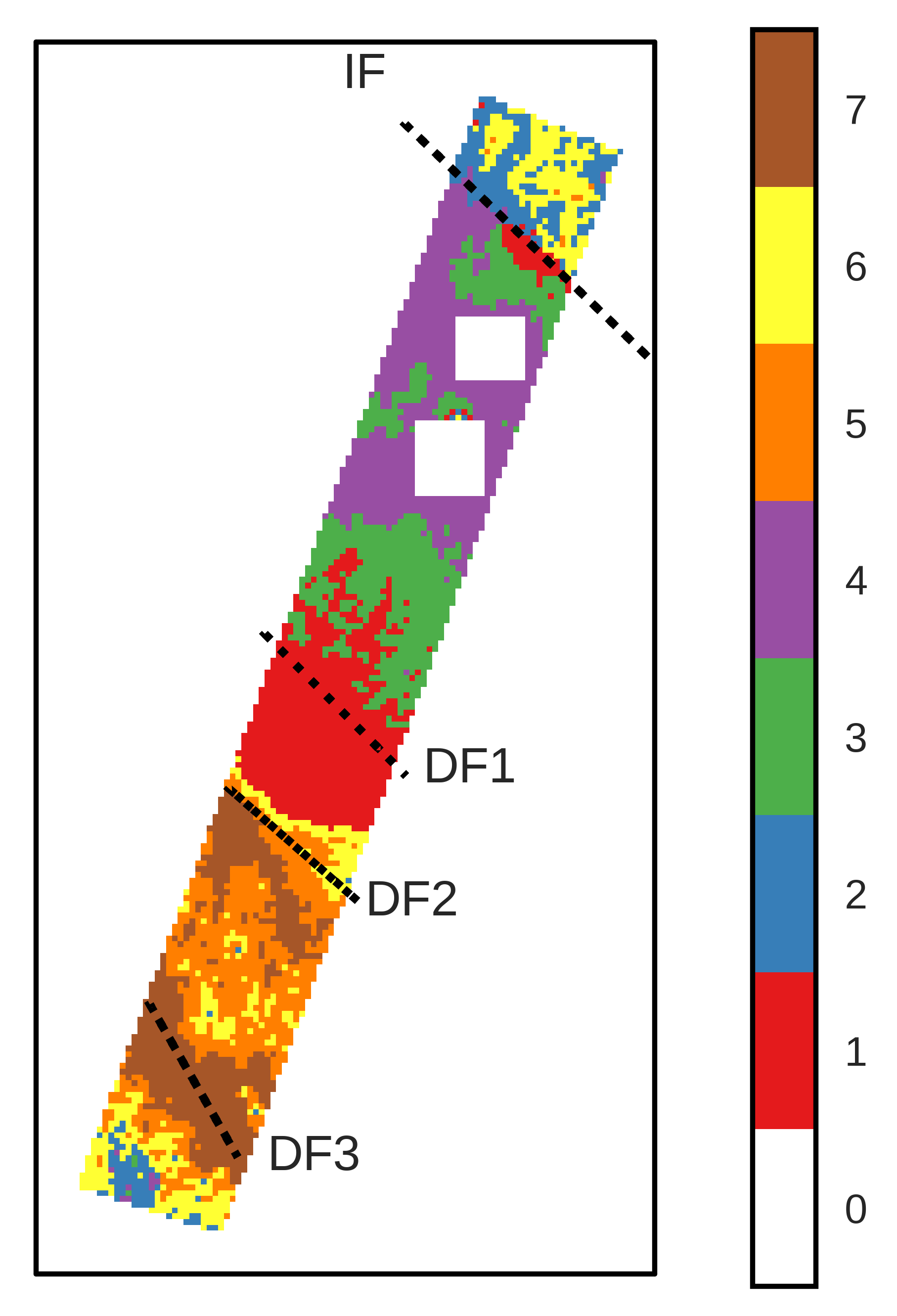}
 }

    \caption{The average spectral profile (left) and spatial footprint (right) for clusters determined in the $5.95-6.6$~\mum region. Each cluster is labelled with a number (in an arbitrary manner). Shading and  normalization (left), and labels and masked pixels (right) are the same as in Fig.~\ref{fig:results_62}.}
    \label{fig:results_62_all}
\end{figure*}

\begin{figure*}
    \centering
    \resizebox{.9\hsize}{!}{%
    \includegraphics[width = 3cm, height = 1.5cm]{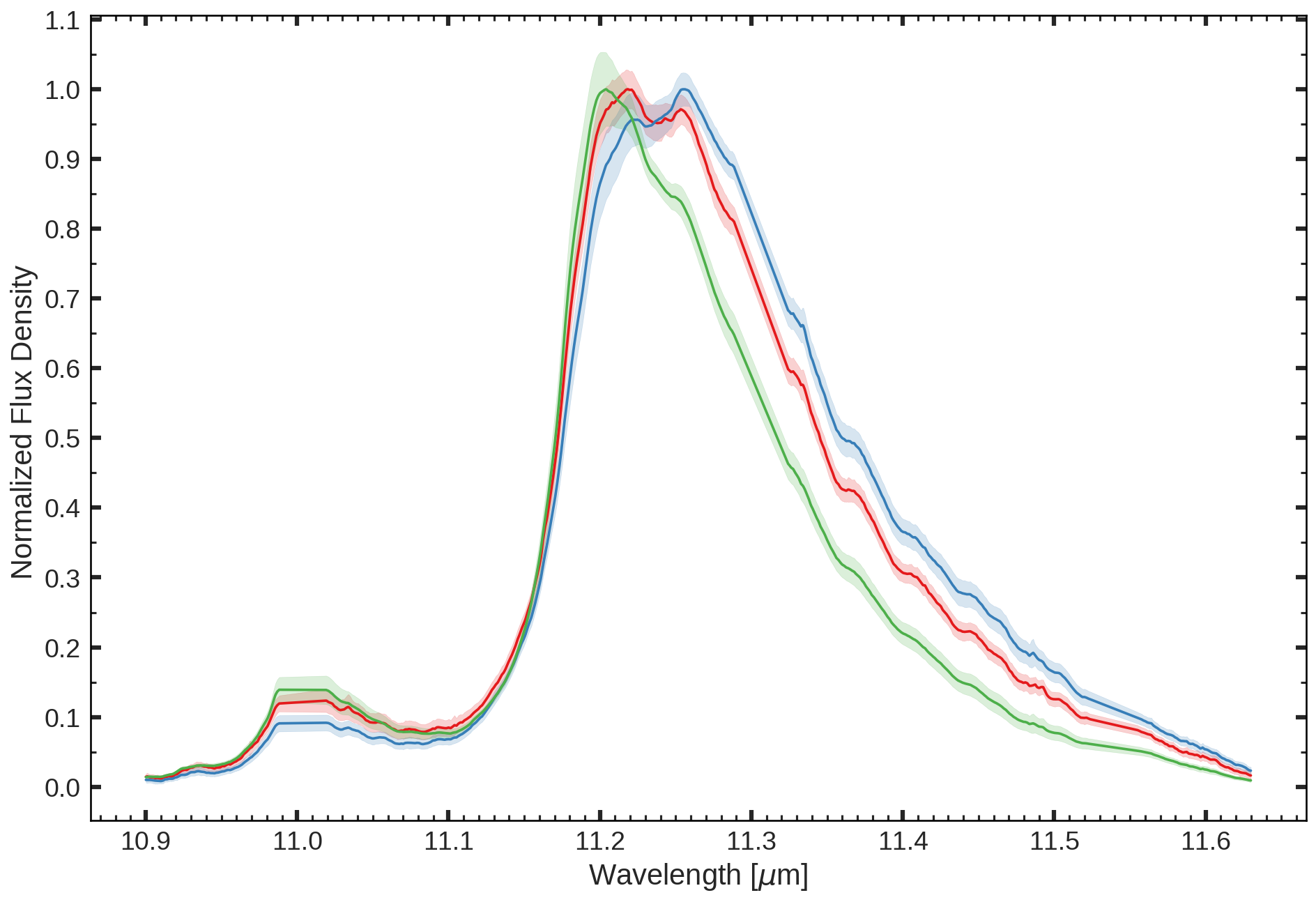}
    \includegraphics[width = 1cm, height = 1.5cm]{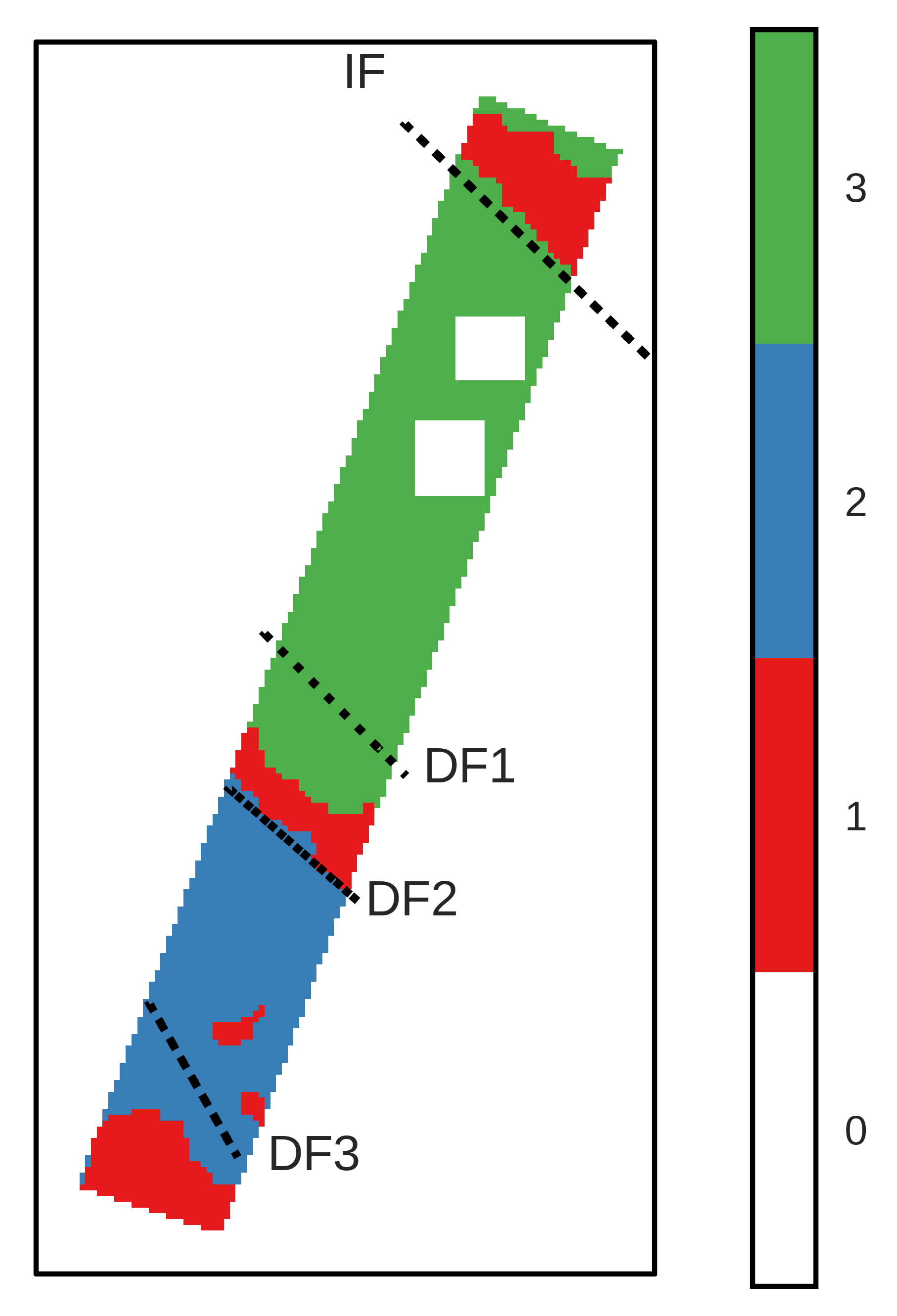}
 }
    \resizebox{.9\hsize}{!}{%
    \includegraphics[width = 3cm, height = 1.5cm]{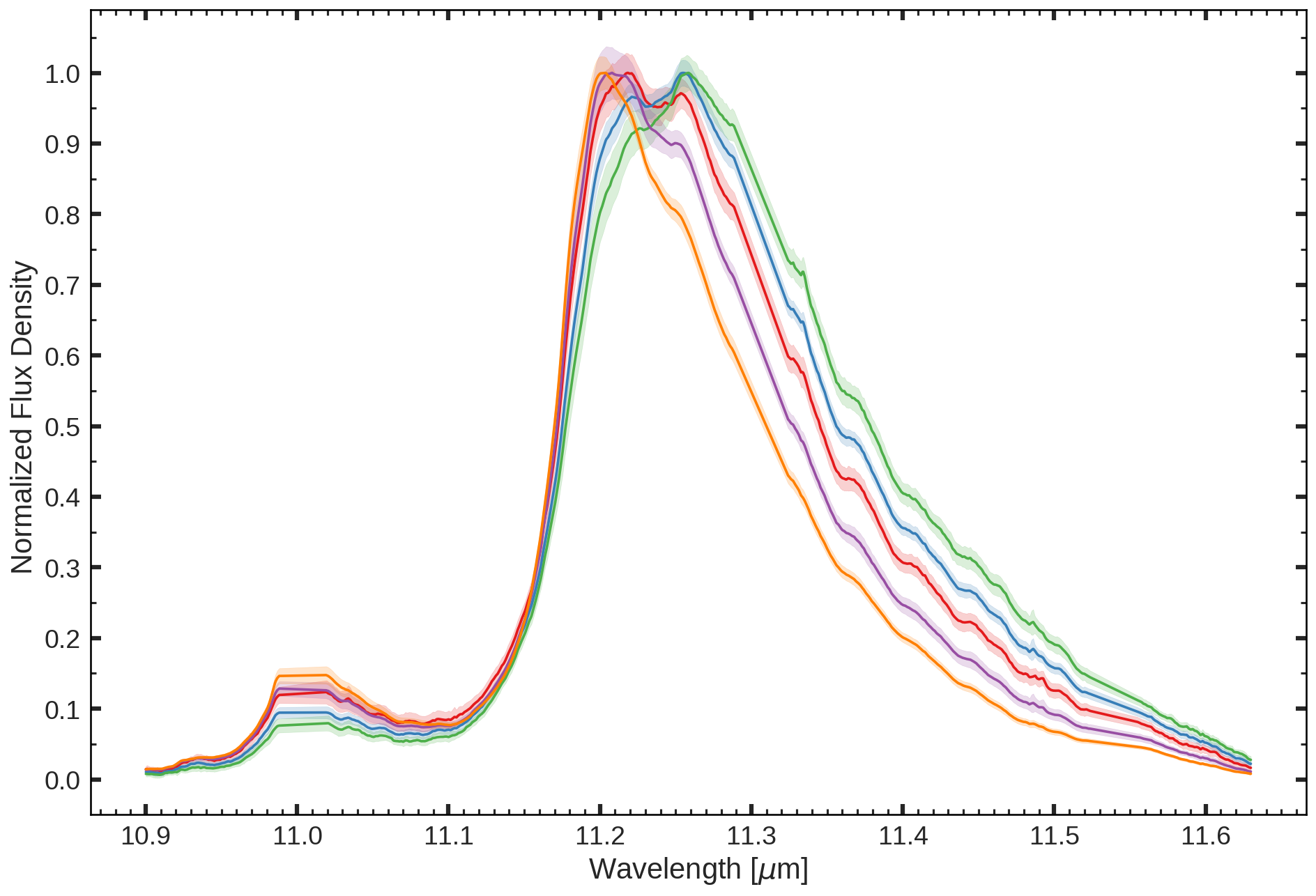}
    \includegraphics[width = 1cm, height = 1.5cm]{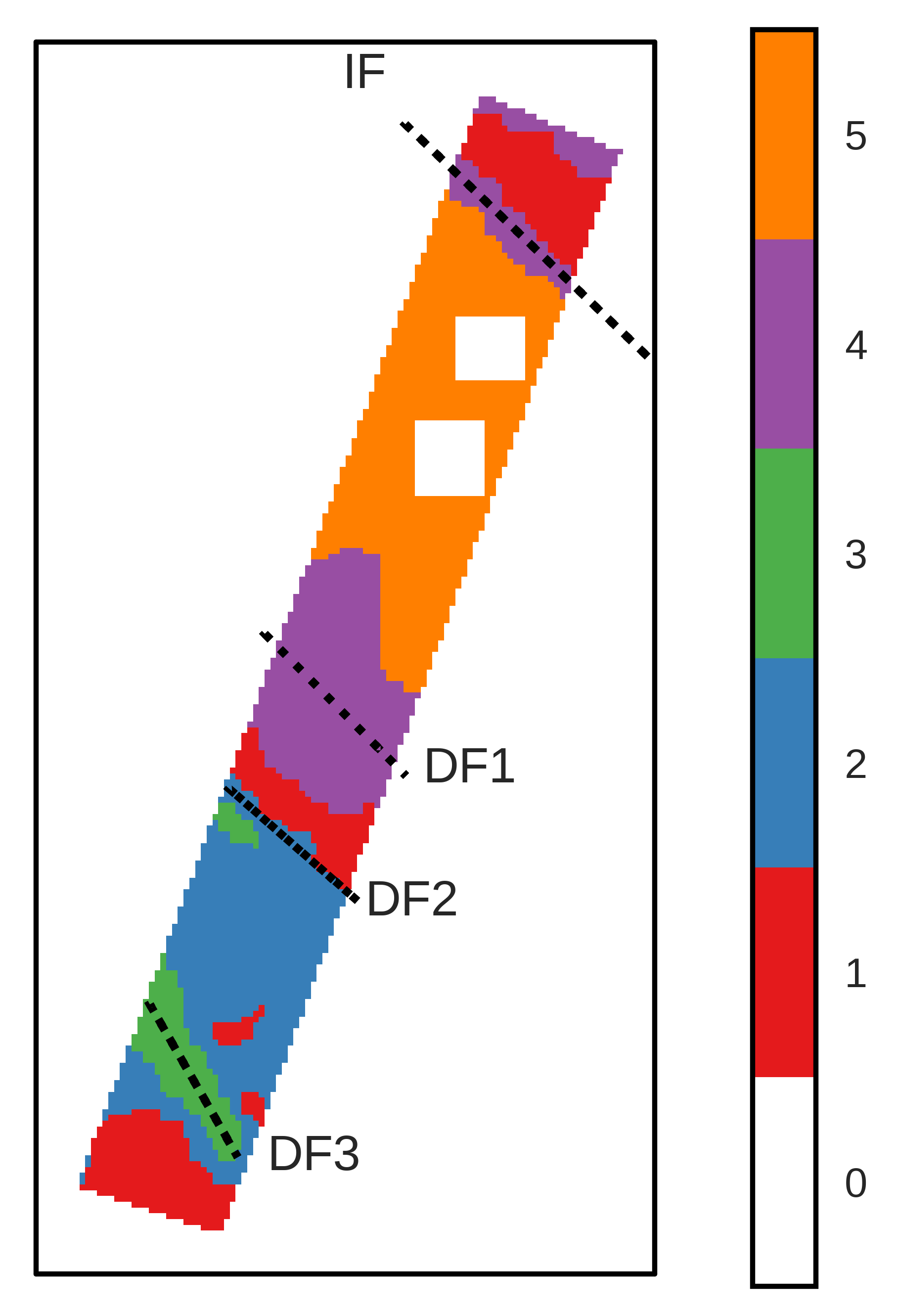}
 }
    \resizebox{.9\hsize}{!}{%
    \includegraphics[width = 3cm, height = 1.5cm]{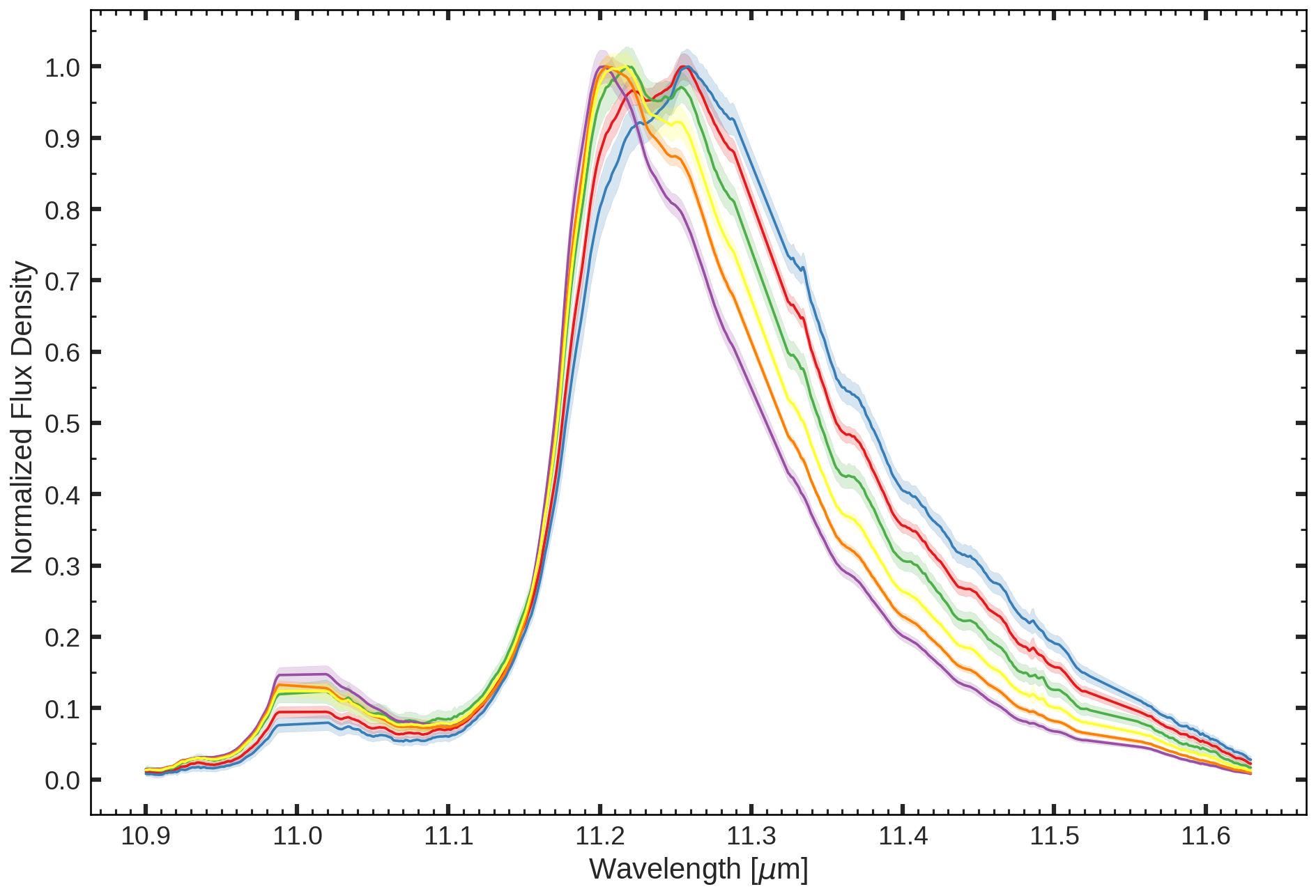}
    \includegraphics[width = 1cm, height = 1.5cm]{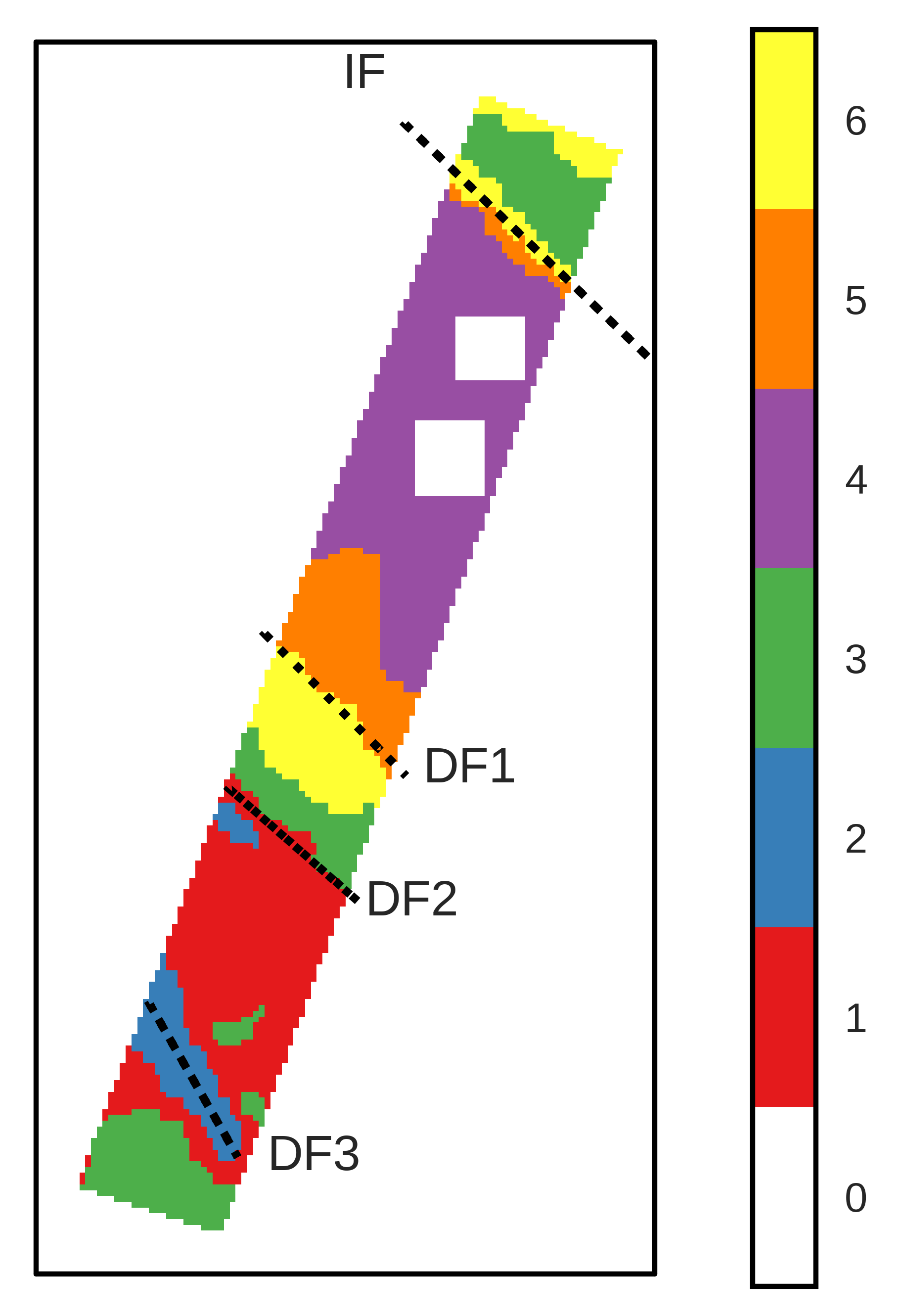}
 }
    \resizebox{.9\hsize}{!}{%
    \includegraphics[width = 3cm, height = 1.5cm]{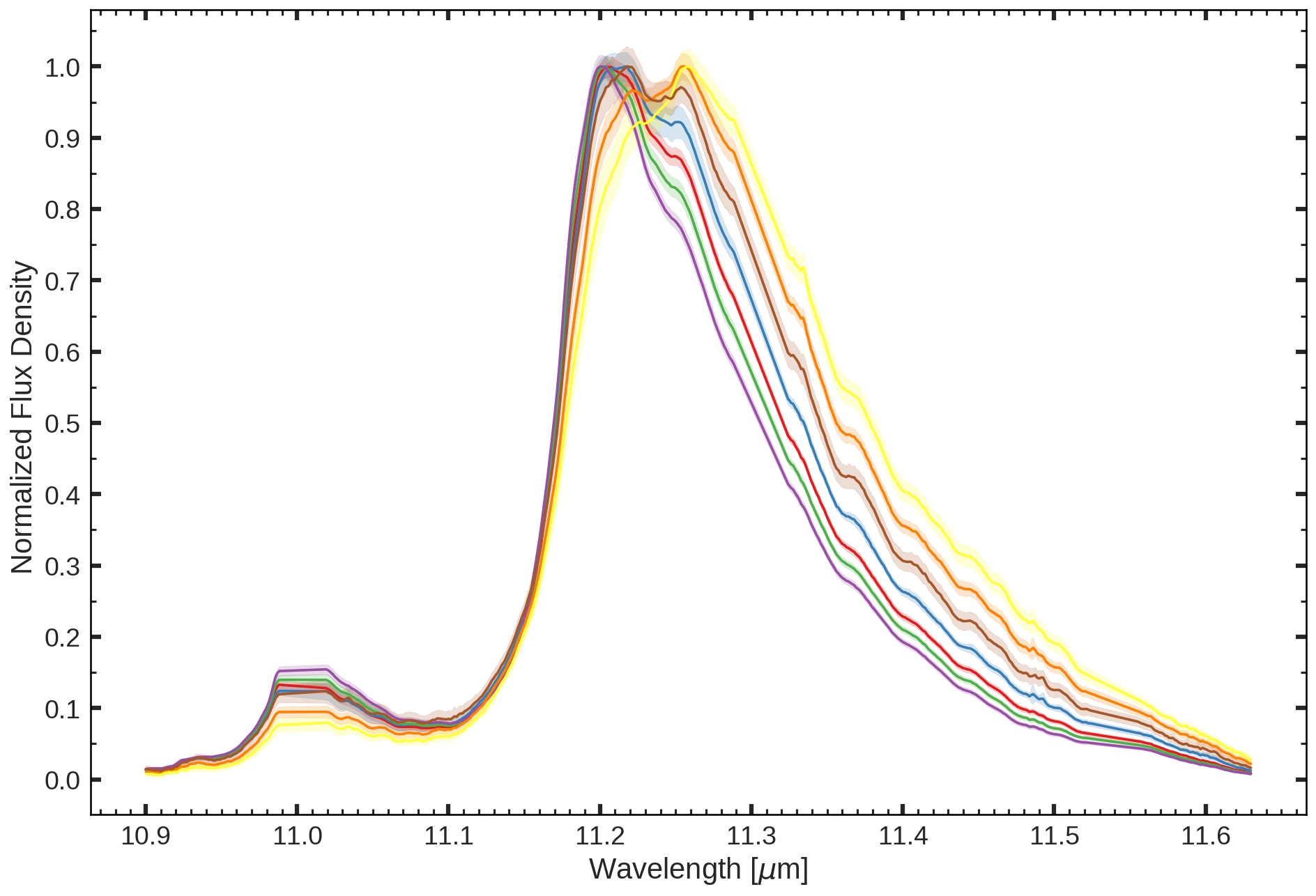}
    \includegraphics[width = 1cm, height = 1.5cm]{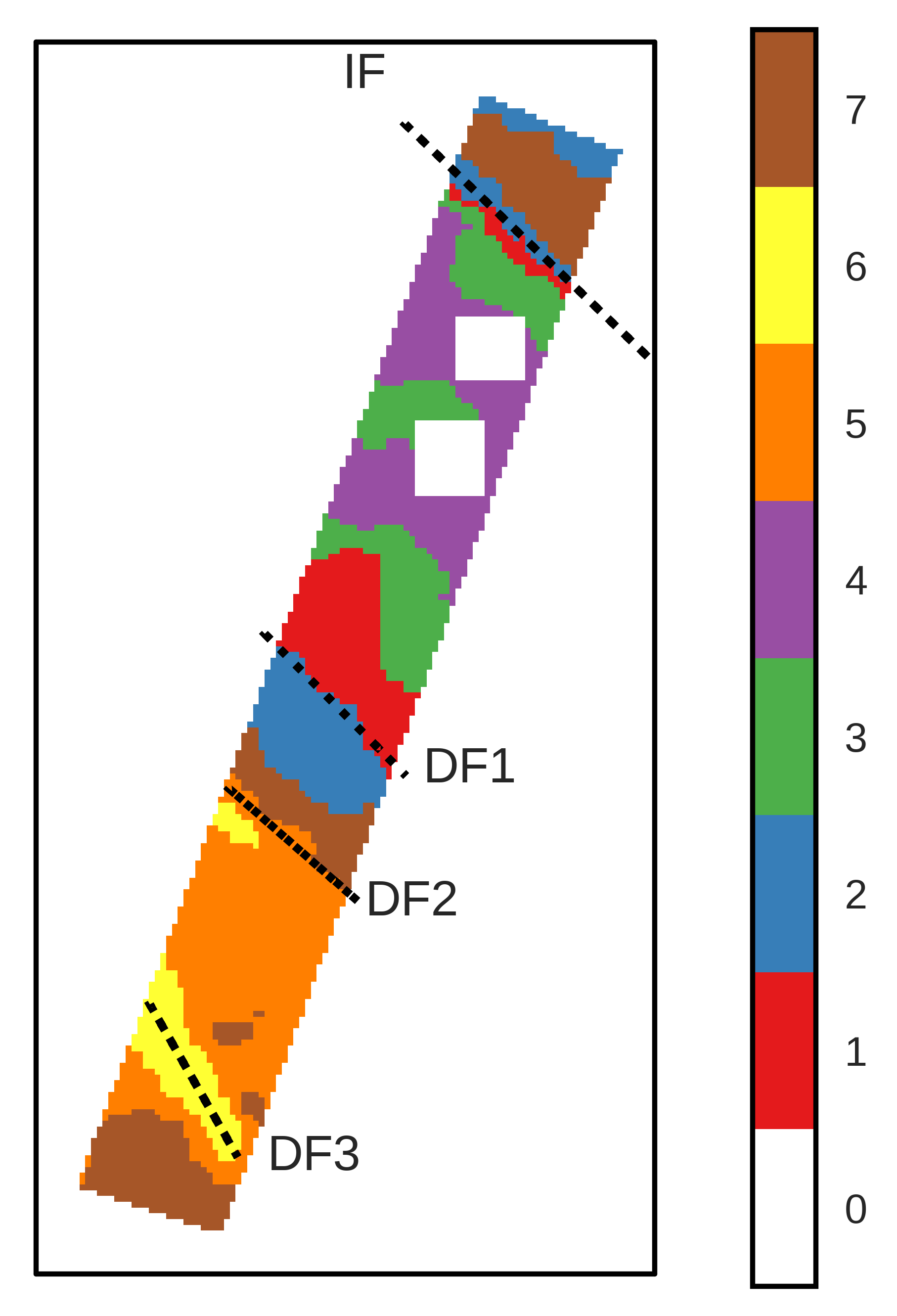}
 }

    \caption{The average spectral profile (left) and spatial footprint (right) for clusters determined in the $10.9-11.63$~\mum region. Each cluster is labelled with a number (in an arbitrary manner). Shading and  normalization (left), and labels and masked pixels (right) are the same as in Fig.~\ref{fig:results_112}.}
    \label{fig:results_112_all}
\end{figure*}

The averaged spectral profiles for clusters calculated on each of the $3.2-3.6$, $5.95-6.6$, $7.25-8.95$ and $10.9-11.63$~\mum wavelength regions, along with the corresponding spatial maps for the cluster zones, are given in Figs.~\ref{fig:results_33_all}, \ref{fig:results_62_all}, \ref{fig:results_79} and  \ref{fig:results_112_all}, respectively.

\begin{figure*}
    \centering
    \resizebox{.9\hsize}{!}{%
    \includegraphics[width = 3cm, height = 1.5cm]{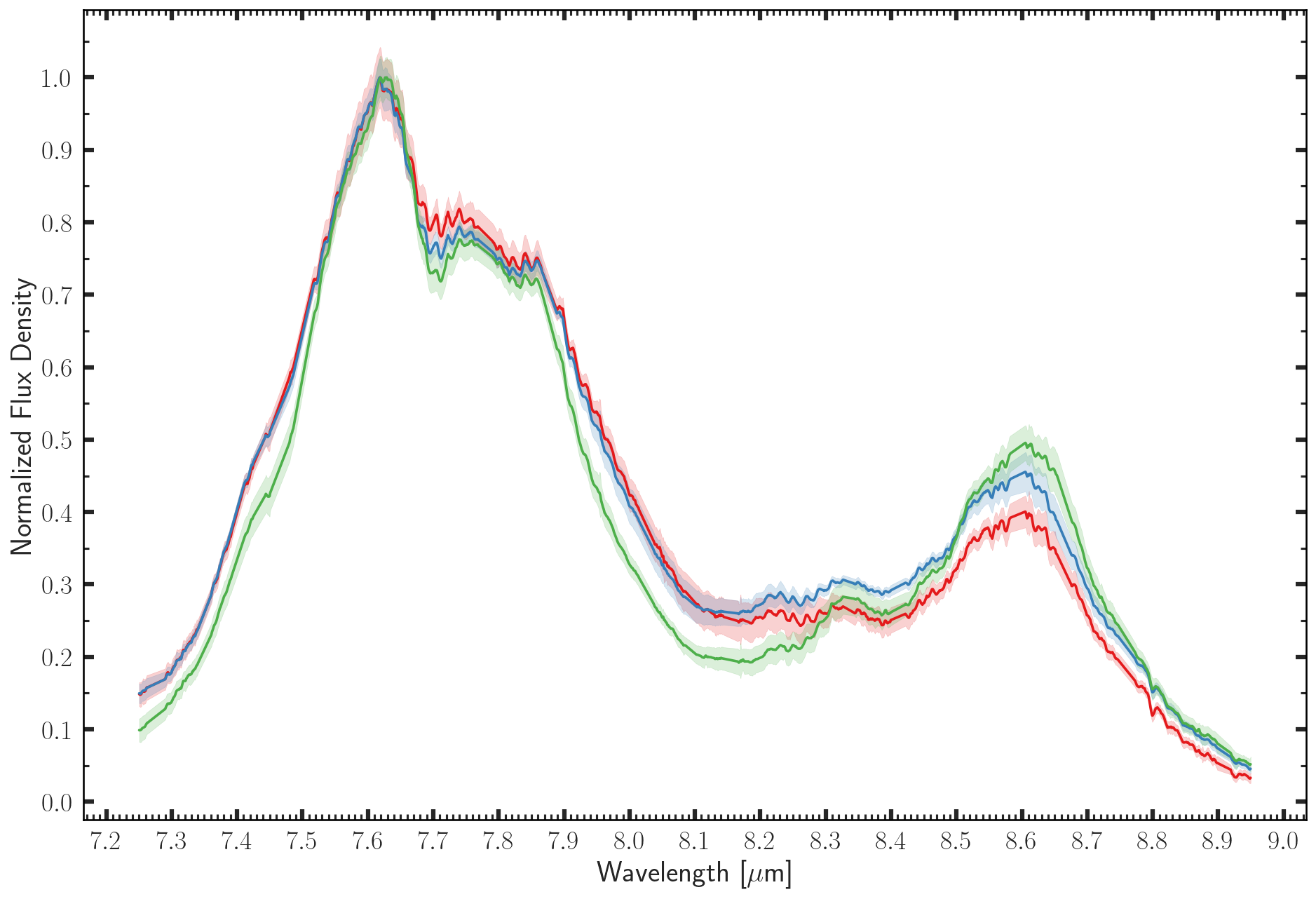}
    \includegraphics[width = 1cm, height = 1.5cm]{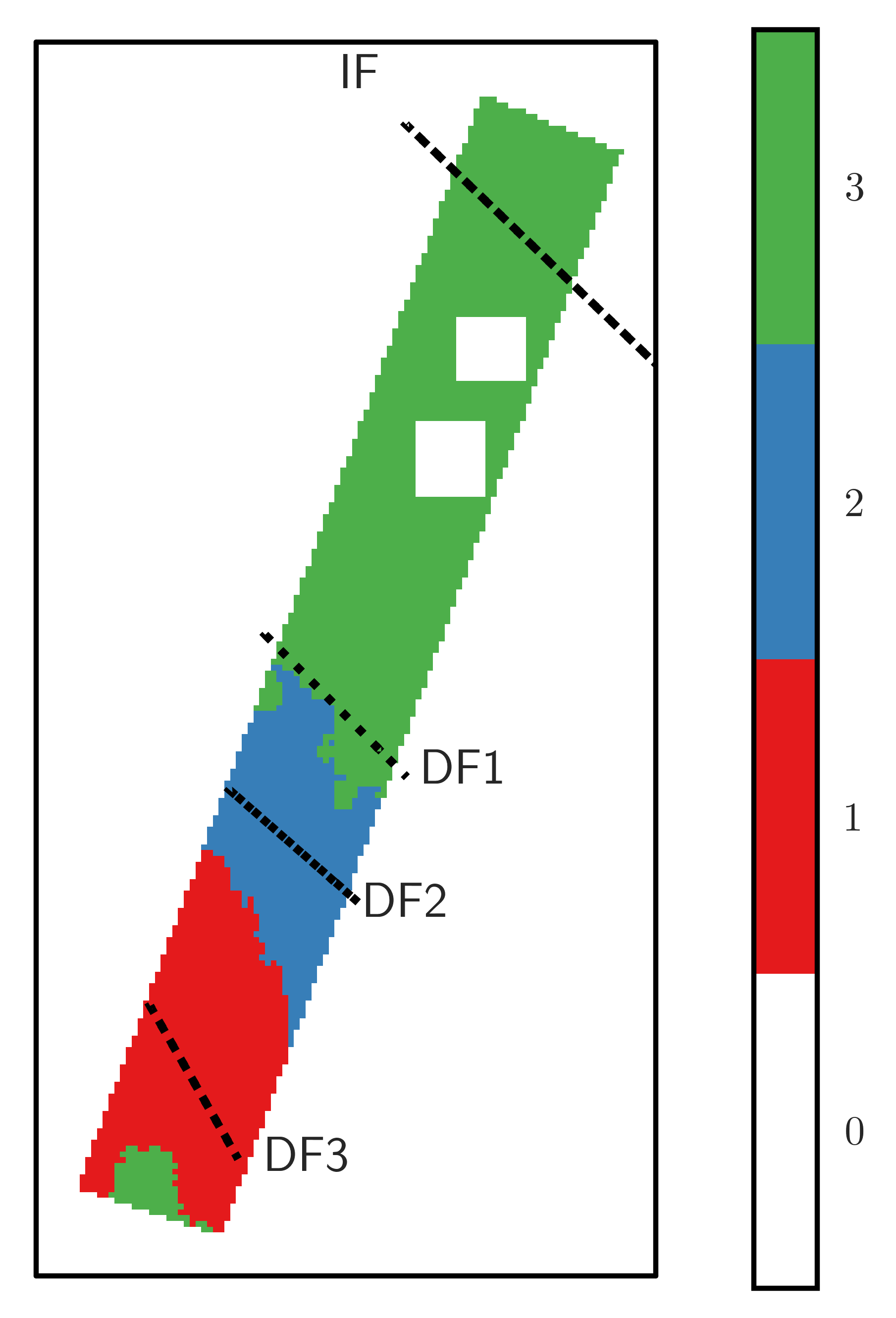}
 }
    \resizebox{.9\hsize}{!}{%
    \includegraphics[width = 3cm, height = 1.5cm]{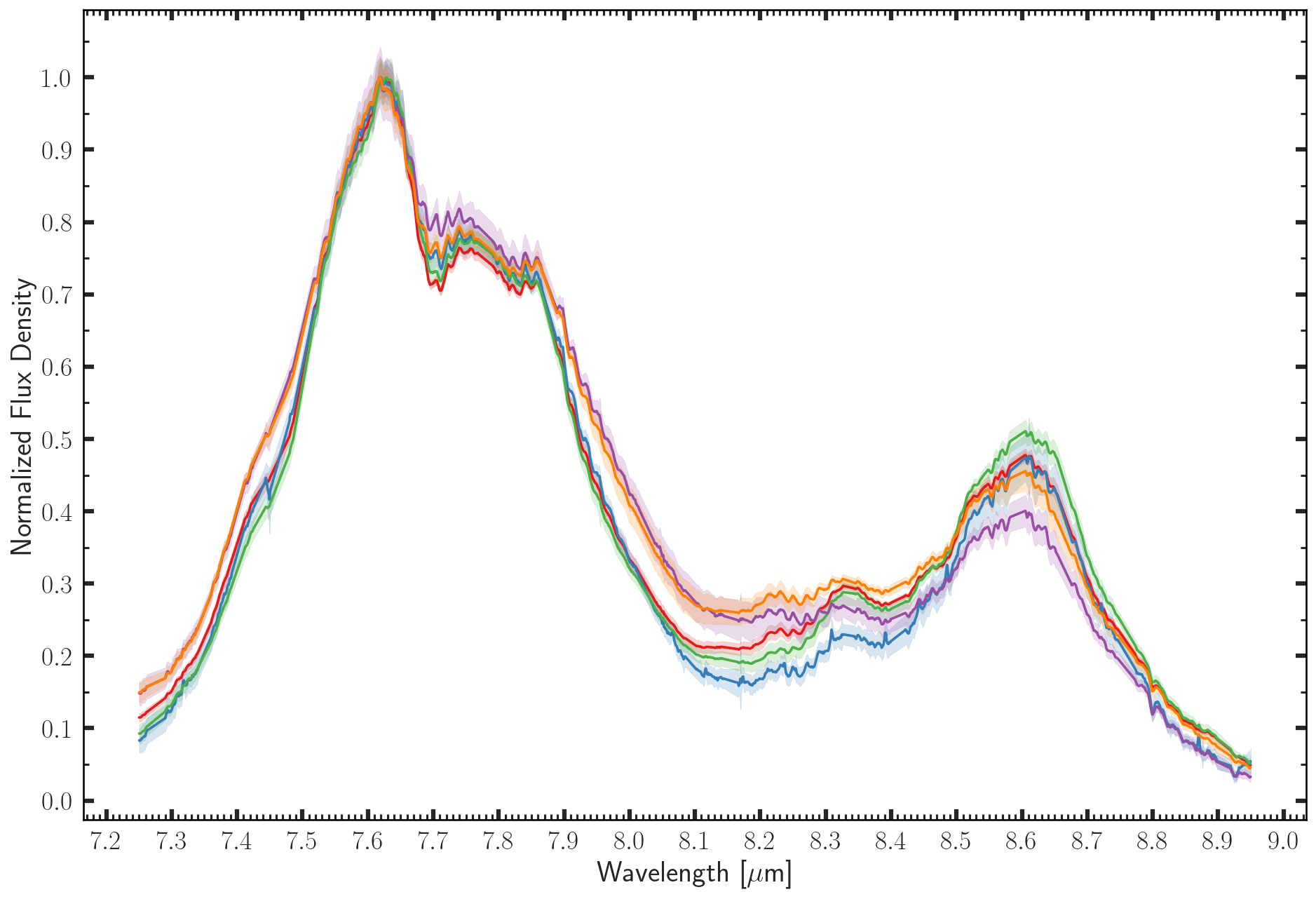}
    \includegraphics[width = 1cm, height = 1.5cm]{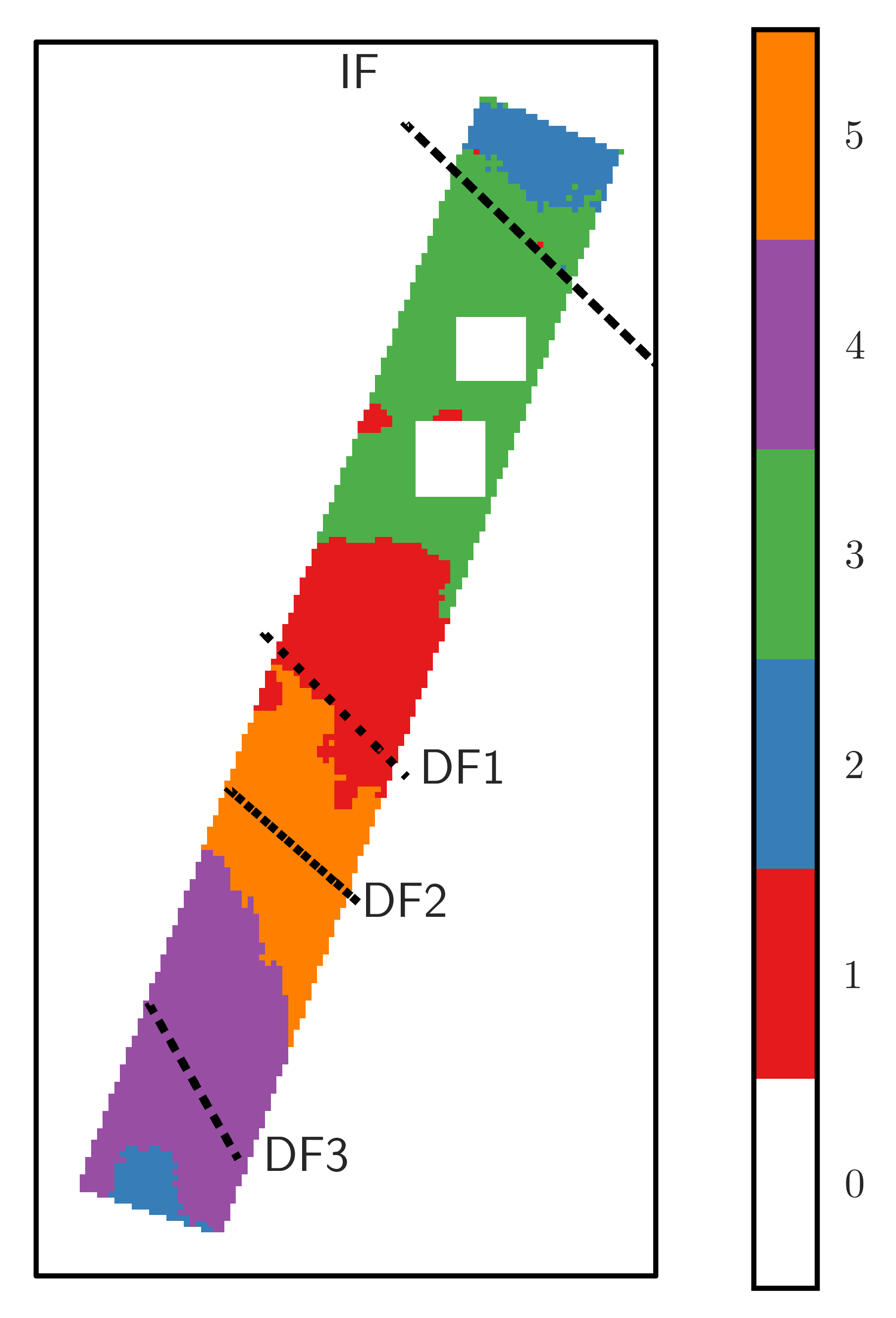}
 }
    \resizebox{.9\hsize}{!}{%
    \includegraphics[width = 3cm, height = 1.5cm]{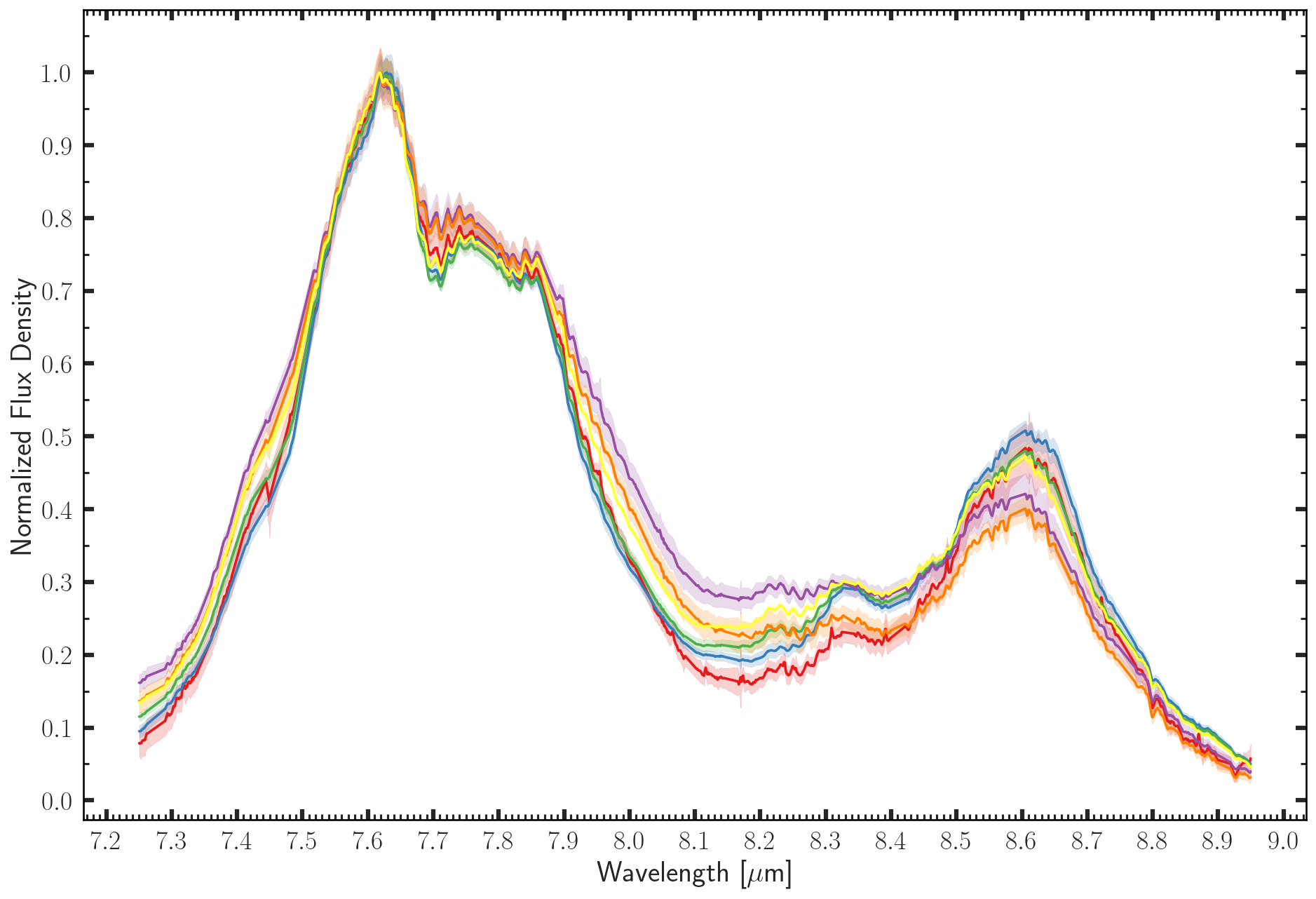}
    \includegraphics[width = 1cm, height = 1.5cm]{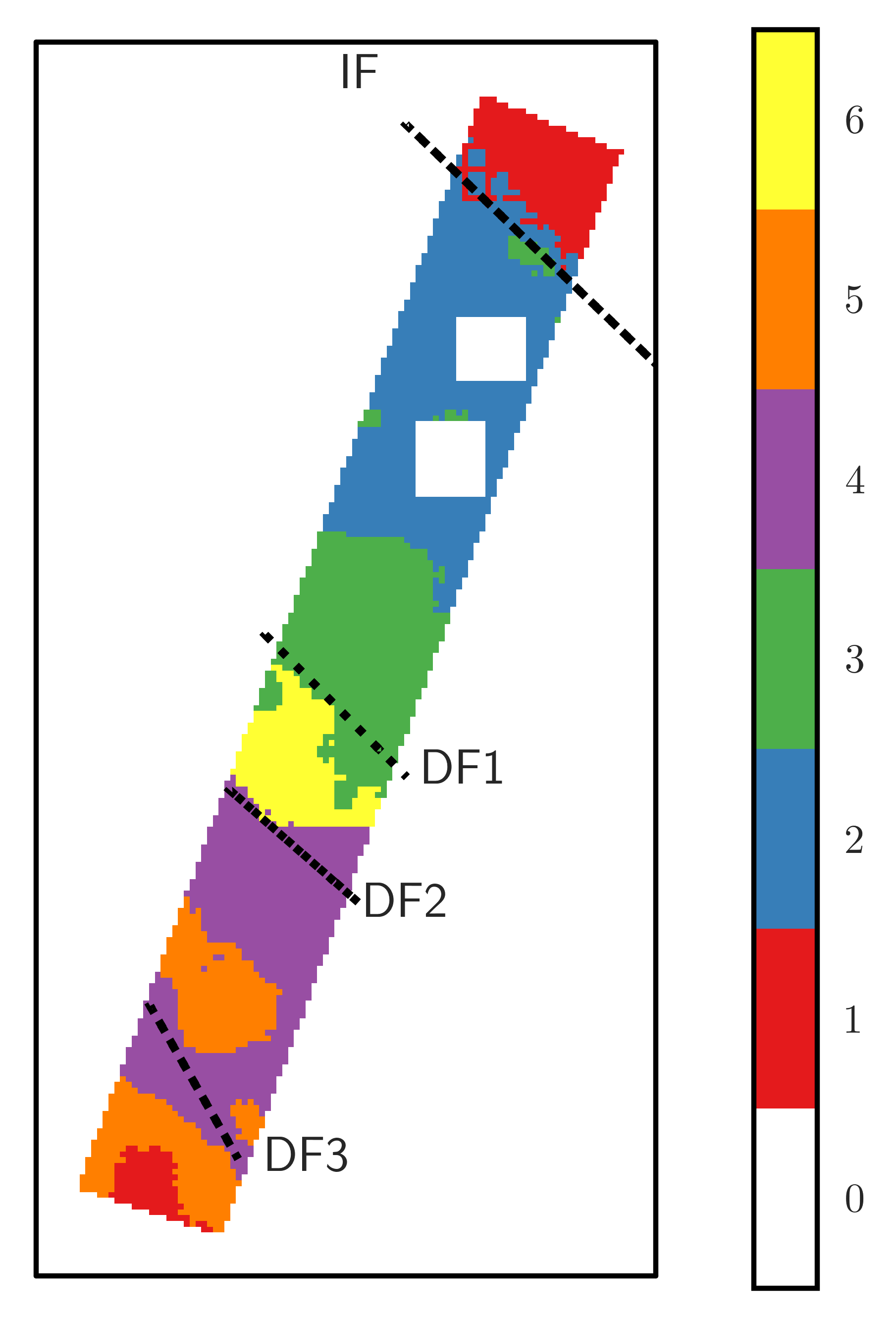}
 }
    \resizebox{.9\hsize}{!}{%
    \includegraphics[width = 3cm, height = 1.5cm]{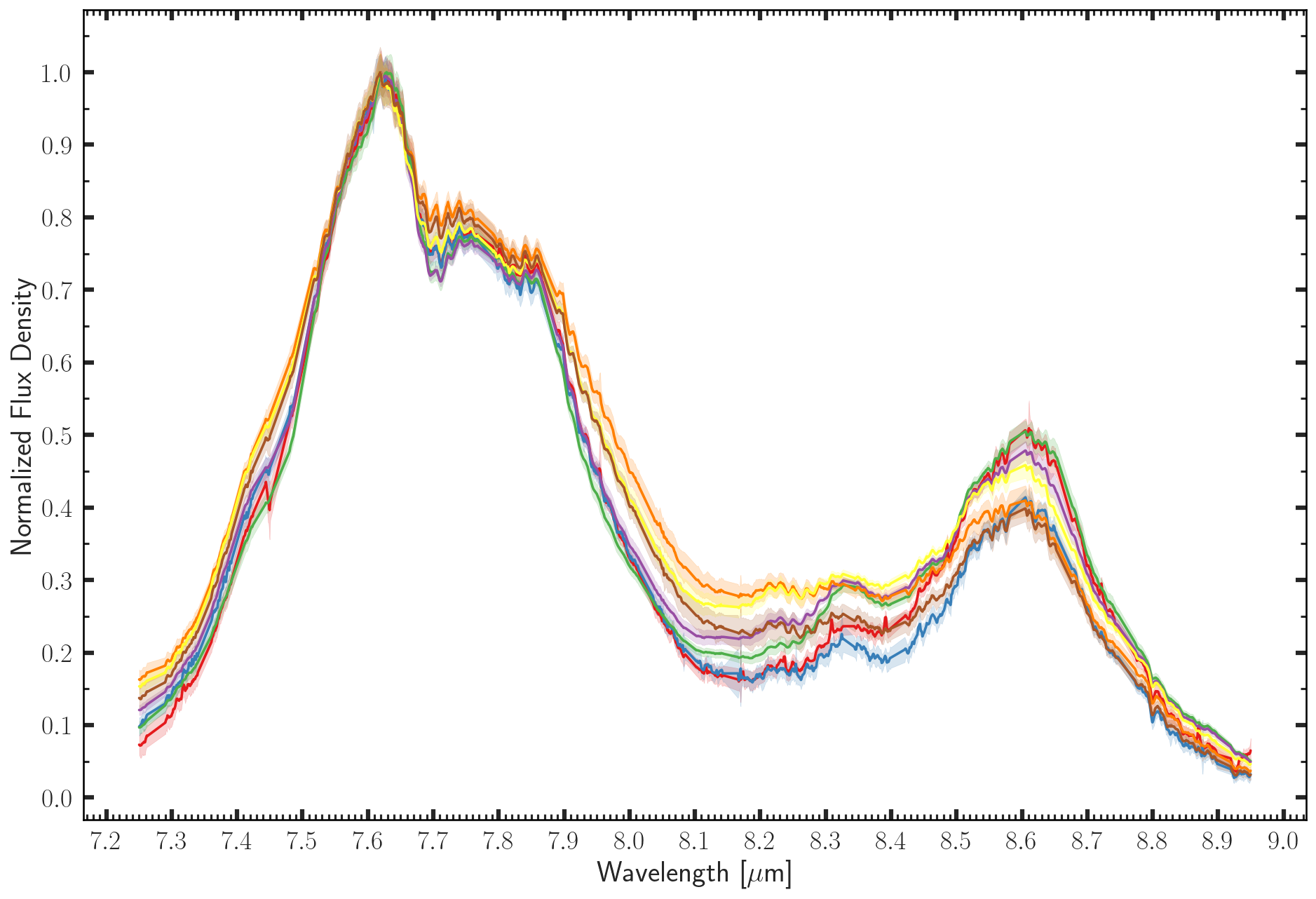}
    \includegraphics[width = 1cm, height = 1.5cm]{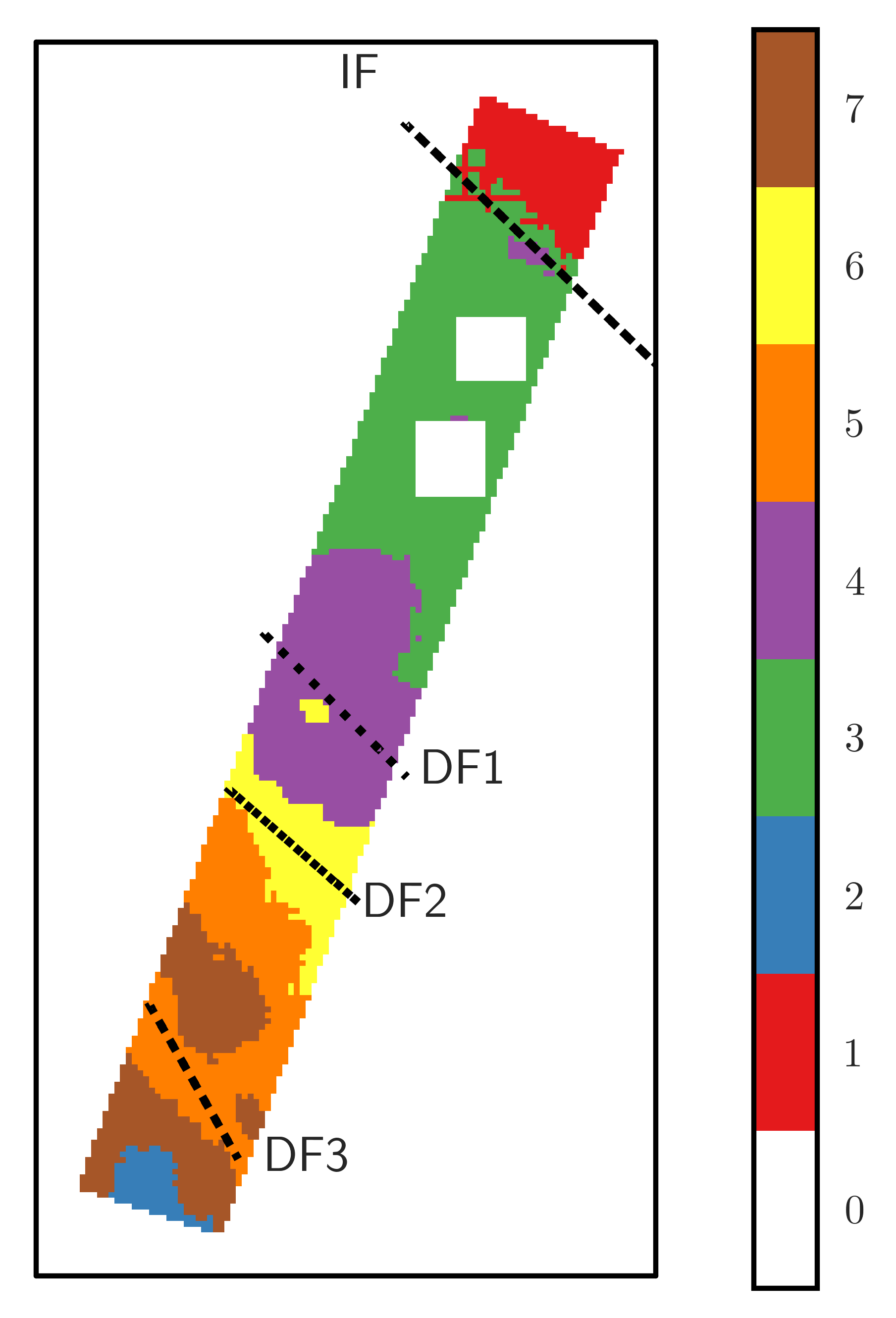}
 }

    \caption{The average spectral profile (left) and spatial footprint (right) for clusters determined in the $7.25-8.95$~\mum region. Each cluster is labelled with a number (in an arbitrary manner). Shading and  normalization (left), and labels and masked pixels (right) are the same as in Fig.~\ref{fig:results_79}.}
    \label{fig:results_79_all}
\end{figure*}

\begin{figure*}[t]
  \centering
    \resizebox{.9\hsize}{!}{%
    \includegraphics[width = 1.5cm, height = 2cm]{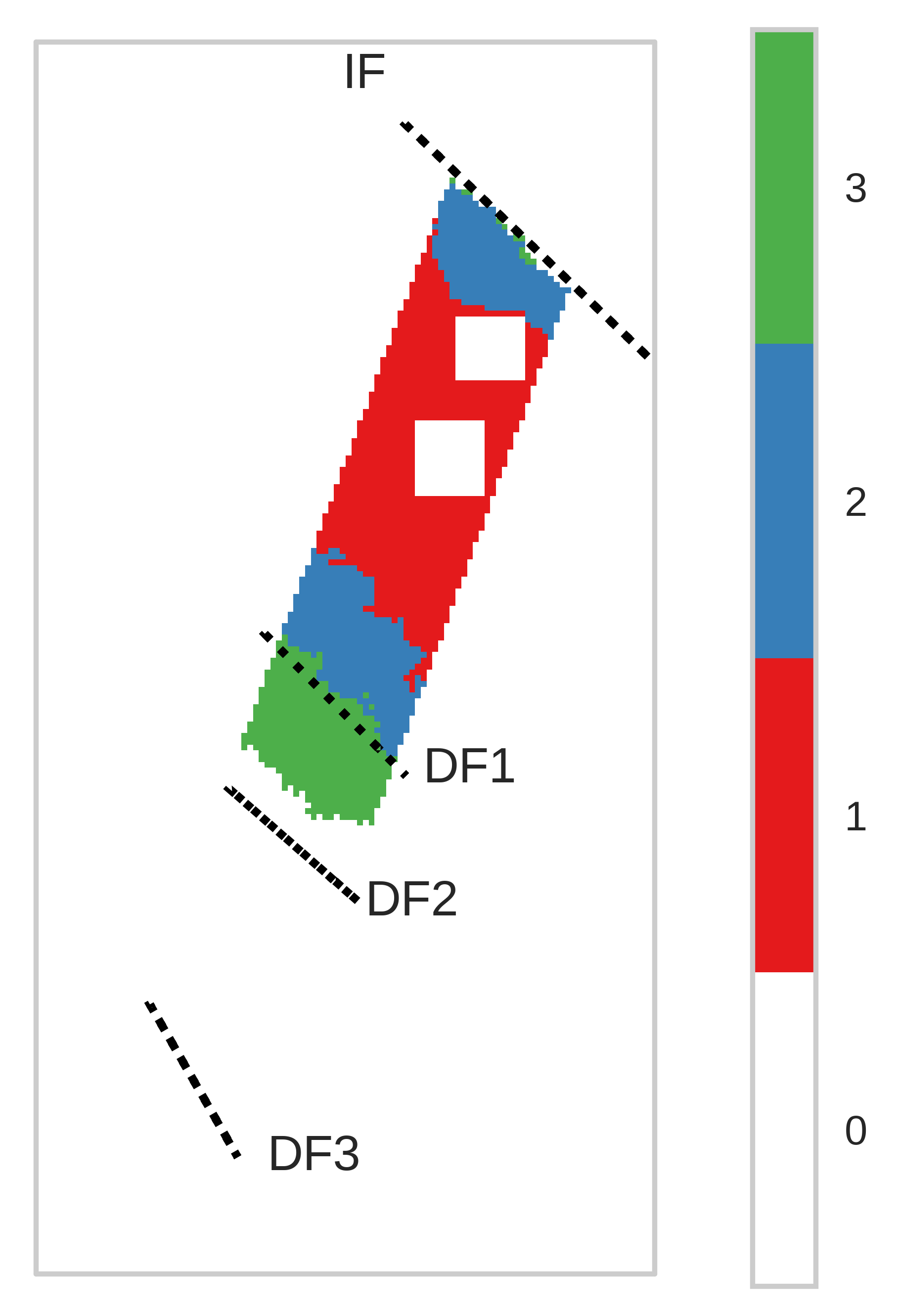}
    \includegraphics[width = 1.5cm, height = 2cm]{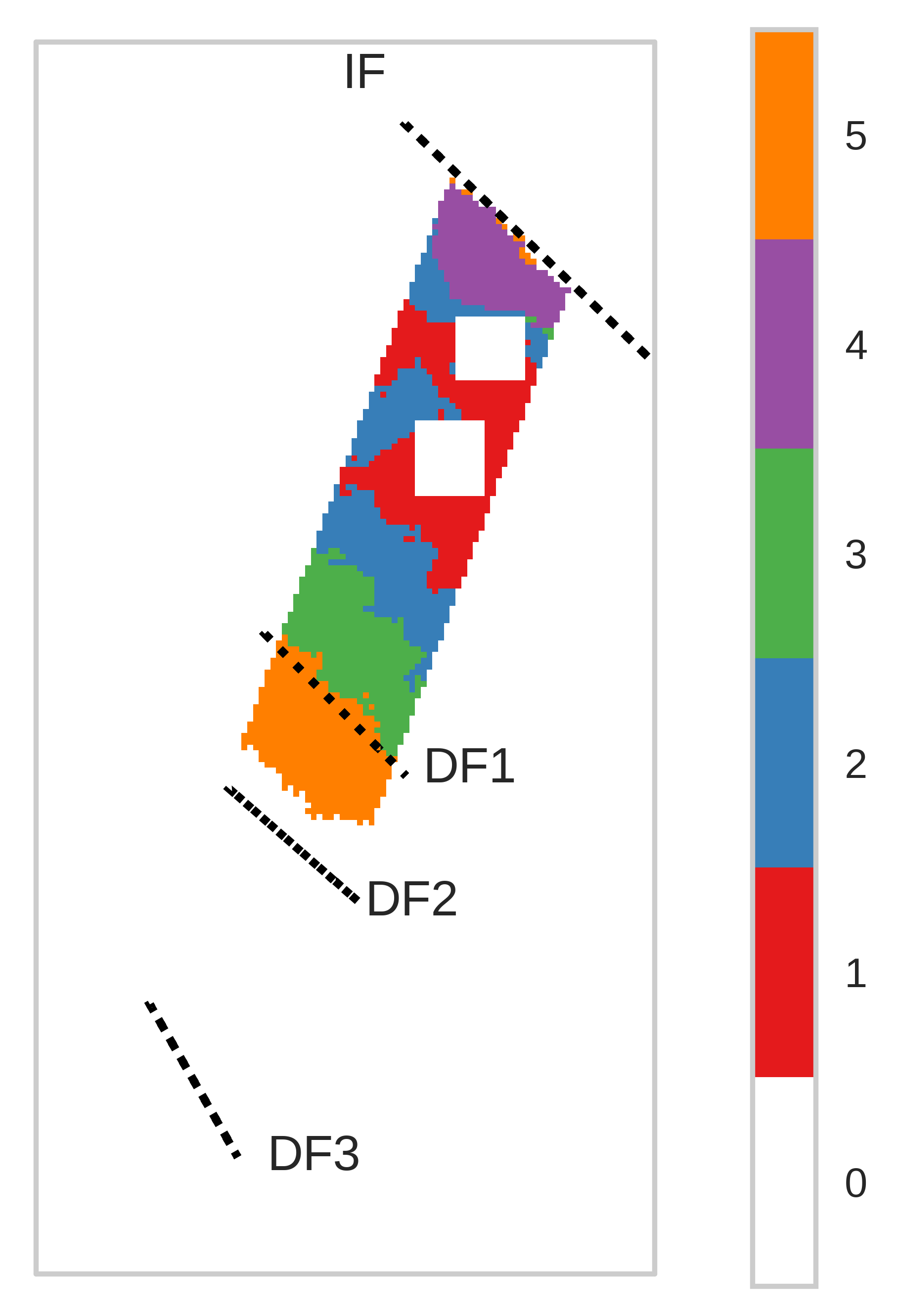}
    }
    \resizebox{.9\hsize}{!}{%
    \includegraphics[width = 1.5cm, height = 2cm]{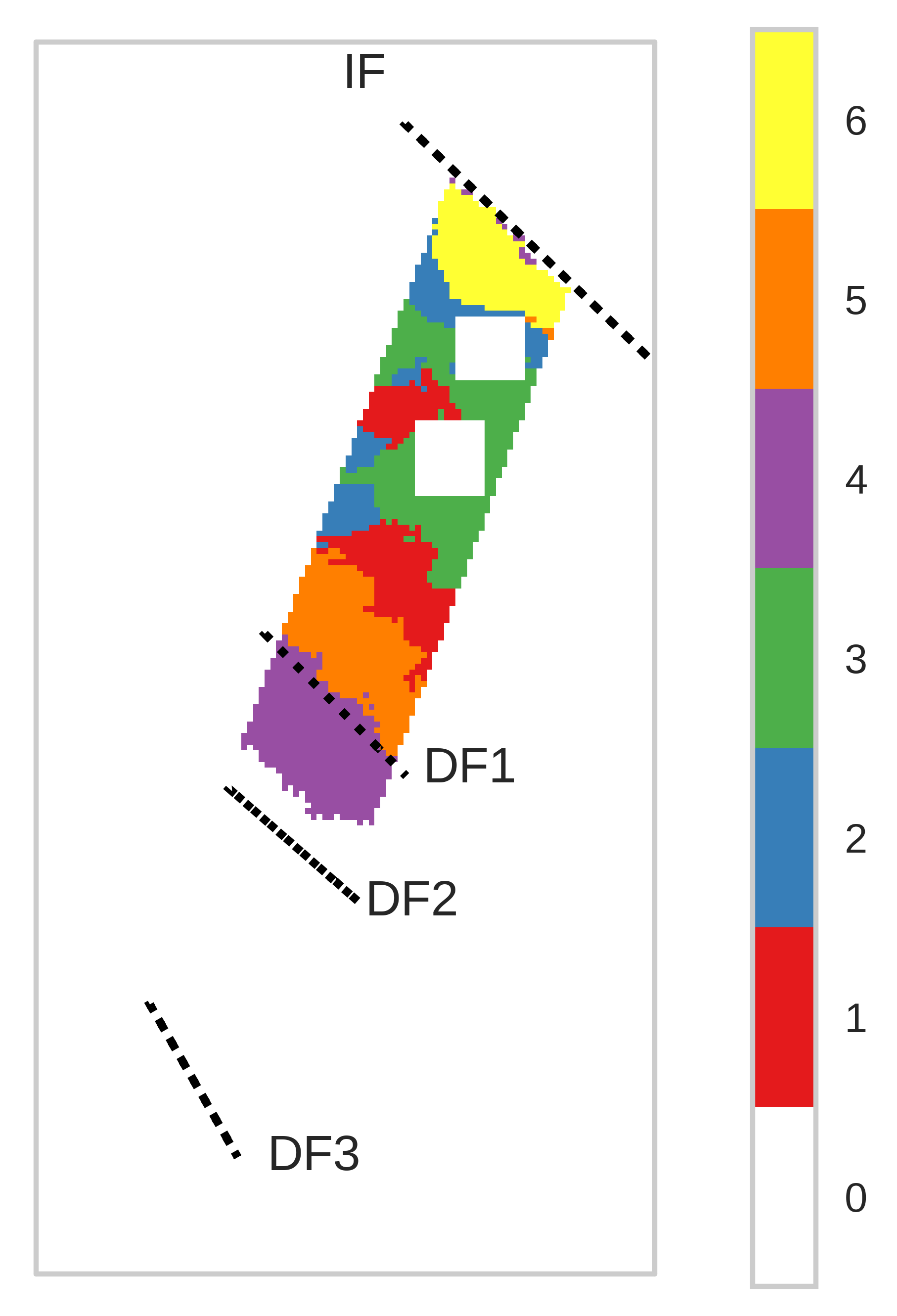}
    \includegraphics[width = 1.5cm, height = 2cm]{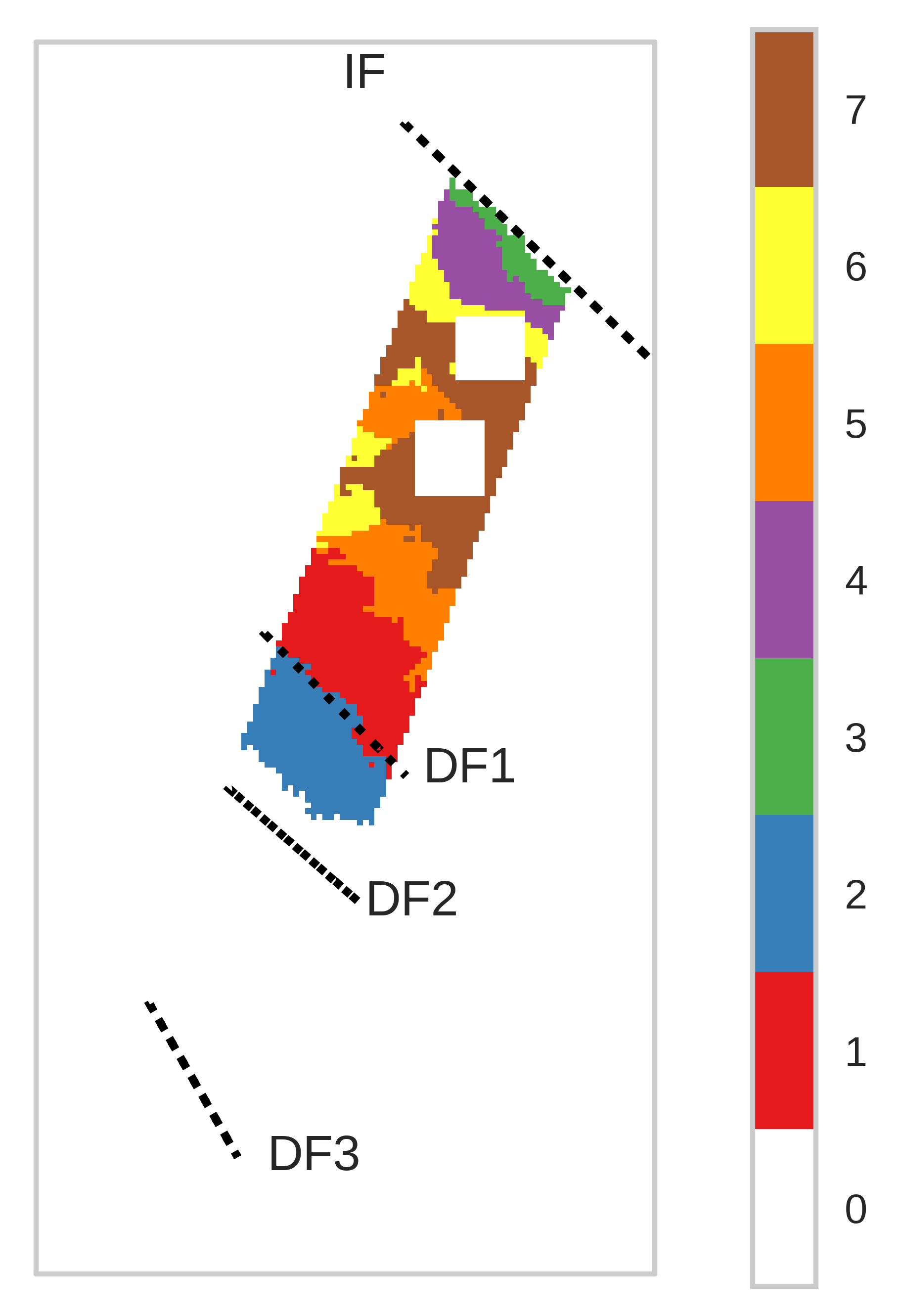}
    }
\caption{The spatial footprint of the MIRI MRS FOV is color-coded by cluster assignment for 3 (top left), 5 (top right), 6 (bottom left) and 7 (bottom right) clusters applied to the $4.9-13.2$~\mum region. Cluster assignment is calculated on the secondary round of clustering applied only to those pixels belonging to cluster 3 in the initial round of clustering (see Fig.~\ref{fig:double_clustering}). Cluster and FOV labels are the same as in Fig.~\ref{fig:double_clustering}.}
\label{fig:double_clustering_all}
\end{figure*}

\begin{figure*}
    \centering
    \resizebox{.9\hsize}{!}{%
    \includegraphics[width = 3cm, height = 2cm]{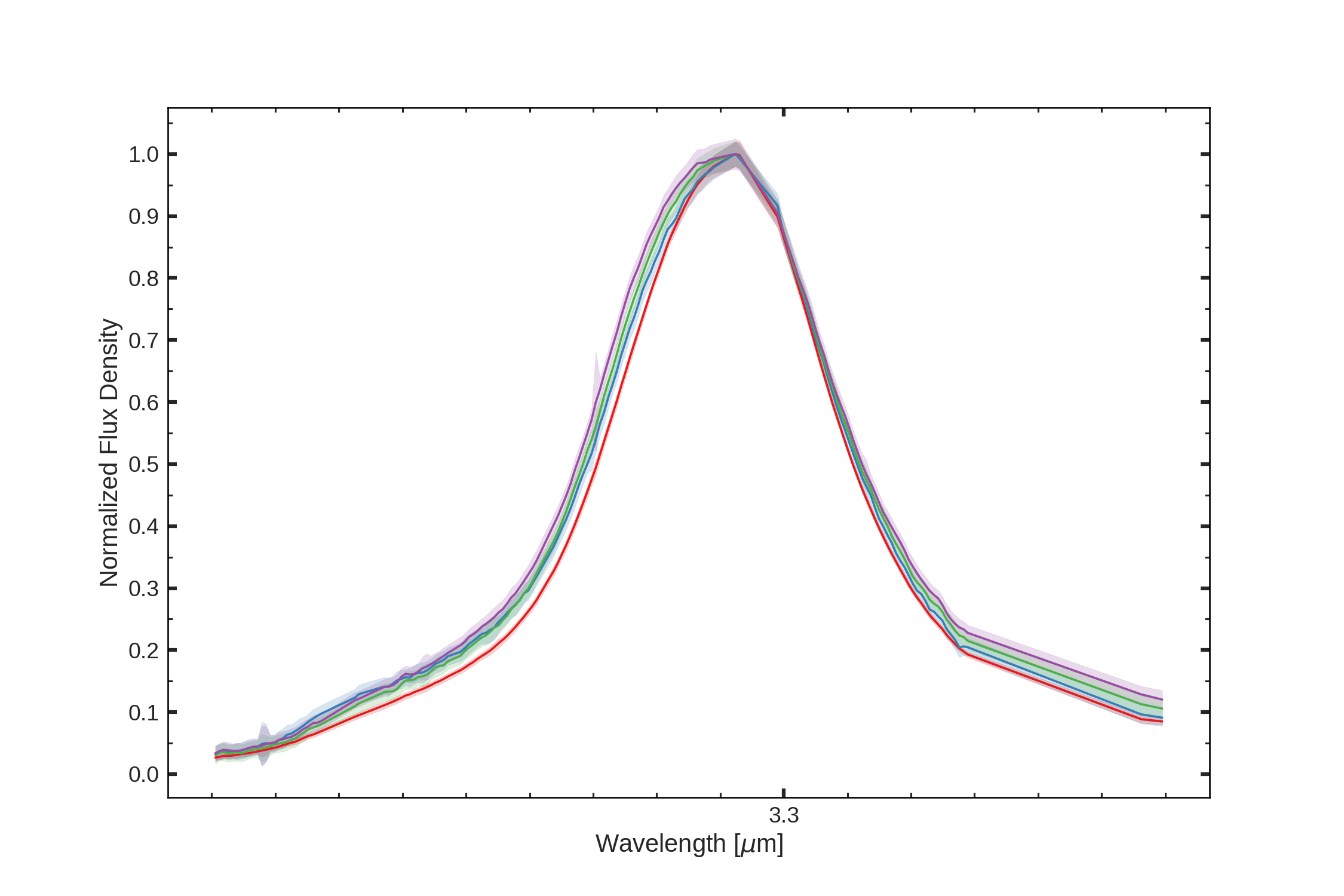}
    \includegraphics[width = 1cm, height = 1.7cm]{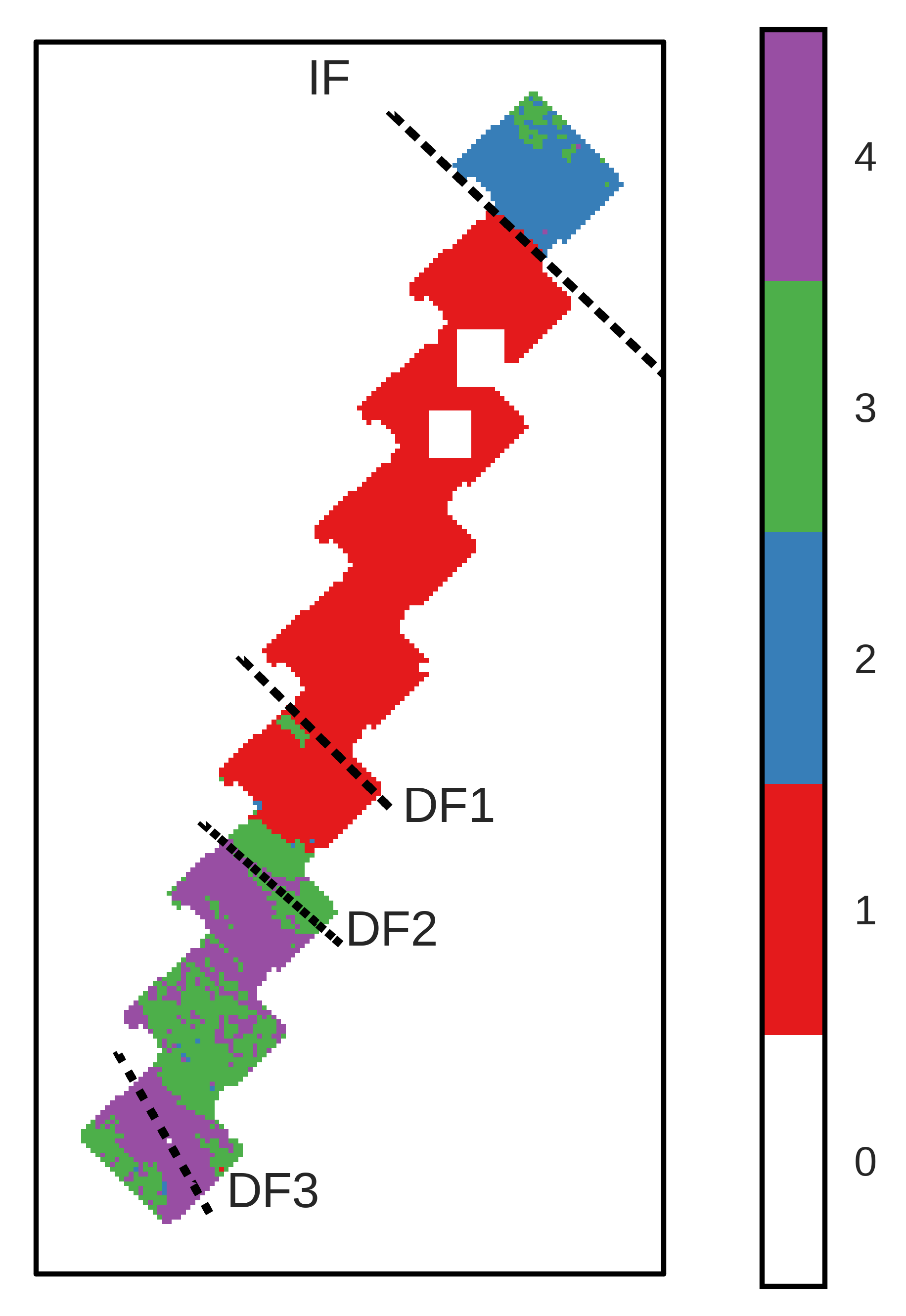}
 }

     \caption{The average spectral profile (left) and spatial footprint (right) for clusters determined in the $3.2-3.36$~\mum region. Each cluster is labelled with a number (in an arbitrary manner). Shading and  normalization (left), and labels and masked pixels (right) are the same as in Fig.~\ref{fig:results_33}.}
    \label{fig:33_exclusive}
\end{figure*}

The mean spectral profile and corresponding cluster zone map for the clustering application on the $3.29$~\mum feature (from $3.2-3.36$~\mum) are given in Figure \ref{fig:33_exclusive}.\par 

\section{Further Comparison of Clustering Results}\label{app:comparisons}

\begin{figure*}
    \centering
    \resizebox{.99\hsize}{!}{%
    \includegraphics[width = 6cm, height = 4cm]{figures/79_average_specs_n_4.png}
    \includegraphics[width = 2.5cm]{figures/79_cluster_zones_n_4.png}
 }
    \resizebox{.99\hsize}{!}{%
    \includegraphics[width = 6cm, height = 4cm]{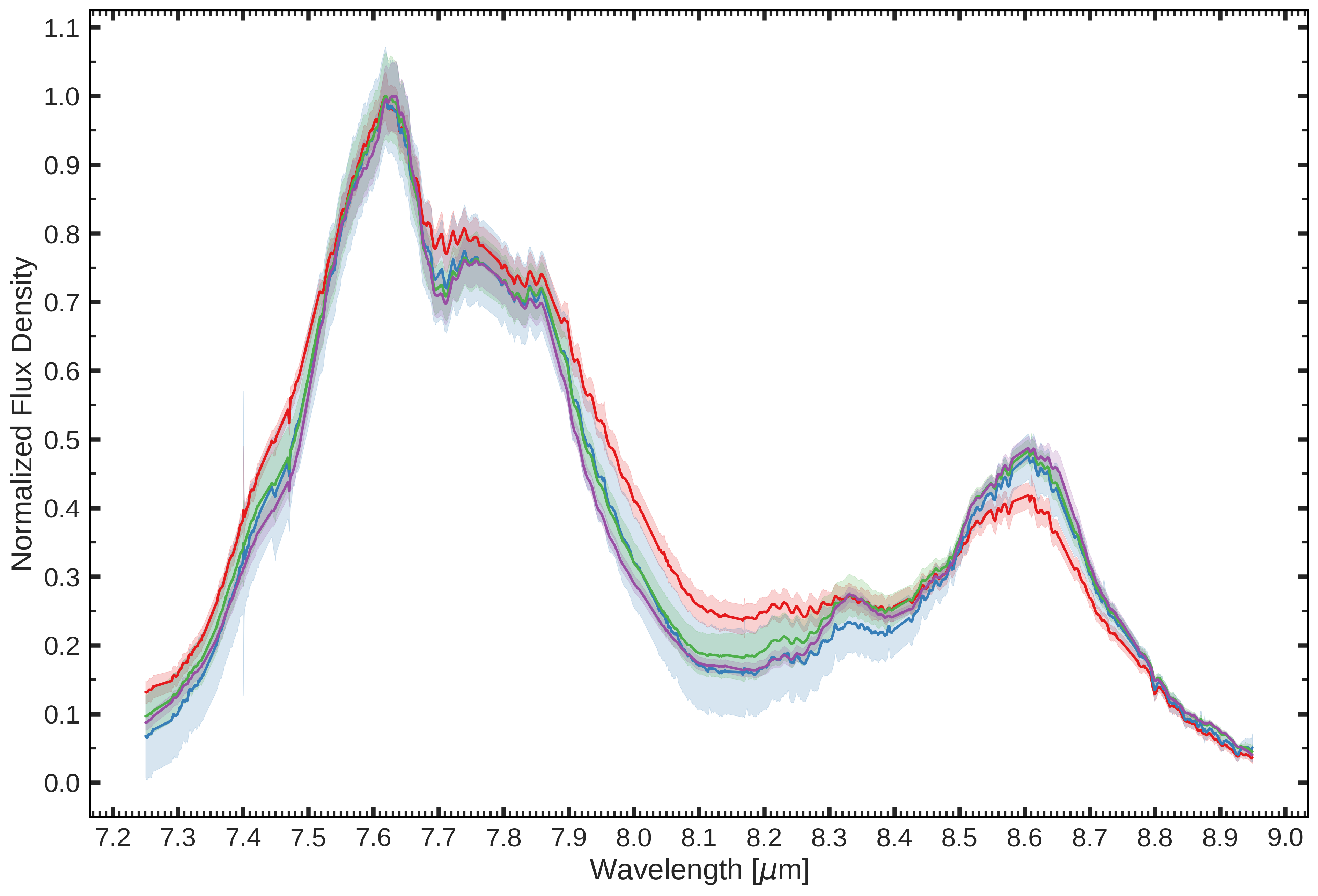}
    \includegraphics[width = 2.5cm]{figures/112_cluster_zones_n_4.png}

}
 
    \caption{The average spectral profiles (left) for the $7.25-8.95$~\mum region obtained from clustering assignment based on the $7.25-8.95$~\mum region and the corresponding cluster zones (right) are given in the top two figures. The average spectral profiles (left) for the $7.25-8.95$~\mum region obtained from clustering assignment based on the $10.9-11.63$~\mum region and the corresponding cluster zones (right) are given in the bottom two figures. Cluster and FOV labels are the same as in Fig.~\ref{fig:discussion_112_62}.}
    \label{fig:discussion_112_79}
\end{figure*}

\begin{figure*}
    \centering
    \resizebox{.99\hsize}{!}{%
    \includegraphics[width = 6cm, height = 4cm]{figures/33_avg_specs_n4.png}
    \includegraphics[width = 2.5cm]{figures/33_cluster_zones_n_4.png}
 }
    \resizebox{.99\hsize}{!}{%
    \includegraphics[width = 6cm, height = 4cm]{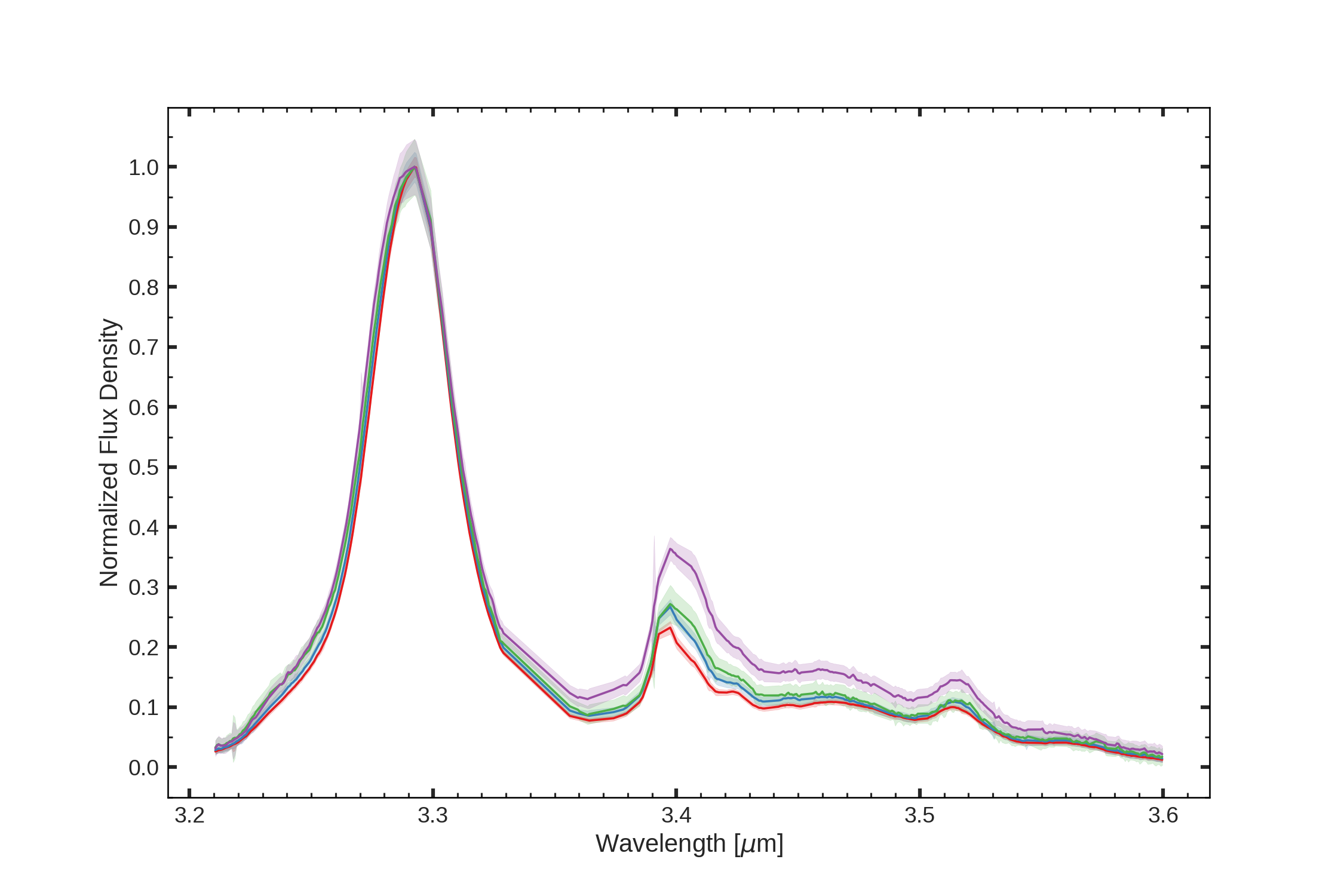}
    \includegraphics[width = 2.5cm]{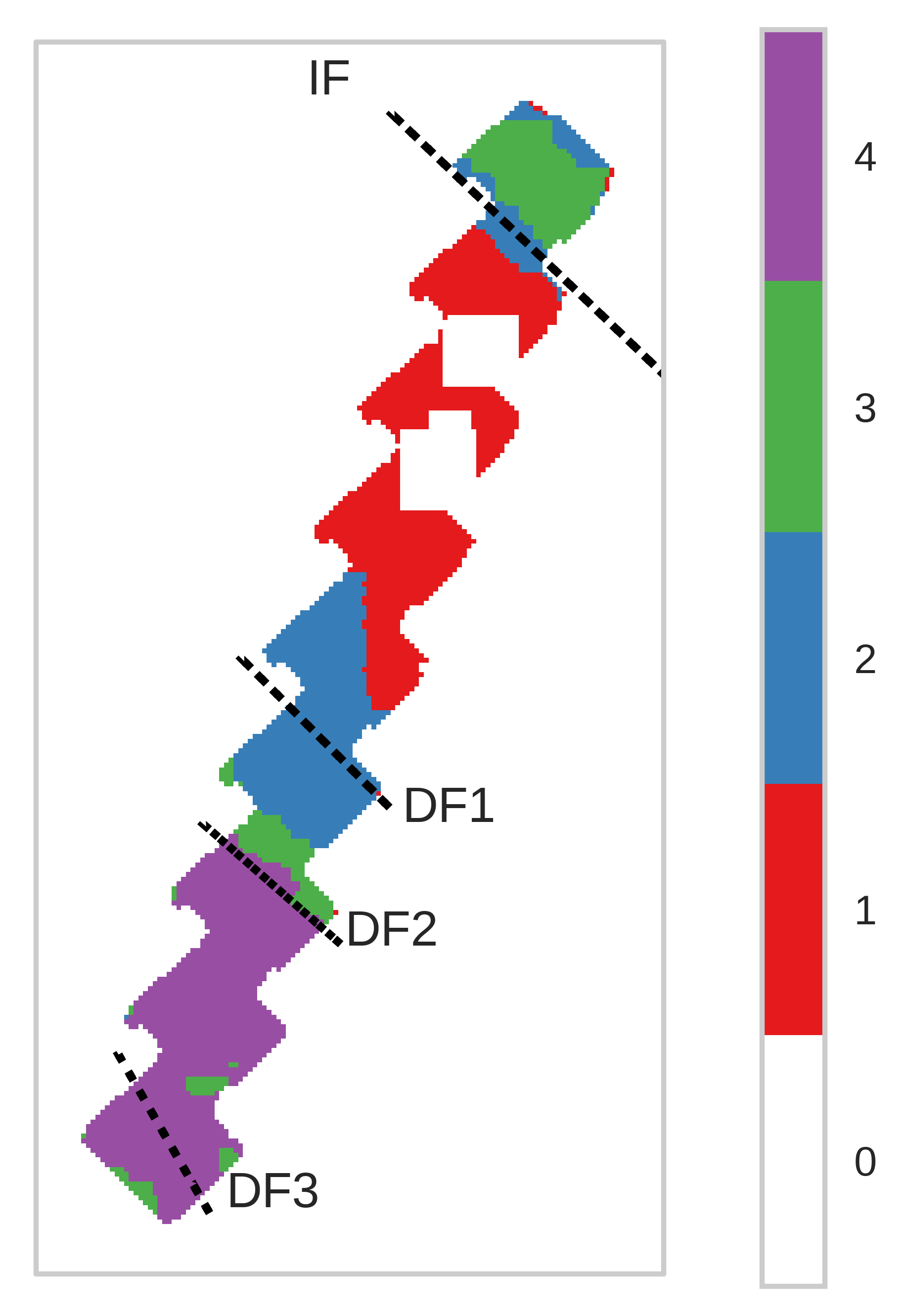}

}
 
    \caption{The average spectral profiles (left) for the $3.2-3.6$~\mum region obtained from clustering assignment based on the $3.2-3.6$~\mum region and the corresponding cluster zones (right) are given in the top two figures. The average spectral profiles (left) for the $3.2-3.6$~\mum region obtained from clustering assignment based on the $10.9-11.63$~\mum region and the corresponding cluster zones (right) are given in the bottom two figures. Note that due to differences in the spatial resolution of the NIRSpec and MIRI MRS datastes, there are small discrepancies in the pixel masks, such are apparent in the lower right panel. Cluster and FOV labels are the same as in Fig.~\ref{fig:discussion_112_62}.}
    \label{fig:discussion_112_33}
\end{figure*}

Comparison of the mean spectral profiles and cluster zone maps for each of the $7.25-8.95$ and $3.2-3.6$~\mum regions with those based on clusters from the $10.9-11.63$~\mum region are given in Figures ~\ref{fig:discussion_112_79} and ~\ref{fig:discussion_112_33}, respectively.\par 

\end{appendix}

%-------------------------------------------------------------------

\end{document}